\documentclass[a4paper,11pt]{article}
\pdfoutput=1
\usepackage{jheppub}
\usepackage[T1]{fontenc}
\usepackage{multirow}
\usepackage{booktabs}
\usepackage{mathtools}
\usepackage{adjustbox}
\usepackage{array}
\usepackage{caption}
\usepackage{slashed}
\usepackage{graphicx}
\usepackage{subcaption}
\usepackage{float}
\usepackage{dirtytalk}
\usepackage{amsmath} 
\usepackage{amssymb}
\usepackage{xcolor}
\usepackage{hyperref}

\usepackage{mathtools}

\begin{document}
\title{Complementary signatures of $\alpha-$attractor inflation in CMB and cosmic string Gravitational Waves}
\author[a,b]{Mainak Baidya,}
\author[c]{Anish Ghoshal,}
\author[d]{David F. Mota}

\affiliation[a]{Indian Institute of Science Education and Research (IISER) Kolkata, \\ Mohanpur, West Bengal 741246, India}
\affiliation[b]{Harish-Chandra Research Institute (HRI), A CI of Homi Bhabha National Institute,\\ Chhatnag Road, Jhunsi, Prayagraj, Uttar Pradesh 211019, India }
\affiliation[c]{ Institute of Theoretical Physics, Faculty of Physics, \\ University of Warsaw,
ul. Pasteura 5, 02-093 Warsaw, Poland}
\affiliation[d]{Institute of Theoretical Astrophysics, University of Oslo,\\
P.O. Box 1029 Blindern, N-0315 Oslo, Norway.}

\emailAdd{mainakb1843@gmail.com}
\emailAdd{anish.ghoshal@fuw.edu.pl}
\emailAdd{mota@uio.no}

\abstract{When cosmic strings are formed during inflation, they regrow to reach a scaling regime, leaving distinct imprints on the stochastic gravitational wave background (SGWB). Such signatures, associated with specific primordial features, can be detected by upcoming gravitational wave observatories, such as the LISA and Einstein Telescope (ET). Our analysis explores scenarios in which cosmic strings form either before or during inflation. We examine how the number of e-folds experienced by cosmic strings during inflation correlates with the predictions of inflationary models observable in cosmic microwave background (CMB) measurements. This correlation provides a testable link between inflationary physics and the associated gravitational wave signals in a complementary manner.
Focusing on $\alpha$-attractor models of inflation, with the Polynomial $\alpha$-attractor serving as an illustrative example, we find constraints, for instance, on the spectral index $n_s$ to $0.962 \lesssim n_s \lesssim 0.972$ for polynomial exponent $n=1$, $0.956 \lesssim n_s \lesssim 0.968$ for $n=2$, $0.954 \lesssim n_s \lesssim 0.965$ for $n=3$, and $0.963 \lesssim n_s \lesssim 0.964$ for $n=4$, which along with the GW signals from LISA, are capable of detecting local cosmic strings that have experienced $\sim 34 - 47$ e-folds of inflation consistent with current Planck data and are also testable in upcoming CMB experiments such as LiteBIRD and CMB-S4. 

}
\maketitle

\section{Introduction}\label{sec:intro}


Inflation \cite{Guth:1980zm, PhysRevLett.48.1220, STAROBINSKY198099} is a period of superluminal expansion in the early universe driven by a scalar field, the inflaton. It not only provides a solution to the classical problems of the standard hot Big Bang cosmology, but also provides a compelling framework to address the initial conditions of the cosmos, including the generation of primordial density fluctuations. These fluctuations evolve and imprint themselves as anisotropies in the Cosmic Microwave Background (CMB), which makes the study of these inflationary observables a direct probe of the very early universe~\cite{Aghanim:2018eyx,Akrami:2018odb}. 

A certain framework of inflationary models are often considered as particularly important for future observations of CMBR is known as the $\alpha$ attractors. A key prediction of such classes of inflationary models in this framework is the presence of primordial tensor fluctuations. The ratio of the tensor and scalar power spectra amplitude, which is depicted as $r$ is related to the slow roll parameter $\epsilon$ at the time of the Hubble horizon exit of the modes of interest in CMB measurements. In addition, the scalar power spectrum spectral tilt $n_s$ is determined using a linear combination of $\epsilon$ and $\eta$. This points to a possible hierarchy between the two slow-roll parameters in the theory; inflationary models that can achieve this are only consistent with the present data. Such models correctly predict the observed tilt and are consistent with the absence of observational evidence for CMB BB-modes or primordial Gravitational Waves (i.e. gives only upper bound on $r$).
Motivated by this result, several proposed models are consistent with the present data, for example, the Starobinsky model involving a quadratic $\mathcal{R}^2$ in Ricci scalar\cite{Starobinsky:1980te}, the GL model \cite{Goncharov:1983mw,Goncharov:1984jlb,Linde:2014hfa}, and the Standard Model Higgs inflation \cite{Futamase:1987ua,Salopek:1988qh,Bezrukov:2007ep}. In fact there exist a very large family of different inflationary theories predicting the same characteristic form of observables \cite{Kallosh:2013hoa,Kallosh:2013pby,Kallosh:2013lkr,Ferrara:2013rsa, Kallosh:2013maa, Kallosh:2013tua, Ellis:2013xoa,Buchmuller:2013zfa,Kallosh:2013yoa,Galante:2014ifa,Kallosh:2015zsa,Kumar:2015mfa,Kallosh:2019eeu,Kallosh:2019hzo,Ferrara:2016fwe,Kallosh:2017ced,Gunaydin:2020ric,Kallosh:2021vcf, LiteBIRD:2022cnt, Kallosh:2022feu, Odintsov:2022bpg, Odintsov:2020thl, Odintsov:2016vzz,Bhattacharya:2022akq}. These models consequently also serve as important science targets for future cosmological experiments such as Planck \cite{Aghanim:2018eyx}, SPT-3G \cite{SPT-3G:2021wgf}, CMB Stage 4 \cite{CMB-S4:2016ple}, CMB-Bharat \footnote{CMB-Bharat Collaboration, \url{http://cmb-bharat.in}}, CMB-HD \cite{Sehgal:2019ewc, CMB-HD:2022bsz}, PICO \cite{Alvarez:2019rhd}, LiteBIRD \cite{LiteBIRD:2022cnt} and CORE \cite{CORE:2016ymi, CORE:2017oje}. These models are collectively known as $\alpha$-attractors, where $\alpha$ is a parameter that affects the CMB observables. Hereafter, we consider such models of cosmic inflation throughout the paper and try to establish whether certain degeneracies in the predictions of such models can be broken by using the formation of cosmic strings during inflation.

Any period of inflation is known to dilute away any  cosmological relic that may be present before, such as superheavy massive particles or most types of topological defects, such as monopoles or domain walls~\cite{Guth:1980zm,Linde:1981mu,Albrecht:1982wi}. However, cosmic strings are an exception to this matter~\cite{Cui:2018rwi,Ghoshal:2023sfa,Ferrer:2023uwz,Gouttenoire:2019kij,Datta:2025yow}. Cosmic strings can be fundamental objects~\cite{Copeland:2003bj,Dvali:2003zj,Polchinski:2004ia,Jackson:2004zg,Tye:2005fn} 
or one dimensional topological defect configurations of quantum fields such as those originated from a $U(1)$ symmetry
breaking~\cite{Nielsen:1973cs,Kibble:1976sj}; however, at very large  distances, they are usually characterized almost completely by their energy per unit length, also known as the cosmic string tension (which is determined by the scale of $U(1)$ symmetry breaking: the larger the symmetry breaking scale, the larger the string tension) $\mu$~\cite{Vilenkin:2000jqa}. As usual, the cosmic string network reaches a scaling regime very quickly after its formation; during this period, the total energy density of the network follows the dominant cosmological energy density touted with a relative fraction on the order of $G\mu$~\cite{Albrecht:1984xv,Bennett:1987vf,Allen:1990tv}. Here, $G = 1/8\pi M_{Pl}^2$ is the Newtonian gravitational constant and $M_{Pl}$ is the reduced Planck mass.

Let us carefully understand the scenario in which cosmic strings are created during cosmic inflation. To sustain the scaling regime of the cosmic string network, once attained shortly after formation, energy must be transferred from the network to the radiation species. In the case of local or fundamental strings, this transfer occurs via the emission of gravitational waves through closed string loops, as shown in Refs. ~\cite{Vilenkin:1981bx,Vachaspati:1984gt,Turok:1984cn,Burden:1985md,Olum:1999sg,Moore:2001px,Matsunami:2019fss}.  Therefore, this means that the most promising and intriguing observational signal of such a scaling network of strings can be possibly the stochastic gravitational wave background (SGWB) arising from the combined and unresolved emission of GWs or graviton production by closed string loops over the entire history of cosmic expansion ~\cite{Allen:1996vm,Blanco-Pillado:2013qja,Blanco-Pillado:2017rnf,Blanco-Pillado:2017oxo,Ringeval:2017eww,Abbott:2017mem,Caprini:2018mtu,Auclair:2019wcv,Auclair:2019jip}. 
With this in mind, one may consequently utilize such characteristic primordial features in the GW frequency spectrum from the cosmic string network to probe and test a myriad of inflationary models, just as it acts as a probe of the post-inflationary expansion history of the universe, as detailed in Ref. ~\cite{Cui:2017ufi,Cui:2018rwi,Chang:2019mza,Gouttenoire:2019rtn,Gouttenoire:2019kij,Ghoshal:2023sfa,Datta:2025yow,Ferrer:2023uwz}.

Let us now discuss the situation in which the universe undergoes an exponential increase in size. Initially, the cosmic string number density became exponentially diluted. However, if one closely follows the evolution of energy densities, in the post-inflationary inflation the energy density in
long cosmic strings only falls off as $a^{-2}$, where $a(t)$ is cosmological scale factor in FRW metric, 
which is much slower than the $a^{-4}$ dilution of radiation and $a^{-3}$ of matter that 
are expected to dominate the energy budget of the cosmos later on in the cosmic history timeline.  
What this tells us is the fact that since the radiation and matter dilutes faster than strings, given a sufficient period of time a network of cosmic strings may regrow 
and enters the scaling regime once again.  In such a string scenario, GW spectrum may get suppressed in frequency (moderate or high or ultrahigh) for high scale inflation Ref.~\cite{Guedes:2018afo} and at the lower frequencies probed suitable to be detected by pulsar timing arrays 
such as the PPTA ~\cite{Shannon:2015ect}. This allows large string tensions $G\mu$ to be consistent 
with the existing bounds ~\cite{Kamada:2012ag,Kamada:2014qta,Ringeval:2015ywa}. Recently Ref.~\cite{Gouttenoire:2019kij,Gouttenoire:2019rtn} analysed the impact of an intermediate secondary inflationary era on the SGWB produced from such cosmic strings experiencing inflation.

We will show that for given inflationary models and their corresponding predictions in the CMB, if cosmic strings are produced during inflation, then depending on the exact time of its formation, one may have correlated CMB versus GW predictions with detectable characteristic GW spectral shapes in upcoming GW experiments such as LISA and ET. This also serves as a complementary study of comic strings with respect to CMB and GW detectors on the one hand, while it may lead to the breaking of degeneracies in inflationary model CMB predictions. 

\textit{The paper is organized as follows:} In Section \ref{Sec. 2}, we give an overview of inflationary dynamics, with particular emphasis on the $\alpha$-attractor models of inflation. We describe the T-Model and Polynomial Attractor model in detail, which will be our primary focus in this paper. In Section \ref{CS in inf}, we explore a scenario of cosmic string formation during inflation. We introduced the turning-point frequency within the framework of the VOS model of cosmic strings and derived an explicit expression for it. We also show the impact of T-Model inflation on the gravitational wave spectra of the local and global cosmic strings. In Section \ref{Complementary CS GW discussion}, we discuss our results, and provide a summary and conclusion of our work in Section \ref{Sec. 6}. In Appendix \ref{Sec. 3} we describe in detail the nature of the gravitational wave spectrum from the local and global cosmic strings.

\medskip

\section{Inflationary Models and Predictions} \label{Sec. 2}

In this section, we describe the inflationary dynamics and study key observables, such as the spectral tilt ($n_s$) and tensor-to-scalar ratio ($r$), which encode the nature of primordial fluctuations produced during inflation.

We consider a single-field inflation model driven by a potential $V(\phi)$, where $\phi$ is the inflaton field.
In a homogeneous and isotropic FLRW universe with metric,
\begin{equation}
    ds^2=-dt^2+a^2(t)\delta_{ij}dx^idx^j,
\end{equation}
where $a\equiv a(t)$ is a function of cosmic time called the scale factor, which parameterizes the expansion of the universe. The Friedmann and Klein-Gordon equations govern the evolution of the inflation field, 
\begin{equation}
    H^2=\frac{1}{3M_{\text{Pl}}^2}\left[\frac{\dot{\phi}^2}{2}+V(\phi)\right],
\end{equation}
\begin{equation}
    \ddot{\phi}+3H\dot{\phi}+\frac{dV(\phi)}{d\phi}=0,
\end{equation}
where dots denote derivatives with respect to cosmic time and $H$ is the Hubble parameter defined as $H=\frac{\dot{a}}{a}$. 
$M_{Pl}\approx2.43\times10^{18}$ GeV is the reduced Planck’s mass. By definition, an inflationary expansion requires $\ddot{a}>0$ and it is important to determine the duration of this expansion. This is captured in the slow-roll approximation, which is given by two dimensionless parameters \cite{Afzal:2024xci}:
\begin{equation}
    \epsilon_{V}\equiv\frac{M_{\mathrm{Pl}}^2}2\left(\frac{V^\prime(\phi)}{V(\phi)}\right)^2,
\end{equation}
and
\begin{equation}
    \eta_{V}\equiv M_\mathrm{Pl}^2\frac{V^{\prime\prime}(\phi)}{V(\phi)},
\end{equation}
where primes denote derivatives with respect to $\phi$. Inflation lasts as long as $\epsilon_{V}\ll1$ and $|\eta_{V}|\ll1$. The spectral index ($n_s$) and tensor-to-scalar ratio ($r$) are defined using the slow-roll parameters:
\begin{equation} \label{ns pred}
    n_s=1-6\epsilon_{V}+2\eta_{V},
\end{equation}
and
\begin{equation} \label{r pred}
    r=16\epsilon_{V}.
\end{equation}

Both $n_s$ and $r$ are powerful tools for testing the validity of a wide range of inflationary models using observational data. 

In addition to leading-order contributions, one can also define higher-order derivatives. 
The derivatives are taken with respect to the natural logarithm of the wave number $k$, which denotes the different scales of primordial fluctuations.   
\begin{equation}
    n_s^{\prime}\equiv\frac{dn_s}{d\ln k}=16 \epsilon_{V} \eta_{V}-24 \epsilon_{V}^{2}-2 \delta^{2},
\end{equation}
is called the running of the spectral index and,
\begin{equation}
    n_s^{\prime\prime}\equiv\frac{d^2n_s}{d\ln k^2}=-192 \epsilon_{V}^{3}+192 \epsilon_{V}^{2} \eta_{V}-32 \epsilon_{V} \eta_{V}^{2}-24 \epsilon_{V}\delta^{2}+2 \eta_{V}\delta^{2}+2\sigma^{3}
\end{equation}
is the running of the running, where,
\begin{equation}
    \delta^2=M_{\mathrm{Pl}}^4\left(\frac{V^\prime V^{\prime\prime\prime}}{V^2}\right),
\end{equation}
and
\begin{equation}
    \sigma^3=M_{\mathrm{Pl}}^6\left(\frac{(V^\prime)^2 V^{\prime\prime\prime\prime}}{V^3}\right).
\end{equation}

Although the running and running of the running terms are generally small, their precise measurement could provide critical tests for inflationary models, particularly in the context of future high-precision CMB experiments. 

The current observational constraints on $n_s$, $n_s^{\prime}$ and $n_s^{\prime\prime}$ are given by Planck 2018 \cite{Planck:2018vyg}: $n_s=0.964\pm 0.004$ (at $68$\% C.L.), $n_s^{\prime} = 0.001\pm0.009$ and $n_s^{\prime\prime}=0.009\pm0.012.$. The value of the tensor-to-scalar ratio, $r$, is constrained by the Planck, WMAP, and BICEP/Keck Observations \cite{BICEP:2021xfz}: $r_{0.05}<0.036$ (at 95\% C.L.). All these observations were made at the pivot scale, $k_\star=0.05 \mathrm{Mpc}^{-1}$. Computing the slow-roll parameters at a fiducial value of the e-fold, $N_e$, defined as the number of e-folds between the horizon exit of the pivot CMB mode and the end of inflation, allows consistent approximations for $n_s$ and $r$ in cases when the slow roll approximation is valid. $N_e$ is defined as: 
\begin{equation}
    N_e=\frac{1}{M_{Pl}^{2}}\int_{\phi_{\mathrm{end}}}^{\phi}\frac{V}{V^{\prime}}d\phi=\frac{1}{M_{Pl}}\int_{\phi}^{\phi_{\mathrm{end}}}\frac{|d\phi|}{\sqrt{2\epsilon_{V}}}
\end{equation}
where $\phi_{end}$ denotes the value of the inflaton field at the end of inflation when $\epsilon=1$.

The slow-roll parameters calculated at pivot scale $k_*$ are given as:
\begin{equation}
    n_s=1-6\epsilon_*+2\eta_*,r=16\epsilon_*,\alpha=16\epsilon_*\eta_*-24\epsilon_*^2-2\xi_*^2.
\end{equation}
The subscript $*$ represents the values of the parameters at the pivot scale $k_* = 0.05$$Mpc^{-1}$. 
The number of e-foldings for the pivot scale $k_* = 0.05$$Mpc^{-1}$ can be written as \cite{Liddle:2003as}:
\begin{equation}
    N_*\simeq61.5+\frac{1}{2}\mathrm{ln}\frac{\rho_*}{M_{\mathrm{Pl}}^4}-\frac{1}{3(1+\omega_{reh})}\mathrm{ln}\frac{\rho_{end}}{M_{\mathrm{Pl}}^4}+\left(\frac{1}{3(1+\omega_{reh})}-\frac{1}{4}\right)\mathrm{ln}\frac{\rho_{reh}}{M_{\mathrm{Pl}}^4}.
\end{equation}
where, $\rho_{*}=V(\phi_{*})$ is the energy density of the Universe when the pivot scale exits the horizon, $\rho_{end}=V(\phi_{end})$ is the energy density at the end of inflation and $\rho_{reh}=(\pi^{2}/30)g_{*}T_{reh}^{4}$ is the energy
density at the reheating time and $\omega_{reh}$ is the effective equation-of-state parameter from the end of inflation until reheating. The effective number of massless degrees of freedom during the reheating epoch is taken to be $g_{*}=106.75$.  
In our analysis, we will consider instant reheating where $\omega_{reh}=1/3$\footnote{Instantaneous reheating is where all the energy of the potential is converted into radiation very quickly. A simple condition can be understood for the assumption behind instantaneous reheating in terms of the parameters that $\tau_\phi \leq t_p$, where $t_p$ is the age of the universe during the reheating, leads to $\varphi_{\rm end} \geq O(10^{18})$ GeV or $\rho_{\rm end} \geq O(10^{13})$ GeV, which is typical energy scales of inflation at the end, for $m_\phi=0.1 \varphi_{\rm end}$. Values of $\varphi_{\rm end}$ smaller than these should be analyzed in more detail, and may deviate from instantaneous reheating which is beyond the scope of the present paper. Eqn (2.14) still holds to be true a value different from $w_{\rm reh} = 1/3$ will change the no. of e-folds which predict eventually the exact GW spectrum shape since it depends on total $N_{\rm efolds}$ suffered by the cosmic string networks. In cases where reheating is not instantaneous $w_{\rm reh}$ represents the average value of the barotropic parameter $w$ over the entire period of reheating.  It is needless to mention that the reheating sector can be completely separate from the inflaton, for instance, in various mechanisms of reheating where a separate species is responsible for reheating. like for examples via mechanisms like gravitational particle production also known as gravitational reheating~\cite{Ford:1986sy,Chun:2009yu} or instant preheating \cite{Felder:1998vq,Dimopoulos:2017tud}, curvaton reheating \cite{Feng:2002nb,BuenoSanchez:2007jxm}, Ricci reheating \cite{Dimopoulos:2018wfg,Opferkuch:2019zbd,Bettoni:2021zhq,Laverda:2023uqv,Figueroa:2024asq}, reheating via evaporation of primordial black holes \cite{Dalianis:2021dbs,Bhaumik:2022pil,Bhaumik:2022zdd,Ghoshal:2023fno}.}.

A diverse array of inflationary models satisfies these constraints, certain models involving UV-completions with the prospects of unifying high-energy physics with cosmological observations. In the next section, we look into a particular class of such models—the $\alpha$-attractors \cite{Kallosh:2013yoa, Kallosh:2022feu} which include the so-called E-Model, T-Model \cite{Kallosh:2013yoa}, and Polynomial $\alpha$-Attractors \cite{Kallosh:2022feu, Bhattacharya:2022akq}. 

\subsection{$\alpha$-attractors}

\begin{figure}[!h] 
    \centering
    \begin{subfigure}[t]{0.495\textwidth} 
        \centering
        \includegraphics[width=\linewidth]{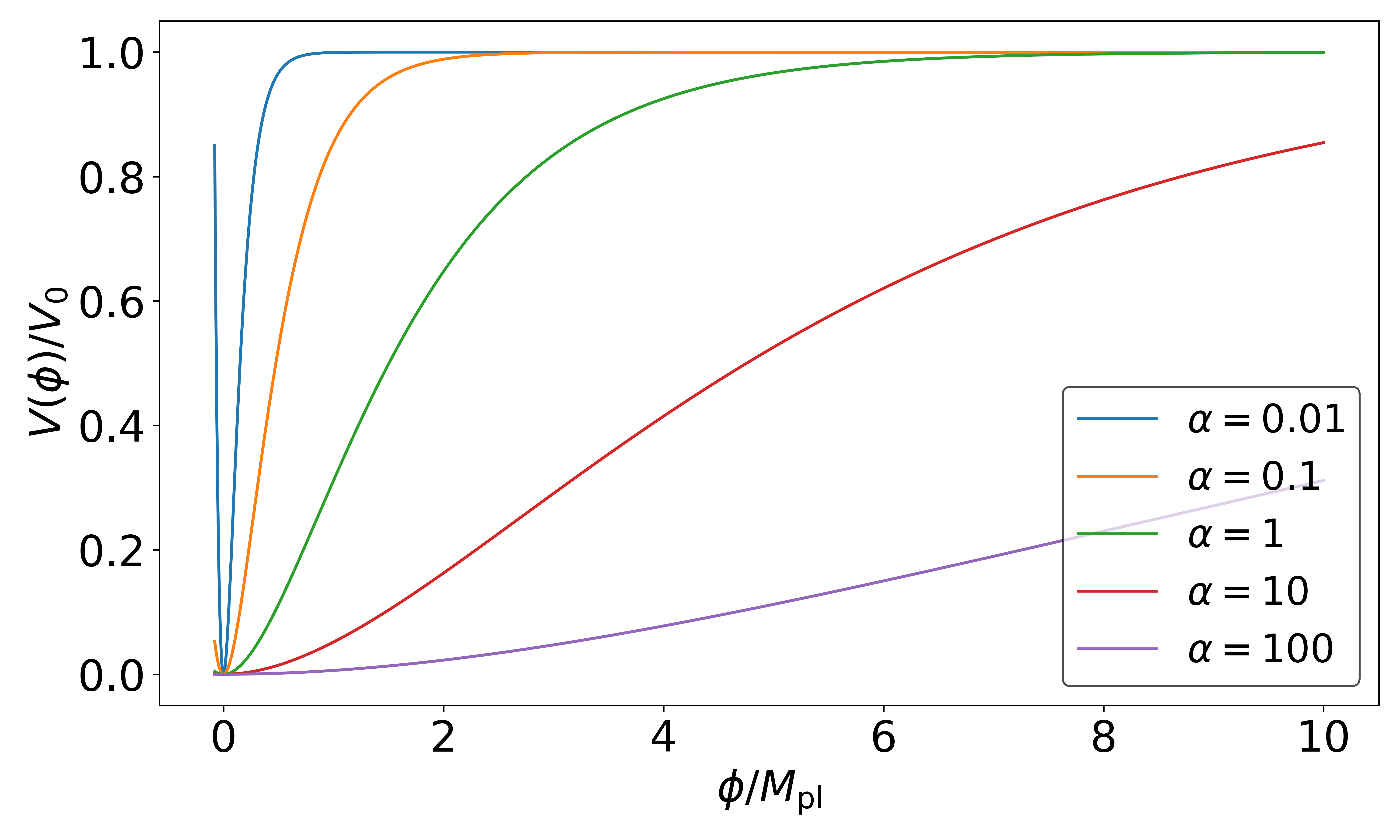}
        \caption{\it E-Model of $\alpha$-attractor with n=1}
        \label{fig:E-Model fig1}
    \end{subfigure}
    \hfill
    \begin{subfigure}[t]{0.495\textwidth} 
        \centering
        \includegraphics[width=\linewidth]{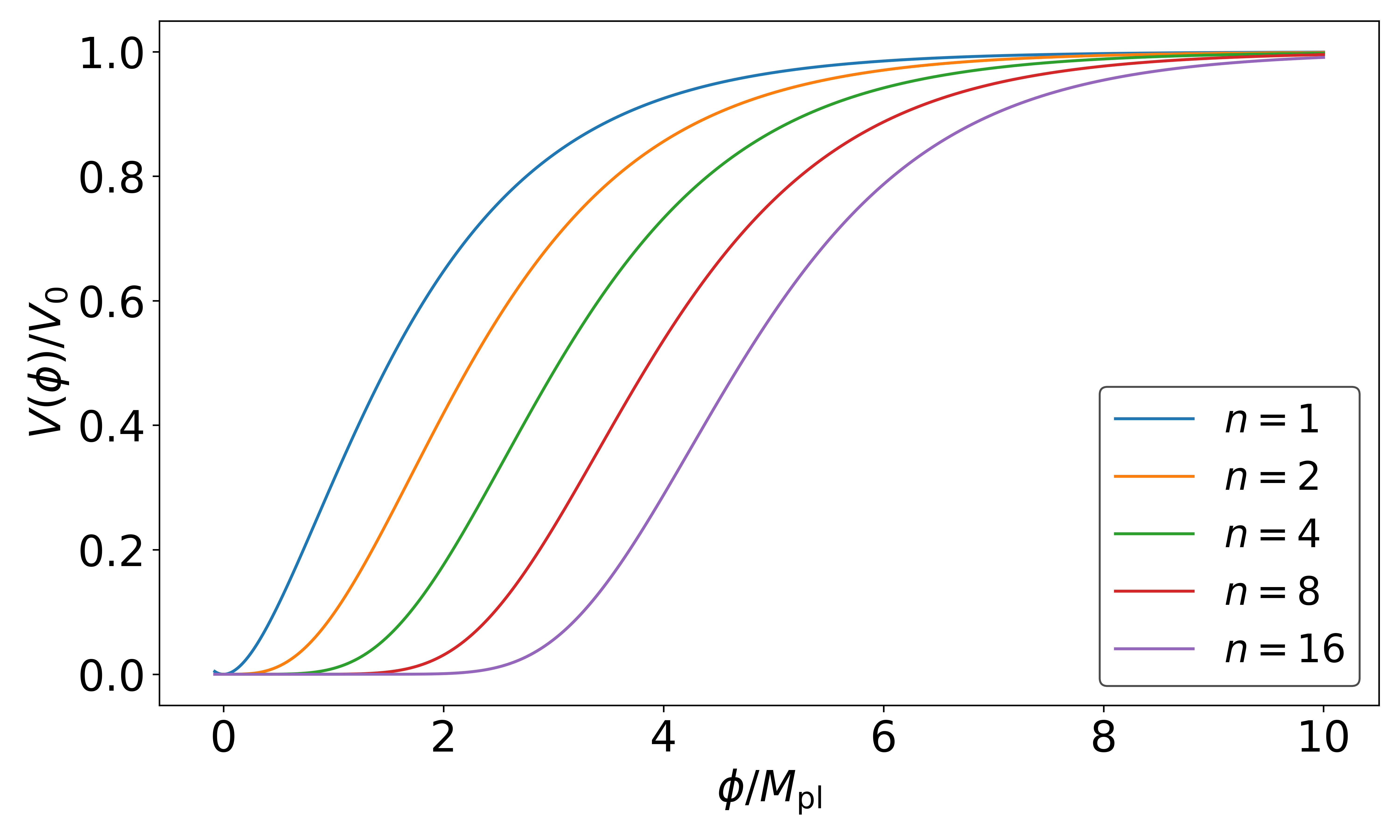}
        \caption{\it E-Model of $\alpha$-attractor with $\alpha$=1}
        \label{fig:E-Model fig2}
    \end{subfigure}
    \caption{\it Potential profiles of the E-model $\alpha$-attractors given by Eq. (\ref{E-Model eq1}), illustrating the dependence on the parameters $\alpha$ and $n$. \textbf{(a)} The potential $V(\phi)/V_0$ is plotted as a function of $\phi/M_{Pl}$ for fixed $n=1$ and varying $\alpha$. \textbf{(b)} The potential is shown for fixed $\alpha=1$ and varying $n$. These plots highlight the sensitivity of the inflationary potential's shape to changes in $\alpha$ and $n$ - the key parameters in the $\alpha$-attractor framework.}
    \label{fig:E-Model Plot1}
\end{figure}

Several inflationary models (like the Starobinsky Model, chaotic inflation model, $\lambda \phi^4$ with non-minimal coupling to gravity  \cite{Starobinsky:1980te, Kallosh:2013pby, Kallosh:2013lkr, Salopek:1988qh, Ferrara:2013rsa, Kallosh:2013maa, Kallosh:2013tua, Ellis:2013xoa, Buchmuller:2013zfa, Dimopoulos:2020pjx}) have a common prediction for the inflationary observables $n_s$ and $r$ in the limit of large $N_e$. 
\begin{equation} \label{pred1}
    \begin{matrix}n_s\approx1-\frac{2}{N_e}\end{matrix},
\end{equation}
and
\begin{equation} \label{pred2}
    r \approx \frac{12}{N_{e}^{2}},
\end{equation}
The predictions (\ref{pred1}) and (\ref{pred2}) are a nearly universal feature of a broad class of inflationary models with spontaneously broken conformal or super conformal invariance \cite{Kallosh:2013hoa, Kallosh:2013daa}. One can go a step further and generalize these models by introducing a parameter $\alpha$, inversely related to the curvature in the field space of the inflaton. These represent a class of inflationary models called single-field $\alpha$-attractors, consistent with observations (as shown in \cite{Planck:2015sxf}) and have a general prediction for inflationary observables:
\begin{equation} \label{nsr-1}
    \begin{matrix}n_s&\approx1-\frac{2}{N_e}\end{matrix}
\end{equation}
and
\begin{equation} \label{nsr-2}
    r \approx \frac{12\alpha^{2}}{N_{e}^{2}}
\end{equation}
 in the limit of large $N_e$ and small $\alpha$. The most popular example is the E-Model $\alpha$- attractor, given by,
\begin{equation} \label{E-Model eq1}
V(\phi)=V_0\left(1-e^{-\sqrt{\frac{2}{3 \alpha}} \frac{\phi}{M_{p l}}}\right)^{2 n}.
\end{equation}
Fig. \ref{fig:E-Model Plot1} gives an idea of how the potential would look like for a fixed value of $n=1$ and $\alpha=1$. $V_0$ is the overall normalization given by,
\begin{equation}
    V_0 = \frac{3\alpha m^2}{4},
\end{equation}
where $m$ is the mass of the inflaton field. $M_{Pl}$ is the Planck Mass, and the exponent $n$ can take any positive value. Substituting $n=1$ and $\alpha=1$ in Eq. (\ref{E-Model eq1}) gives us the Starobinsky model of inflation \cite{STAROBINSKY198099}:  
\begin{equation}
    V(\phi)=V_0\left(1-e^{-\sqrt{\frac{2}{3}} \frac{\phi}{M_{p l}}}\right)^{2}.
\end{equation}


\subsubsection{T-Model of $\alpha$-attractor}
\begin{figure}[!h] 
    \centering
    \begin{subfigure}{0.495\textwidth} 
        \centering
        \includegraphics[width=\linewidth]{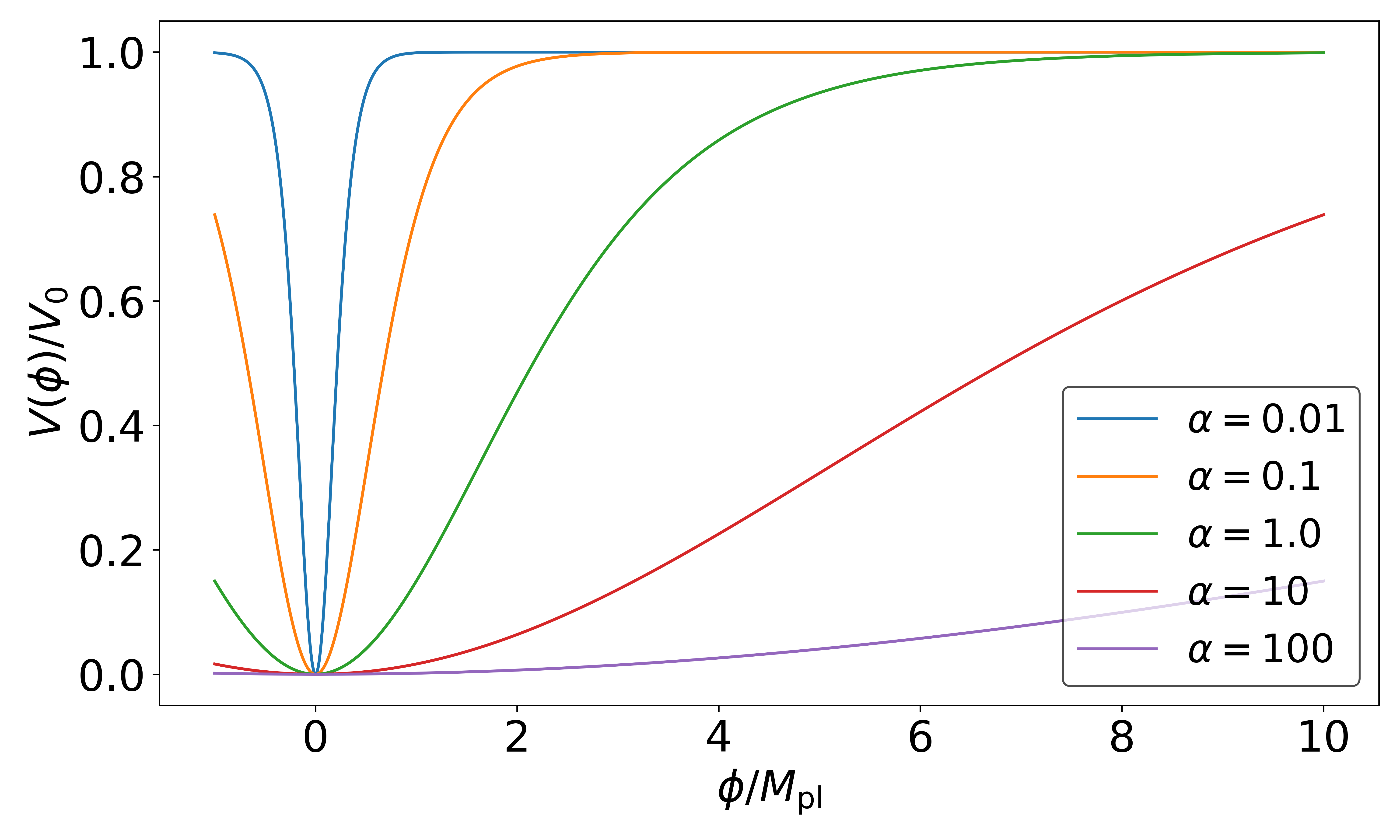}
        \caption{\it T-Model of $\alpha$-attractor with n=1}
        \label{fig:T-Model fig1}
    \end{subfigure}
    \hfill
    \begin{subfigure}{0.495\textwidth} 
        \centering
        \includegraphics[width=\linewidth]{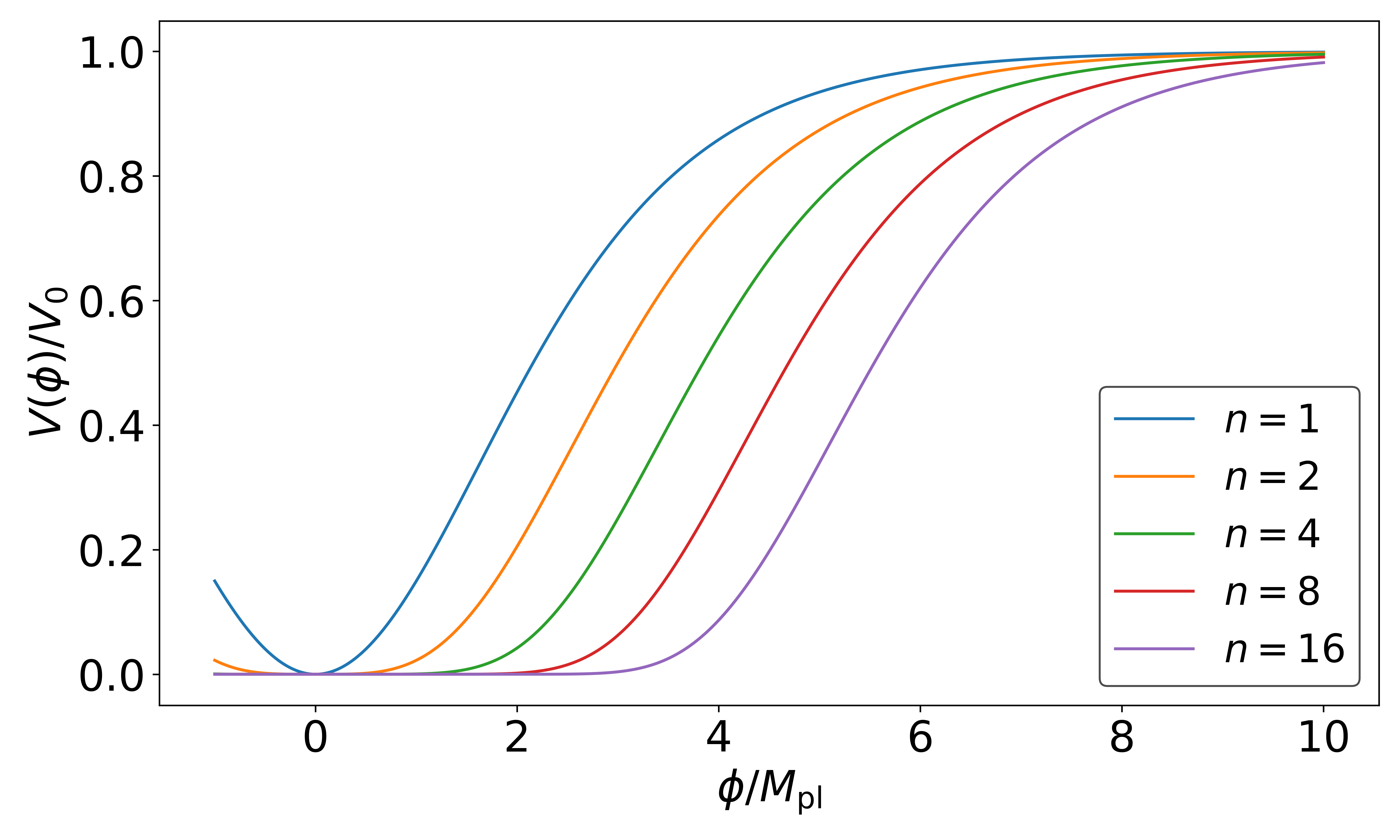}
        \caption{\it T-Model of $\alpha$-attractor with $\alpha$=1}
        \label{fig:T-Model fig2}
    \end{subfigure}
    \caption{\it Potential profiles of the T-model of $\alpha$-attractors, defined in Eq. (\ref{T-Model eq1}). \textbf{(a)} The potential $V(\phi)/V_0$ is plotted as a function of $\phi/M_{Pl}$ for fixed $n=1$ and varying $\alpha$. \textbf{(b)} The potential is shown for fixed $\alpha=1$ and varying $n$.}
    \label{fig:T-Model Plot1}
\end{figure}

The T-Model $\alpha$-attractor, whose potential profiles are shown in Fig. \ref{fig:T-Model Plot1}, is given by,

\begin{equation}\label{T-Model eq1}
V(\phi)=V_0\tanh^{2n}(\frac{\phi}{\sqrt{6 \alpha}M_{p l}}),
\end{equation}

also provide the same predictions as in Eqs. (\ref{nsr-1}) and (\ref{nsr-2}) in the limit of large $N_e$ and small $\alpha$. The full predictions of the T-Model are \cite{Bhattacharya:2022akq}, 
\begin{equation} \label{nsrT1}
    n_s(\alpha,n,N_e)=\frac{1-\frac{2}{N_e}-\frac{3\alpha}{4N_{e}^2}+\frac{1}{2nN_e}\left(1-\frac{1}{N_e}\right)g(\alpha,n)}{1+\frac{1}{2nN_e}g(\alpha,n)+\frac{3\alpha}{4N_{e}^2}},
\end{equation}
and
\begin{equation} \label{nsrT2}
    r(\alpha,n,N_e)=\frac{12\alpha}{N_{e}^2+\frac{N_e}{2n}g(\alpha,n)+\frac{3}{4}\alpha},
\end{equation}
where, 
\begin{equation}\label{nsrT3}
    g(\alpha,n)\equiv\sqrt{3\alpha(4n^{2}+3\alpha)}.
\end{equation}

Similar to the E-Model, $n$ can also take any positive value. Clearly, for large $N_e$ and small $\alpha$, we obtain Eqs. (\ref{nsr-1}) and (\ref{nsr-2}), respectively.

\subsubsection{Polynomial $\alpha$-attractor}
\begin{figure}[!h] 
    \centering
    \begin{subfigure}{0.495\textwidth} 
        \centering
        \includegraphics[width=\linewidth]{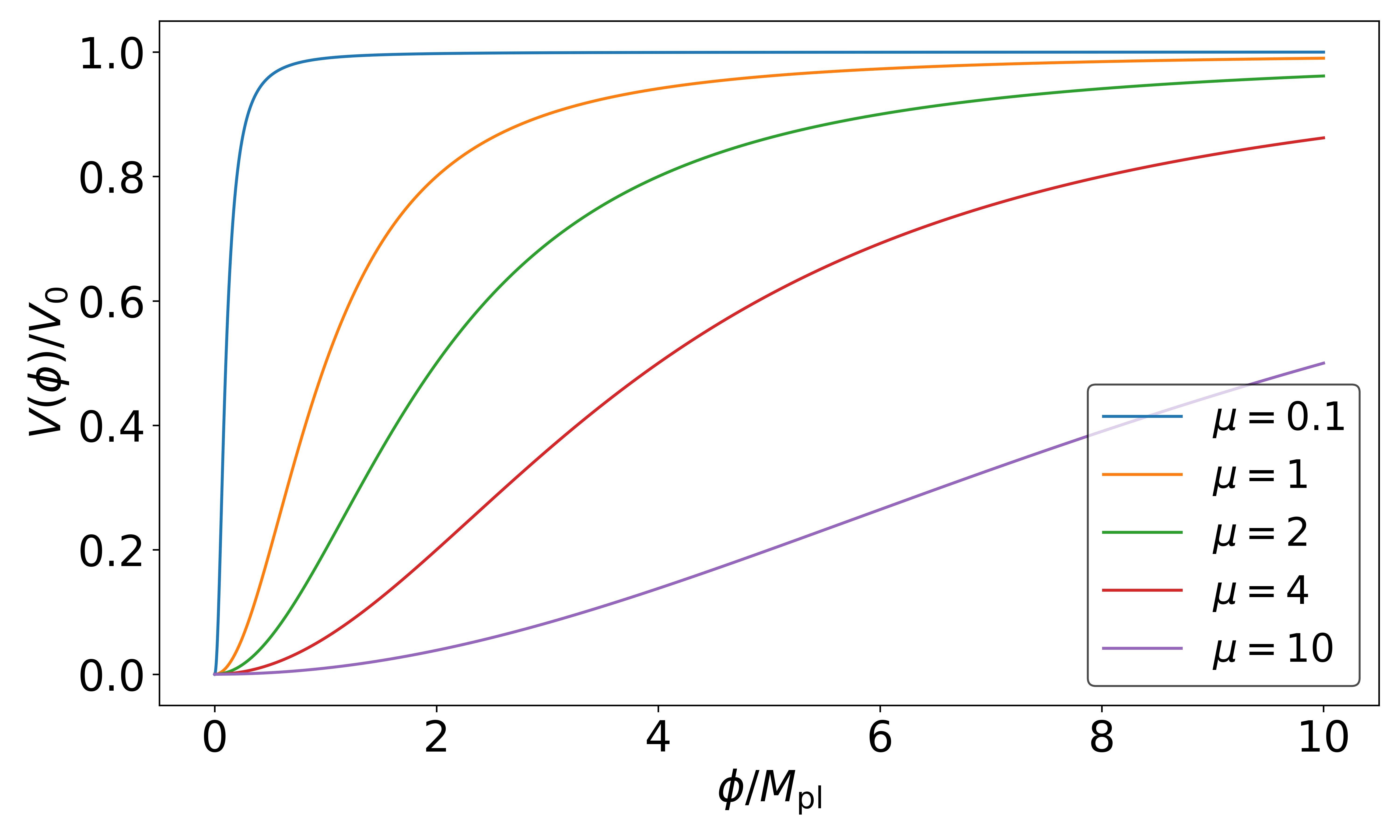}
        \caption{\it Polynomial $\alpha$-attractor with n=1}
        \label{fig:Poly fig1}
    \end{subfigure}
    \hfill
    \begin{subfigure}{0.495\textwidth} 
        \centering
        \includegraphics[width=\linewidth]{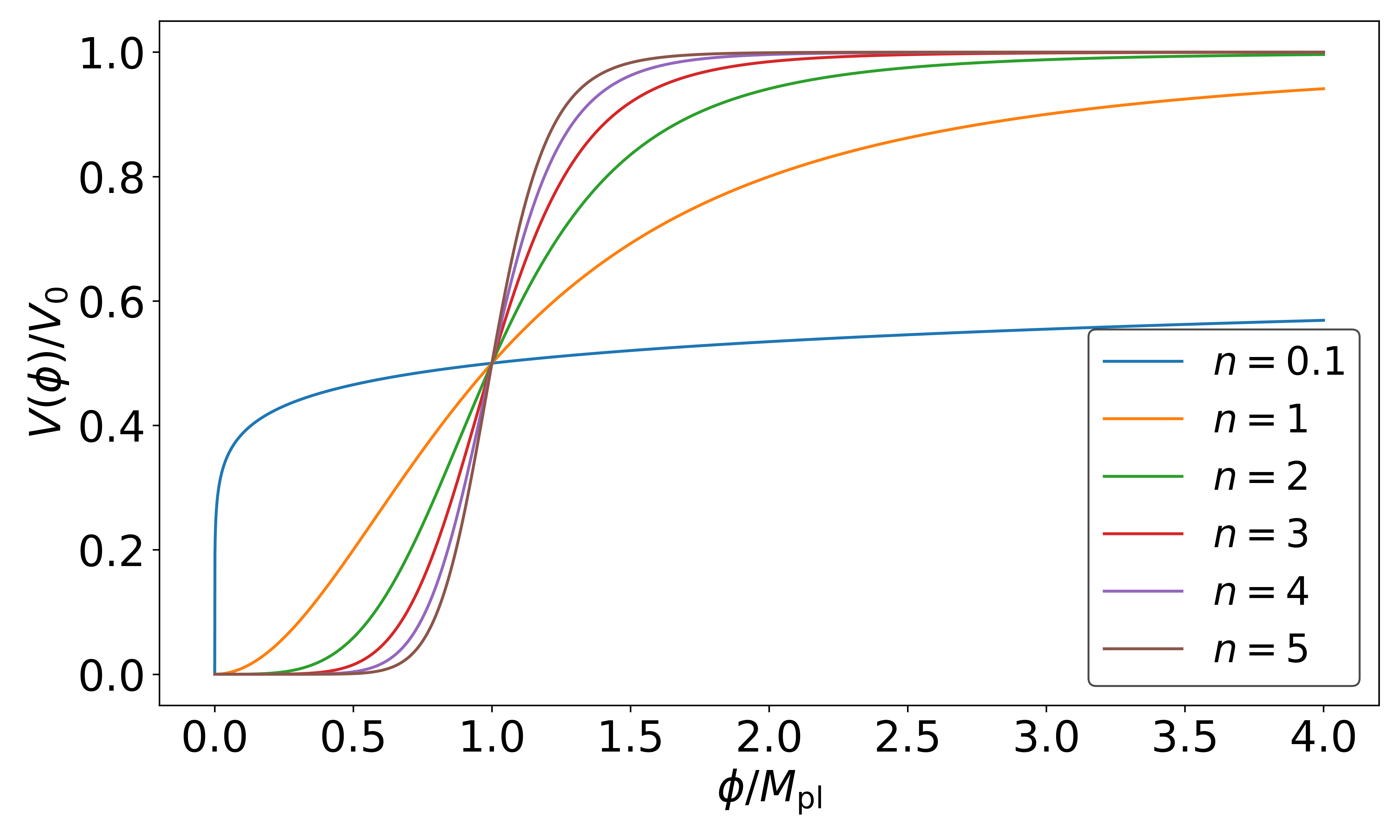}
        \caption{\it Polynomial $\alpha$-attractor with $\mu$=1}
        \label{fig:Poly fig2}
    \end{subfigure}
    \caption{\it Potential profiles of the simplest Polynomial $\alpha$-attractor, defined in Eq. (\ref{Poly Eq1}). \textbf{(a)} The potential $V(\phi)/V_0$ is plotted as a function of $\phi/M_{Pl}$ for fixed $n=1$ and varying $\mu$. \textbf{(b)} The potential is shown for fixed $\mu=1$ and varying $n$.}
    \label{fig:Poly-Model Plot1}
\end{figure}

Another class of inflationary attractors, which was studied fairly recently \cite{Kallosh:2022feu}, are the Polynomial attractors (Fig. \ref{fig:Poly-Model Plot1}), where instead of an `exponential' potential, the inflationary plateau is reached (much more slowly than the exponential scenario) with an inverse power of the inflaton field. 
\begin{equation}
V \sim V_0\left(1-\frac{\mu^n}{\varphi^n}+\ldots\right),
\end{equation} 
where $\mu$ is a free parameter and $n$ can take any positive value. The simplest polynomial attractor model is given by, 
\begin{equation}\label{Poly Eq1}
    V(\phi)=V_0\frac{\phi^{2n}}{\phi^{2n}+\mu^{2n}}.
\end{equation}

The potential profile of Eq. (\ref{Poly Eq1}) is shown in Fig. (\ref{fig:Poly-Model Plot1}). In the slow-roll approximation and limit of large $N_e$, the full $n_s$-$r$ predictions \cite{Bhattacharya:2022akq} can be found using Eqs. \ref{ns pred} and \ref{r pred} :

\begin{equation}\label{nsP1}
    n_s = 1-\frac{n+1}{n+2}\frac{2}{N_e},
\end{equation}
and
\begin{equation}\label{rP1}
    r = 2^{\frac{3+n}{1+n}}\left(\mu^{2n}n\right)^{\frac{1}{1+n}}\left((n+1)N_e\right)^{-\frac{1+2n}{1+n}}.
\end{equation}

\begin{figure}[!h]
    \centering
    \includegraphics[width=0.9\linewidth]{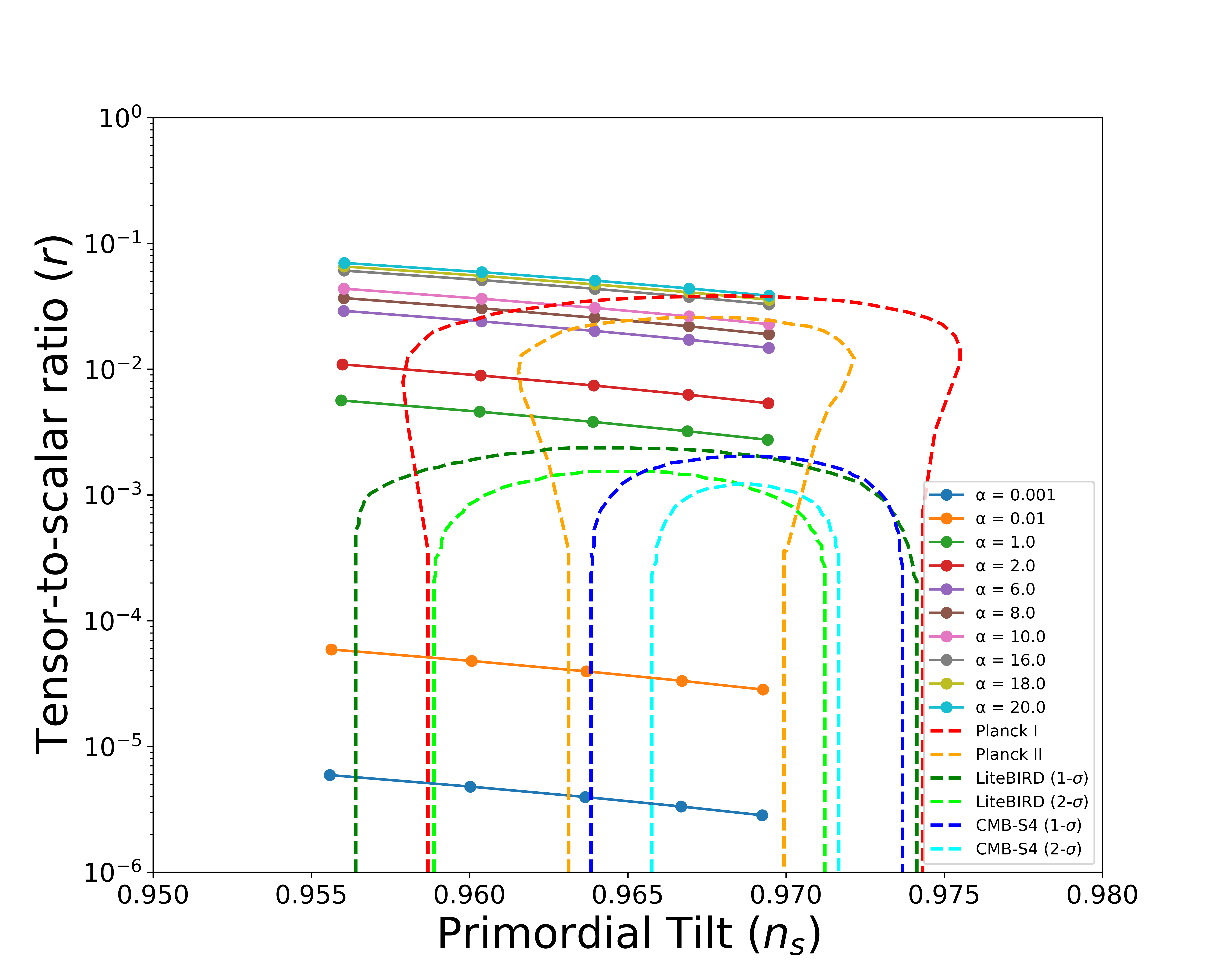}
    \caption{\it The $n_{s}-r$ prediction for T-Model of $\alpha$-attractor with $n=1$. The dots along each line represent a specific value of $N_{e}$. Going from right to left, $N_{e}=65, 60, 55, 50, 45.$\\ Planck I correspond to Planck TT, TE, EE+lowE+lensing+BK18+BAO at $1-\sigma$ and Planck II implies Planck TT, TE, EE+lowE+lensing+BK18+BAO at $2-\sigma$. 
   }.
    \label{fig:nsrTfig1}
\end{figure}

\medskip

\section{Formation of Cosmic Strings during Inflation} \label{CS in inf}

Inflation smoothens out any initial inhomogeneities and drives the universe towards a nearly flat geometry. However, if inflation ends through a phase transition, as in many hybrid inflation models, it can lead to the formation of topological defects, including cosmic strings \cite{Vilenkin:2000jqa}.

In this section, we present a scenario where we explore the possibility of the cosmic strings being formed during inflation. Several models supports the formation of cosmic strings at the later stages of inflation or even end inflation via formation of topological defects in general \cite{PhysRevD.29.1870, Vishniac:1986sk, Kofman:1986wm, PhysRevLett.63.712, Linde:1991km, PhysRevD.54.6083, PhysRevD.49.748, PhysRevD.68.103514,Lazarides:2023rqf,Maji:2024cwv}. Such a situation can have a significant impact on the cosmic string phenomenology, the most significant being on the stretching of the correlation length and suppression of the GW spectra. 

The period of accelerated expansion is expected to rapidly stretch the strings, so that they become frozen in comoving coordinates, with a characteristic length significantly larger than the Hubble radius, at the end of inflation \cite{Guedes:2018afo}. Such a stretching regime cannot be maintained indefinitely after the inflationary stage ends. These strings will eventually re-enter our Hubble sphere, and the string network will approach the evolution of a standard cosmic string network (produced after inflation). The effect of the phase of accelerated expansion delays the attainment of the standard evolution: the more the network is stretched, the longer this delay is. Observational signatures of these GWs would depend on the inflationary dynamics and the specific string tension.

\subsection{Evolution of Cosmic Strings: VOS Model}

The Velocity-dependent One Scale (VOS) Model \cite{PhysRevD.54.2535,PhysRevD.65.043514} is a pair of coupled differential equations characterized by two important features of the cosmic string: 
\begin{enumerate}
    \item The characteristic length, $L=(\mu/\rho)^{1/2}$, where $\mu$ is the string tension and $\rho$ is the average density of the cosmic string network.

    \item The root-mean-square (rms) velocity of the cosmic string network, $\Bar{v}$.
\end{enumerate}
assuming the cosmic strings to be infinitely thin, the VOS equations for local cosmic strings read as \cite{Gouttenoire:2019kij}:
\begin{equation}\label{VOS1}
    \frac{dL}{dt}=HL\left(1+\bar{v}^{2}\right)+\frac{1}{2}\tilde{c} \bar{v},
\end{equation}
\begin{equation}\label{VOS2}
    \frac{d\bar{v}}{dt}=(1-\bar{v}^2)\left[\frac{k(\bar{v})}{L}-2H\bar{v}\right],
\end{equation}
where
\begin{equation}\label{VOS3}
    k(\bar{v})=\frac{2\sqrt{2}}{\pi}(1-\bar{v}^{2})(1+2\sqrt{2}\bar{v}^{3})\frac{1-8\bar{v}^{6}}{1+8\bar{v}^{6}},
\end{equation}
is the curvature parameter and is a measure of the deviation from a straight string, for which $k(\bar{v})=1$. $\tilde{c}=0.23\pm0.04$ is a phenomenological parameter quantifying the loop chopping efficiency. 

For global cosmic strings, the VOS equations are given by \cite{Gouttenoire:2019kij}:
\begin{equation}\label{GVOS1}
    \frac{dL}{dt}=HL\left(1+\bar{v}^2\right)+\left.F(\bar{v})\right|_{\mathrm{global}},
\end{equation}
\begin{equation}\label{GVOS2}
    \frac{d\bar{v}}{dt}=(1-\bar{v}^2)\left[\frac{k(\bar{v})}{L}-\frac{\bar{v}}{l_d}\right],
\end{equation}
where $F(\bar{v})|_{\mathrm{global}}$ is the energy-loss coefficient and $l_d$ is the damping length scale.

Now, if we look at the first term on the RHS of Eqs. (\ref{VOS1}) and (\ref{GVOS1})  - $HL$ essentially describes how the universe's expansion affects the characteristic length of the cosmic string. We are particularly interested in the following two scenarios:
\begin{enumerate}
    \item $\boldsymbol{HL\gg1}$: The characteristic length of the cosmic string is much larger than the Hubble radius, and the universe's expansion dictates the string network. This happens during inflation, when the strings are effectively frozen in place by the rapid expansion and their internal dynamics (intercommutation of strings) become negligible compared to the expanding background. Quantitatively, one can write \cite{PhysRevD.65.043514}:
    \begin{equation}\label{HL1}
        L\propto a.
    \end{equation}

    \item $\boldsymbol{HL=1}$: The characteristic length is comparable to the Hubble radius when the late-time energy density takes over inflation
    \begin{equation}\label{HL2}
        LH\propto t^{(2-n)/n}\quad\text{during the era with} \quad \rho\propto a^{-n}.
    \end{equation}
    For $n>2$, the Hubble horizon will catch up with the string length, allowing them to re-enter the horizon and start loop production through intercommutation. 
\end{enumerate}
In the scenario that we are interested in, the cosmic strings are produced when $HL\gg1$, i.e., at some point during inflation. Let's say these strings re-enter the horizon at some later time, when the temperature of the universe is $T_{re}$.
\begin{equation}
    L_{re} H_{re} = 1
\end{equation}
Using Eq. \ref{HL2}, we can evolve the correlation length $L$, from the end of inflation up to their re-entry as \cite{Gouttenoire:2019kij},
\begin{equation}
    1=L_{\mathrm{re}}H_{\mathrm{re}} =\left(\frac{t_{\mathrm{re}}}{t_{\mathrm{end}}}\right)^{-1/2}L_{\mathrm{end}}H_{\mathrm{end}},
\end{equation}
\begin{equation}
    1=\left(\frac{t_{\mathrm{re}}}{t_{\mathrm{end}}}\right)^{-1/2}\left(\frac{a_{\mathrm{end}}}{a_{\mathrm{start}}}\right)L_{\mathrm{start}}H_{\mathrm{start}},
\end{equation}
\begin{equation} \label{Tre Tend}
    1\simeq\left(\frac{T_{\mathrm{re}}}{T_{\mathrm{end}}}\right)e^{N_{e}}(0.1).
\end{equation}
The next step is to introduce the energy scale of inflation $E_{inf}$ and redefine $T_{re}$. We re-write Eq. (\ref{Tre Tend}) as,
\begin{equation} \label{Tre-Tend 2}
    T_{\mathrm{re}}\simeq T_{\mathrm{end}}\frac{1}{(0.1)\exp(N_e)}.
\end{equation}
We assume that the end of inflation is characterized by the energy scale $E_{\rm inf}$,
which can be related to $T_{end}$ as \cite{kolb2018early}:
\begin{equation}
    \rho_\mathrm{end}\simeq\frac{\pi^2}{30}g_*(T_\mathrm{end})T_\mathrm{end}^4.
\end{equation}
The energy density of inflation can be roughly taken as $\rho_\mathrm{end}\sim E_\mathrm{inf}^4$,
\begin{equation}
    E_{\inf}^4\simeq\frac{\pi^2}{30}g_*(T_{\mathrm{end}})T_{\mathrm{end}}^4,
\end{equation}
\begin{equation}
    T_{\mathrm{end}}\simeq\frac{E_{\mathrm{inf}}}{\left(\frac{\pi^2}{30}g_*(T_{\mathrm{end}})\right)^{1/4}}.
\end{equation}
Approximating, we can write,
\begin{equation}
    T_{\mathrm{end}}\simeq\frac{E_{\mathrm{inf}}}{g_*^{1/4}(T_{\mathrm{end}})}.
\end{equation}
Substituting this value of $T_{\mathrm{end}}$ in Eq. (\ref{Tre-Tend 2}), we get:
\begin{equation}
    T_{\mathrm{re}}\simeq\frac{E_{\mathrm{inf}}}{g_*^{1/4}(T_{\mathrm{end}})} \frac1{(0.1)\exp(N_e)}.
\end{equation}
Substituting this value of $T_{\mathrm{end}}$ in Eq. (\ref{Tre-Tend 2}), we get:
\begin{equation}
    T_{\mathrm{re}}\simeq\frac{E_{\mathrm{inf}}}{g_*^{1/4}(T_{\mathrm{end}})} \frac1{(0.1)\exp(N_e)}.
\end{equation}
Since $g_*(T_{\mathrm{end}})\approx g_*(T_{\mathrm{re}})$ at such high temperature scales, we can replace $g_*(T_{\mathrm{end}})$ with $g_*(T_{\mathrm{re}})$ to obtain:
\begin{equation}
    T_{\mathrm{re}}\simeq\frac{E_{\mathrm{inf}}}{(0.1)g_*^{1/4}(T_{\mathrm{re}})\exp(N_e)}.
\end{equation}

Using this definition of $T_{re}$, we will now introduce an important ingredient, the \textit{turning-point frequency} \cite{Gouttenoire:2019kij}, which is essential in evaluating the GW spectra from a cosmic string impacted by inflation.

\subsection{Turning-Point Frequency}
We refer the reader to the Appendix for a short review of generation of Gravitational Waves from local and global string network. An important observable in our analysis will be the \textit{turning-point frequency} $f_\Delta$, above which the GW spectrum deviates from the one obtained in standard cosmology (Fig. \ref{fig: Standard Local Spectrum} in Appendix \ref{Sec. 3}). At higher frequencies, both the standard and the non-standard scenario  show a characteristic $\Omega_{GW}h^2 \sim f^{-1}$ behaviour, which happens to be a consequence of the lack of production of
the loops that contributes dominantly in this frequency range. But in the non-standard scenario that we are going to discuss ahead, we will find that, GW spectra from cosmic strings that have experienced some e-foldings of inflation, shifts this $\Omega_{GW}h^2 \sim f^{-1}$ behaviour to lower frequencies. \cite{Guedes:2018afo}. Apart from understanding the non-standard scenario due to the impact of inflation that we will describe in this paper, another important non-standard scenario would be the presence of an early matter dominated era, see Refs. \cite{Cui:2017ufi,Ghoshal:2023sfa,Datta:2025yow} for detailed analyses. The only way to distinguish between them is to estimate the inflation affected cosmic string GW spectrum (as the one in Fig.\ref{fig:1e-11 alpha10 n1 N70}) considering at least $10^{4}$ $k$-modes (which is beyond the scope of this paper), which shows a transition from a $f^{-1/3}$ to $f^{-1}$ in the case of inflation \cite{Gouttenoire:2019kij}. After the end of inflation, one must wait for the correlation length to re-enter the horizon in order to reach the scaling regime again. The duration of the transient regime receives an enhancement factor $\exp({N_e})$. The turning-point frequency for local cosmic strings following Nambu-Goto dynamics read as \cite{Gouttenoire:2019kij, Cui:2018rwi,Ghoshal:2023sfa,Datta:2025yow}:


\begin{equation}\label{turn freq}
    f_{\Delta}=(1.5\times10^{-4}\text{Hz})\left(\frac{T_{\mathrm{re}}}{\text{GeV}}\right)\left(\frac{0.1\times50\times10^{-11}}{\alpha \Gamma G\mu}\right)^{1/2}\left(\frac{g_{*}(T_{\mathrm{re}})}{g_{*}(T_{0})}\right)^{1/4},
\end{equation}

\begin{equation}\label{re-entry temp}
    T_\mathrm{re}\simeq\frac{E_\mathrm{inf}}{(0.1) g_*^{1/4}(T_\mathrm{re}) \exp(N_e)}.
\end{equation}
\\
Now, the energy scale of inflation is related to the tensor-to-scalar ratio ($r$) as \cite{Baumann:2014nda},

\begin{equation}\label{einf to r}
    E_\mathrm{inf} = 1.06\times 10^{16}\times \left(\frac{r}{0.01}\right)^{1/4}.
\end{equation}
\\
Eq. (\ref{einf to r}) is an important equation for us, to understand the impact of inflation on cosmic strings. Using Eq. (\ref{einf to r}) and the values of the tensor-to-scalar ratio and spectral index as shown in Fig. \ref{fig:nsrTfig1}, we can calculate the scale of inflation and subsequently calculate the modified GW spectra.

The turning-point frequency for global strings is independent of the string tension \cite{Chang:2021afa}: 
\begin{equation}\label{global turn freq}
    f_{\Delta}=(3.02\times10^{-6}\text{Hz})\left(\frac{T_{\mathrm{re}}}{\text{GeV}}\right)\left(\frac{0.1}{\alpha}\right)\left(\frac{g_{*}(T_{\mathrm{re}})}{g_{*}(T_{0})}\right)^{1/4}.
\end{equation}
Eqs. (\ref{re-entry temp}) and (\ref{einf to r}) also hold for the global string's scenario.

\subsection{Impact of a T-Model Inflation on GW spectrum due to local and global Cosmic String Loops} \label{T-Model inf 1}

We will now use the T-Model of $\alpha$-attractor inflation (Eq. \ref{T-Model eq1}) to understand its impact on the GW spectrum. We will consider three cases where we vary the parameter $\alpha$ of the T-Model. In the discussion that follows, $N_{total}$ represents the total e-folds of inflation that the universe has undergone and $N_{CS}$ denotes the number of e-folds of inflation that our cosmic string has suffered since their formation (during the inflationary epoch). Our choices of string tension $G\mu$, the total e-folds of inflation $N_{total}$ and the potential $V(\phi)$ driving the inflation are independent. 


\begin{itemize}
    \item \textbf{Case I}: $n=1$, $\alpha=10$, $G\mu=10^{-11}$ / $\eta^{local}=3.860\times10^{13}$ GeV, $N_{total}=65$\\
    Substituting $n=1$ and $\alpha=10$ in Eq. (\ref{T-Model eq1}), we get the potential driving the inflation:
    \begin{equation}
    V(\phi)=V_0\tanh^{2}\left(\frac{\phi}{7.746 M_{pl}}\right).
    \end{equation}
    Setting the total e-folds of inflation to 65 fixes the value of the tensor-to-scalar ratio ($r$) and the spectral index ($n_s$) to $r=0.022$ and $n_s=0.969$.

    \begin{figure}[!h]
    \centering
    \includegraphics[width=0.7\linewidth]{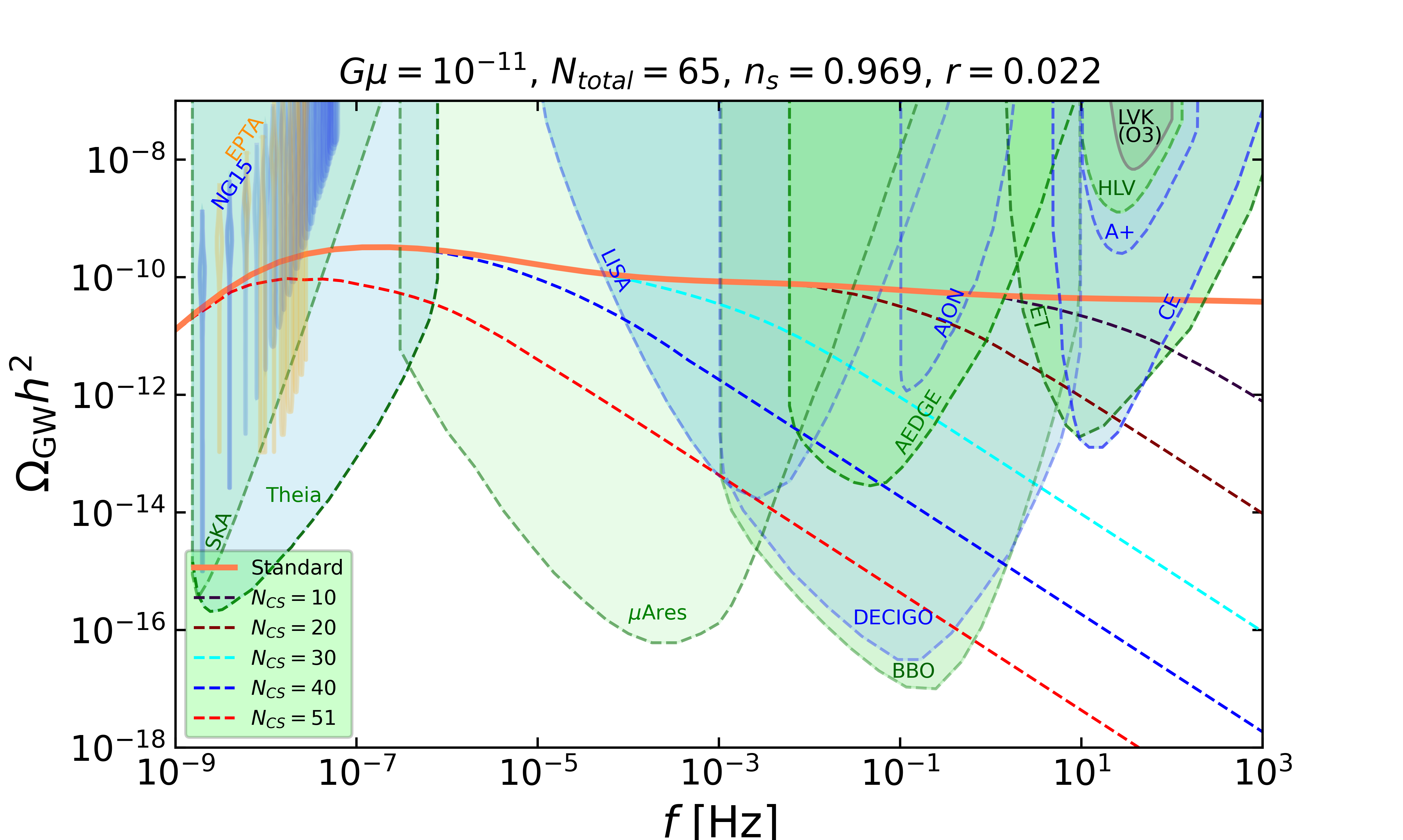}
    \caption{\it GW spectrum due to local cosmic strings of string tension $G\mu=10^{-11}$ or vev $\eta^{local} = 3.860\times10^{13}$ GeV impacted by inflation driven by the T-Model with $n=1$ and $\alpha=10$. Up to 200 $k-$modes were considered for numerically estimating this spectrum. The maximum allowed value of $N_{CS}$ is $\sim51$ e-folds. For any $N_{CS}$ value greater than this, we won't be able to observe any GW spectrum.}
    \label{fig:1e-11 alpha10 n1 N70}
    \end{figure}

    \item \textbf{Case II}:  $n=1$, $\alpha=1$, $G\mu=10^{-11}$ / $\eta^{local}=3.860\times10^{13}$ GeV, $N_{total}=65$\\
    Using Eq. (\ref{T-Model eq1}) with $n=1$ and $\alpha=1$ we get:
    \begin{equation} \label{n=1 alpha=1 TModel}
    V(\phi)=V_0\tanh^{2}\left(\frac{\phi}{2.449 M_{pl}}\right).
    \end{equation}
    We set $N_{total}$ to 65, which fixes the value of the tensor-to-scalar ratio ($r$) and spectral index ($n_s$) to $r=0.002$ and $n_s=0.969$.

    \begin{figure}[H]
    \centering
    \includegraphics[width=0.7\linewidth]{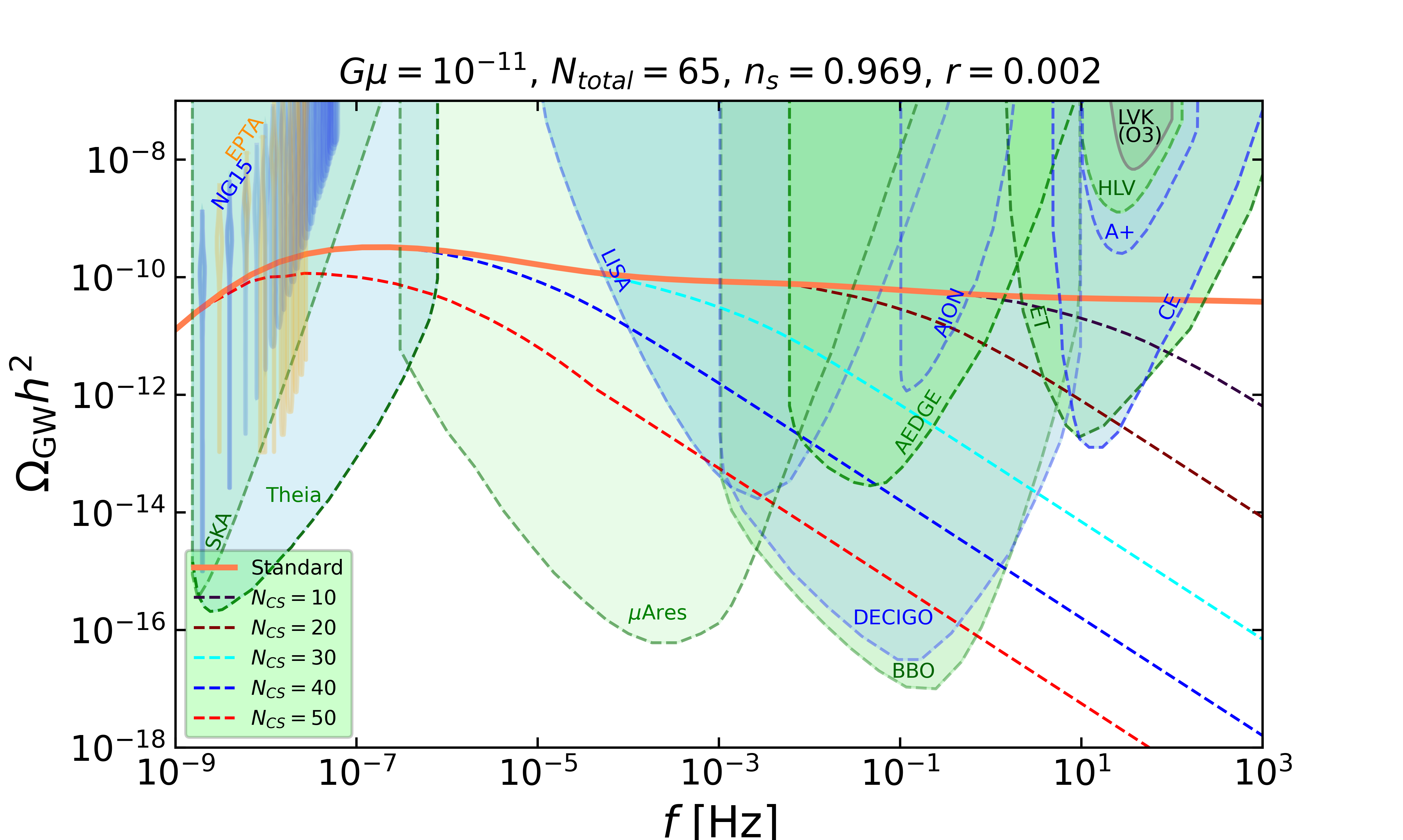}
    \caption{\it GW spectrum due to local cosmic strings of string tension $G\mu=10^{-11}$ or vev  $\eta^{local} = 3.860\times10^{13}$ GeV impacted by the T-Model of inflation with $n=1$ and $\alpha=1$. We considered up to 200 $k-$modes to numerically estimate this spectrum. The maximum allowed value of $N_{CS}$ decreases to $\sim50$ e-folds in this case as we decrease the value of the tensor-to-scalar ratio $r$.}
    \label{fig:1e-11 alpha1 n1 N70}
    \end{figure}

    \item \textbf{Case III}: $n=1$, $\alpha=10^{-4}$, $G\mu=10^{-11}$ / $\eta^{local}=3.860\times10^{13}$ GeV, $N_{total}=65$\\
    We again substitute $n=1$ and $\alpha=10^{-4}$ into Eq. (\ref{T-Model eq1}) to obtain or potential:
    \begin{equation}
    V(\phi)=V_0\tanh^{2}\left(\frac{\phi}{0.024 M_{pl}}\right).
    \end{equation}
    Setting $N_{total}$ to 65 fixes the value of the tensor-to-scalar ratio ($r$) and the spectral index ($n_s$) to $r=2.839\times 10^{-7}$ and $n_s=0.969$.

    \begin{figure}[H]
    \centering
    \includegraphics[width=0.7\linewidth]{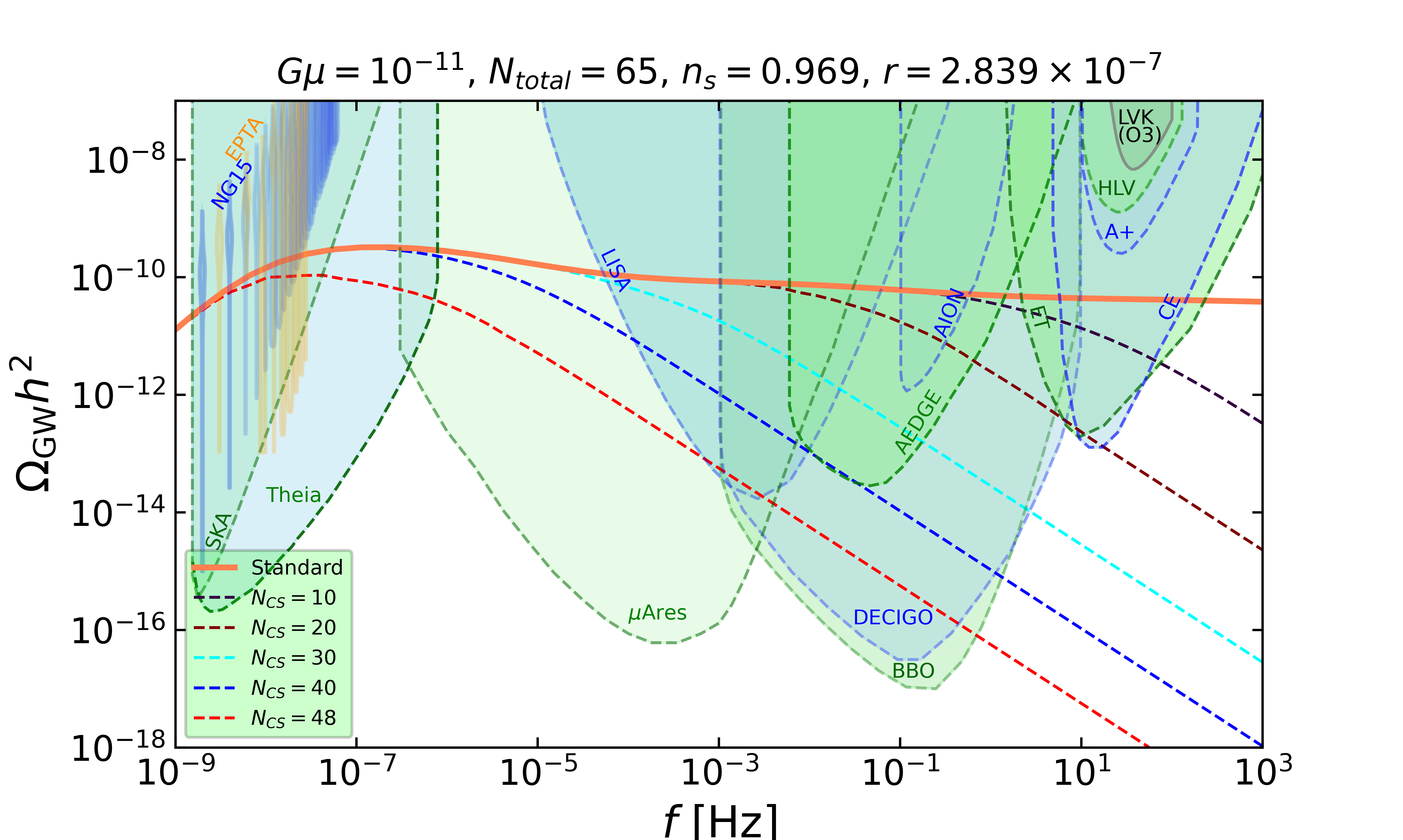}
    \caption{\it GW spectrum due to local cosmic strings of string tension $G\mu=10^{-11}$ or vev $\eta^{local} = 3.860\times10^{13}$ GeV impacted by the T-Model of inflation with $n=1$ and $\alpha=10^{-4}$. We considered up to 200 $k-$modes for numerically estimating this spectrum. The maximum allowed value of $N_{CS}$ further decreases to $\sim48$ e-folds in this case because we again lower the value of the tensor-to-scalar ratio $r$. }
    \label{fig:1e-11 alpha1e-4 n1 N70}
    \end{figure}

\end{itemize}

\begin{figure}[H]
    \centering
    \begin{subfigure}{0.6\textwidth}
        \centering
        \includegraphics[width=\textwidth]{Plots/Fig12.png}
        \caption{\it GW spectrum from local CS ($G\mu=10^{-11}$, $\eta^{\text{local}}=3.86\times10^{13}$ GeV) impacted by T-Model inflation ($n=1$, $\alpha=10^{-4}$), with $N_{\text{CS}} \lesssim 48$ e-folds.}
        \label{fig:Vary Gmu1}
    \end{subfigure}
    \hfill
    \begin{subfigure}{0.6\textwidth}
        \centering
        \includegraphics[width=\textwidth]{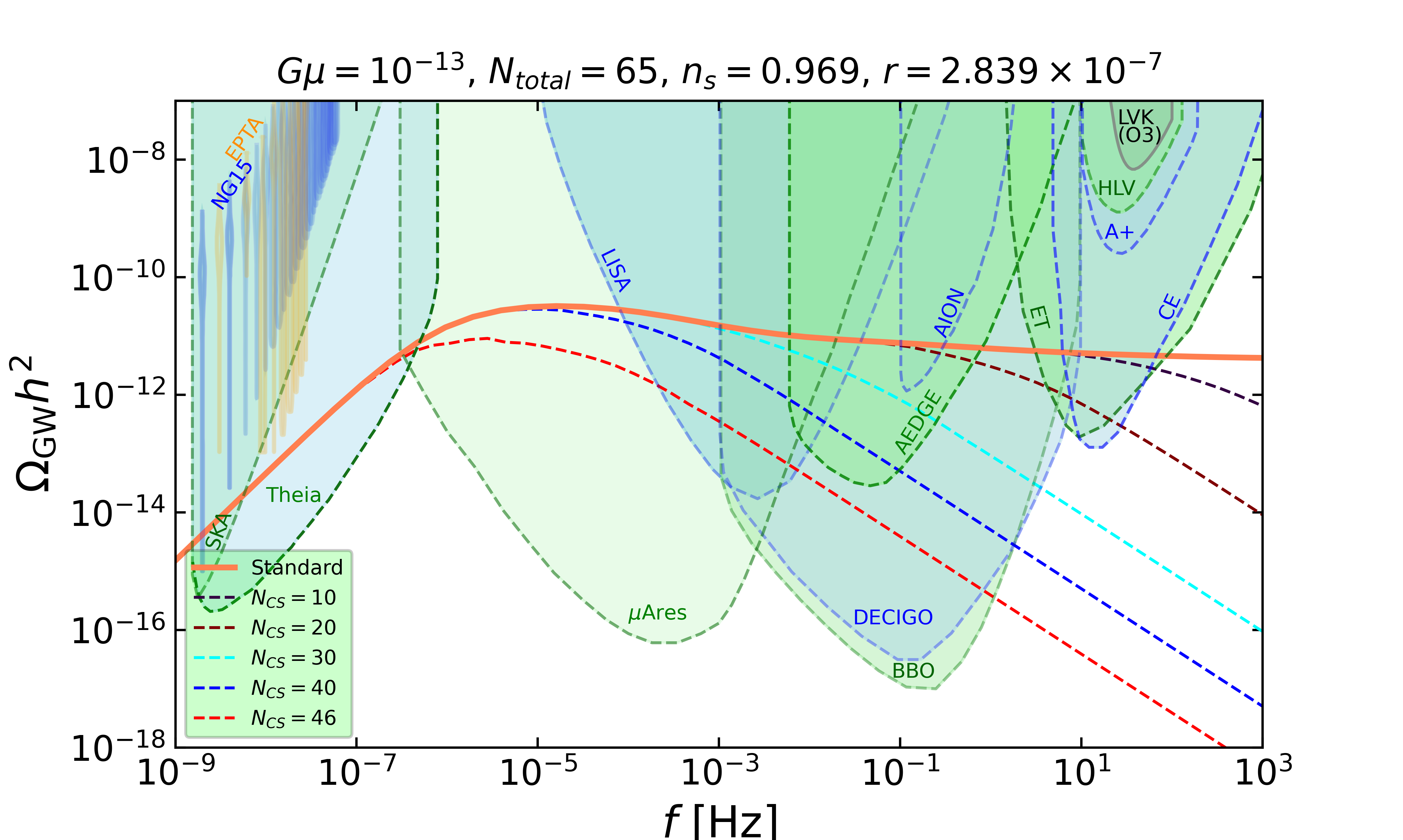}
        \caption{\it GW spectrum from local CS ($G\mu=10^{-13}$, $\eta^{\text{local}}=3.86\times10^{12}$ GeV) impacted by T-Model inflation ($n=1$, $\alpha=10^{-4}$), with $N_{\text{CS}} \lesssim 46$ e-folds.}
        \label{fig:Vary Gmu2}
    \end{subfigure}
    \hfill
    \begin{subfigure}[b]{0.6\textwidth}
        \centering
        \includegraphics[width=\textwidth]{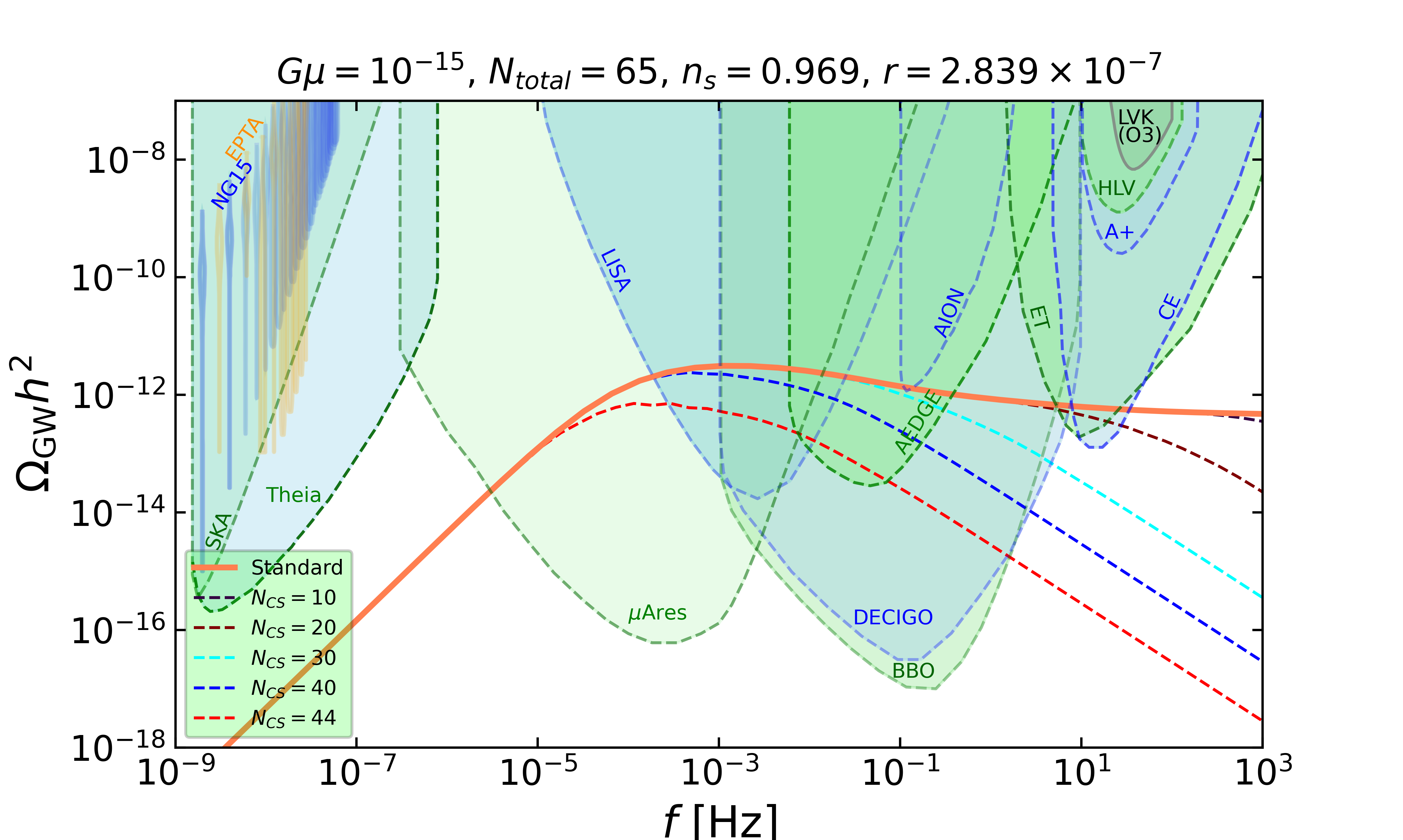}
        \caption{\it GW spectrum from local CS ($G\mu=10^{-15}$, $\eta^{\text{local}}=3.86\times10^{11}$ GeV) impacted by T-Model inflation ($n=1$, $\alpha=10^{-4}$), with $N_{\text{CS}} \lesssim 44$ e-folds.}
        \label{fig:Vary Gmu3}
    \end{subfigure}

    \caption{\it Variation with different $G\mu$ / $\eta^{local}$ keeping $r$ constant.}
    \label{fig: Vary Gmus}
\end{figure}

In figures \ref{fig:1e-11 alpha10 n1 N70}, \ref{fig:1e-11 alpha1 n1 N70}, \ref{fig:1e-11 alpha1e-4 n1 N70} and \ref{fig: Vary Gmus} we have our modified GW spectra from local CS that were formed during inflation and has experienced a certain e-fold of inflation. From the plots, we can determine the points from which the modified spectra deviate from the `standard' spectra, depending on how many e-folds of inflation it has suffered. This is where the turning-point frequency (given by Eq. (\ref{turn freq})) lies. If our CS has suffered lesser e-folds of inflation, then its corresponding GW spectrum (formed due to the emission of GWs after the string network re-enters the horizon) deviates from the `standard' behaviour at higher frequencies. Similarly, if it has suffered a greater amount of inflation, the spectra start to deviate right away from the low-frequency regime. Also, the more inflation the strings suffer, the narrower the flat plateau region becomes in the GW spectrum emitted by these strings. From the figures mentioned above, one can also point out that as the energy scale of inflation $E_{inf}$ (controlled by $r$) decreases, the maximum permissible e-fold of inflation $N_{CS}$ that the string undergoes also decreases. The $f^{-1}$ behaviour at higher frequencies of the non-standard GW spectrum for both local and global CS is a result of the late entry time of the string network in the horizon, which delays the time of loop production. Once the rate of loop production by both the standard and non-standard networks becomes equal as they reach the linear scaling regime, their low-frequency spectra coincide \cite{Guedes:2018afo}. We will emphasize more about the bounds on $N_{CS}$ later on in section \ref{Complementary CS GW discussion}. 

\begin{figure}[H]
    \centering
    \begin{subfigure}{0.7\textwidth}
        \centering
        \includegraphics[width=\textwidth]{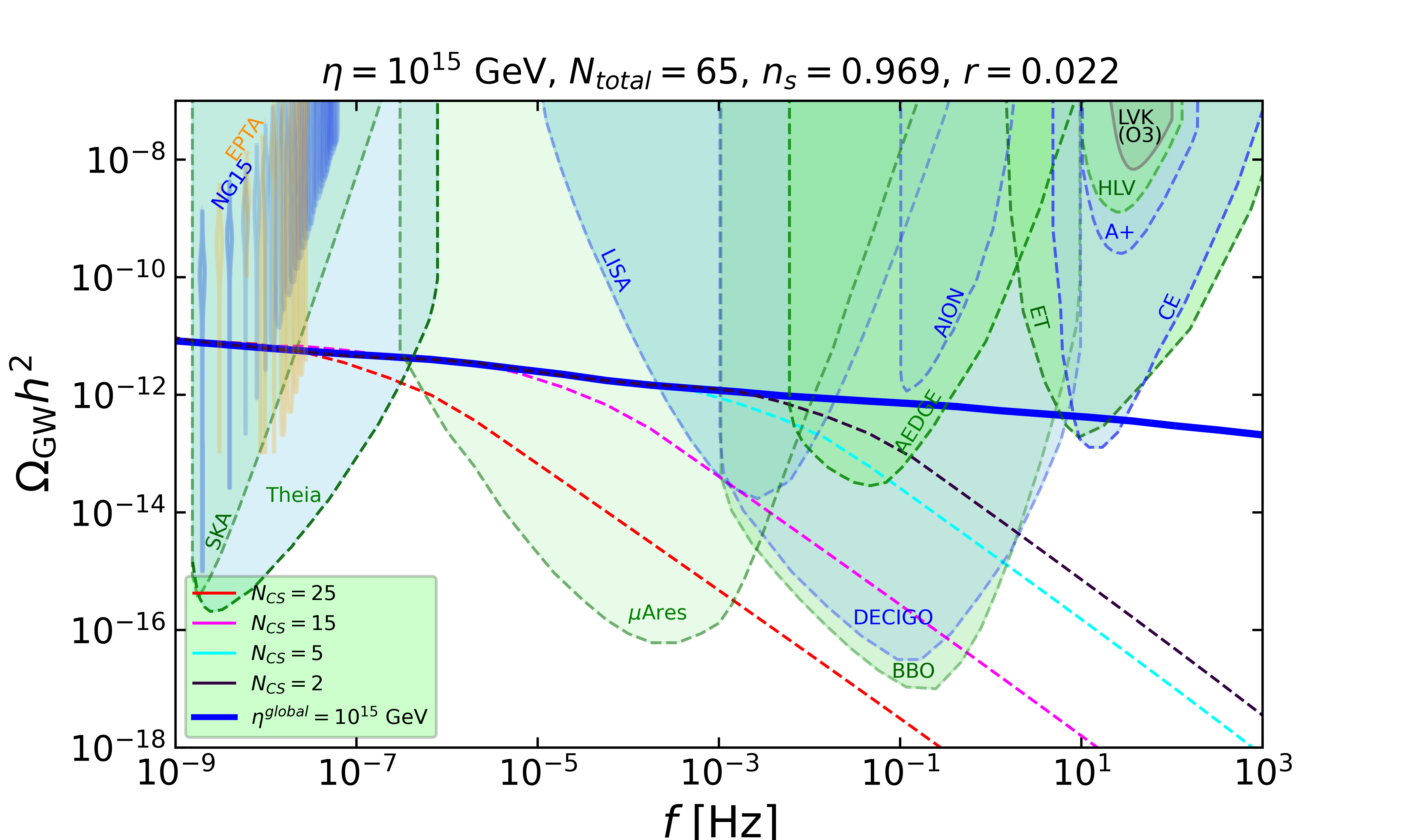}
        \caption{\it $\eta^{global}=10^{15}$ GeV}
        \label{fig:Vary eta 1 GlobalCS}
    \end{subfigure}
    \hfill
    \begin{subfigure}{0.7\textwidth}
        \centering
        \includegraphics[width=\textwidth]{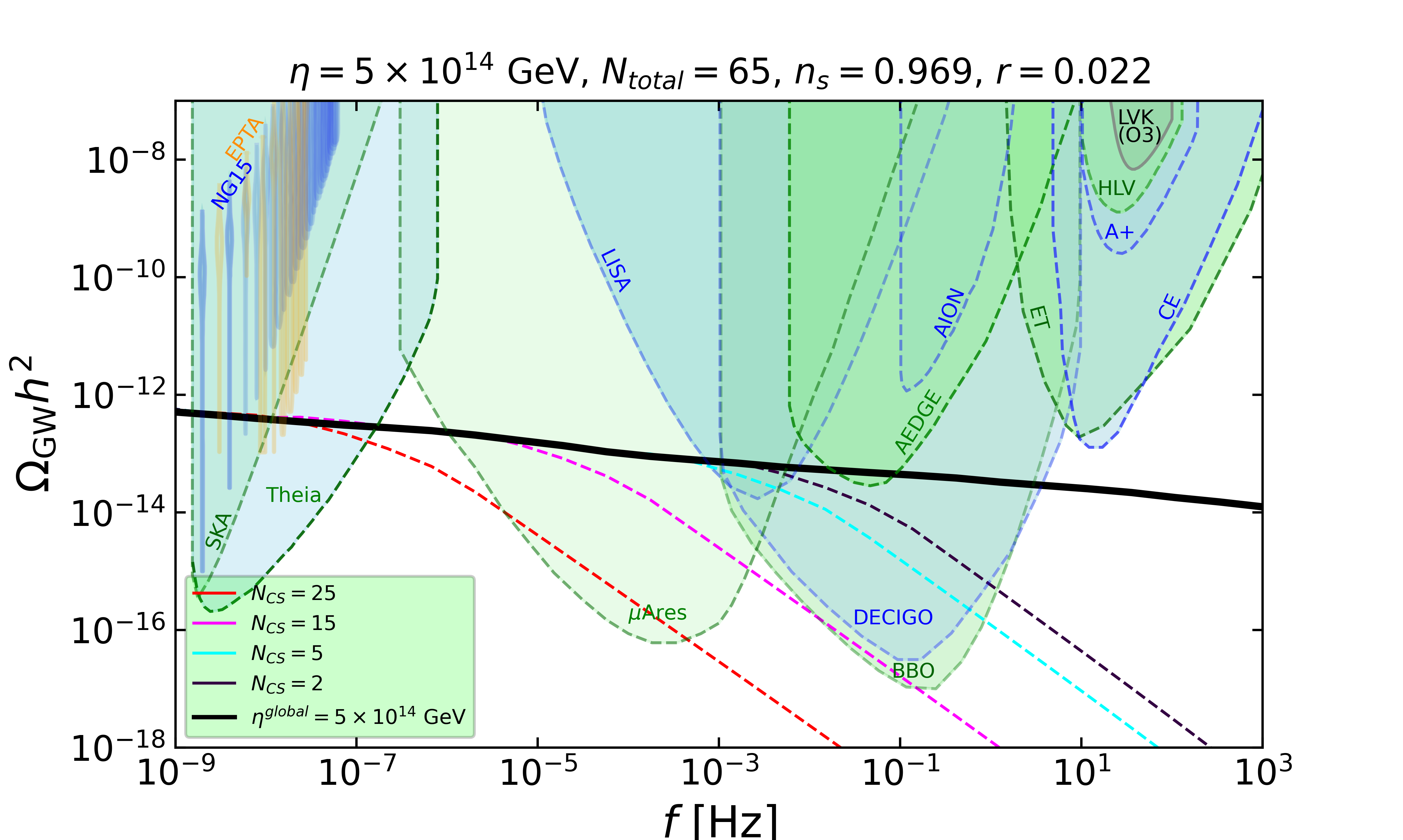}
        \caption{\it $\eta^{global}=5 \times10^{14}$ GeV}
        \label{fig:Vary eta 2 GlobalCS}
    \end{subfigure}
    \hfill

    \caption{\it In a similar manner as described above for local strings, these two plots show the impact of inflation on the GW spectrum due to global cosmic strings. The T-Model of $\alpha-$attractor with n=1 and $\alpha=1$ was considered.}
    \label{fig: Vary etas for GlobalStrings}
\end{figure}

Fig. \ref{fig: Vary etas for GlobalStrings} shows the modified GW spectra from global CS that were formed during inflation and have suffered certain e-folds of inflation. The turning point frequency for the global cosmic string case is independent of the string tension, as pointed out in Eq. (\ref{global turn freq}).

\medskip

\section{Complementarity probe of Inflation via CMB and GW detectors} \label{Complementary CS GW discussion}

Cosmic Strings, both local and global, that were formed during the inflationary phase can serve as a unique relic for exploring both the inflationary phase and the post-inflationary universe. The formation of these topological defects during inflation directly links their properties, such as the number of e-folds they experience ($N_{CS}$), to the inflationary observables like the tensor-to-scalar ratio ($r$) and spectral tilt ($n_s$). We leverage the GW signals emitted by the cosmic strings that are within the reach of the current and future detectors like LISA, SKA and DECIGO/BBO to find a complementary window into the inflationary parameter space. This section explores the distinct GW signatures of local and global cosmic strings produced during inflation and highlights their relevance in constraining inflationary models.


\subsection{Local Cosmic Strings} \label{LCS Comp GW}

We can do some simplification using Eqs. (\ref{turn freq}), (\ref{re-entry temp}) and (\ref{einf to r}) and end up with the following relation between the turning-point frequency and the tensor-to-scalar ratio, $r$:
\begin{equation}\label{complementarity local}
    r = 5.257 \times 10^{-55} \times (f_{\Delta}\cdot \exp N_{e}^{CS})^4 \times \left(\frac{G\mu}{10^{-11}}\right)^2.
\end{equation}

The turning-point frequency, $f_{\Delta}$ (in Hertz) gives information about the critical frequency from where the GW spectra deviate from the predictions obtained in standard cosmology. This critical frequency is influenced by $N_{e}^{CS} \equiv N_{CS}$, which is the number of e-folds of inflation that the cosmic string has suffered. Additionally, the cosmic string tension, $G\mu$ (or alternatively the VEV of the scalar field $\eta^{local}$), controls the amplitude of the GW spectra, thereby impacting its detectability by different GW observatories. The turning-point frequency can fall within the operational frequency range of both present and future GW detectors and can be an excellent probe for understanding the physics of cosmic strings impacted by inflation. Let us understand this in more detail by considering the following GW detectors:

\begin{enumerate}
    \item SKA: $f_{\Delta}^{SKA}\simeq1.54\times10^{-9} $ Hz to $1.69\times10^{-6} $ Hz and $G\mu \lesssim 10^{-13}$.

     \item LISA: $f_{\Delta}^{LISA}\simeq6.45\times10^{-6} $ Hz to $82.7$ Hz and $G\mu \lesssim 10^{-17}$.

    \item DECIGO/BBO: $f_{\Delta}^{DECIGO/BBO}\simeq 10^{-3} $ Hz to $10$ Hz and $G\mu \lesssim 10^{-20}$.

    
\end{enumerate}

Suppose the total e-fold of inflation that the universe has undergone is $N_{total} = 65$ (constrained by $r<0.036$), and consider a CS of string tension $G\mu = 10^{-11}$ being produced at some point during these 65 e-folds of inflation. The string would re-enter the horizon at a later time and will potentially be detected at one of the GW detectors provided it experiences a suitable number of e-folds, $N_{CS}$, such that the turning-point frequency of its GW spectra lies within the working frequency range of a specific observatory, say the Square Kilometre Array (SKA).


Let us consider the T-model of inflation with $n=1$ and $\alpha=1$ (see Eq. (\ref{n=1 alpha=1 TModel})). By setting $G\mu=10^{-11}$ in Eq. (\ref{complementarity local}), we can find all the possible values of $N_{CS}$ \textemdash the number of e-folds of inflation the strings have experienced, so that $f_{\Delta}$ falls within SKA's frequency range. Constrained by the tensor-to-scalar ratio $r<0.036$, we look for all possible integer values of $N_{CS} \in [1,65]$ that satisfy these conditions.

For each of the allowed values of $N_{CS}$, we can calculate the corresponding $n_s$ values using Eq. (\ref{nsrT1}), ensuring that it lies within the aforementioned observational range. Using this approach, we can identify all possible combinations of $n_s, r$ and $N_{CS}$, providing a window into the inflationary dynamics that could produce detectable gravitational waves from cosmic strings. Quantitatively, we can also write:

\begin{equation}
    N_{total} = N_{CS} + N_{form}
\end{equation}
where $N_{form}$ is the e-fold at which the CS forms and $N_{CS}$ as discussed previously is the e-folds that the CS experiences.



\begin{figure}[H] 
    \centering
    \begin{subfigure}{0.4962\textwidth} 
        \centering
        \includegraphics[width=\linewidth]{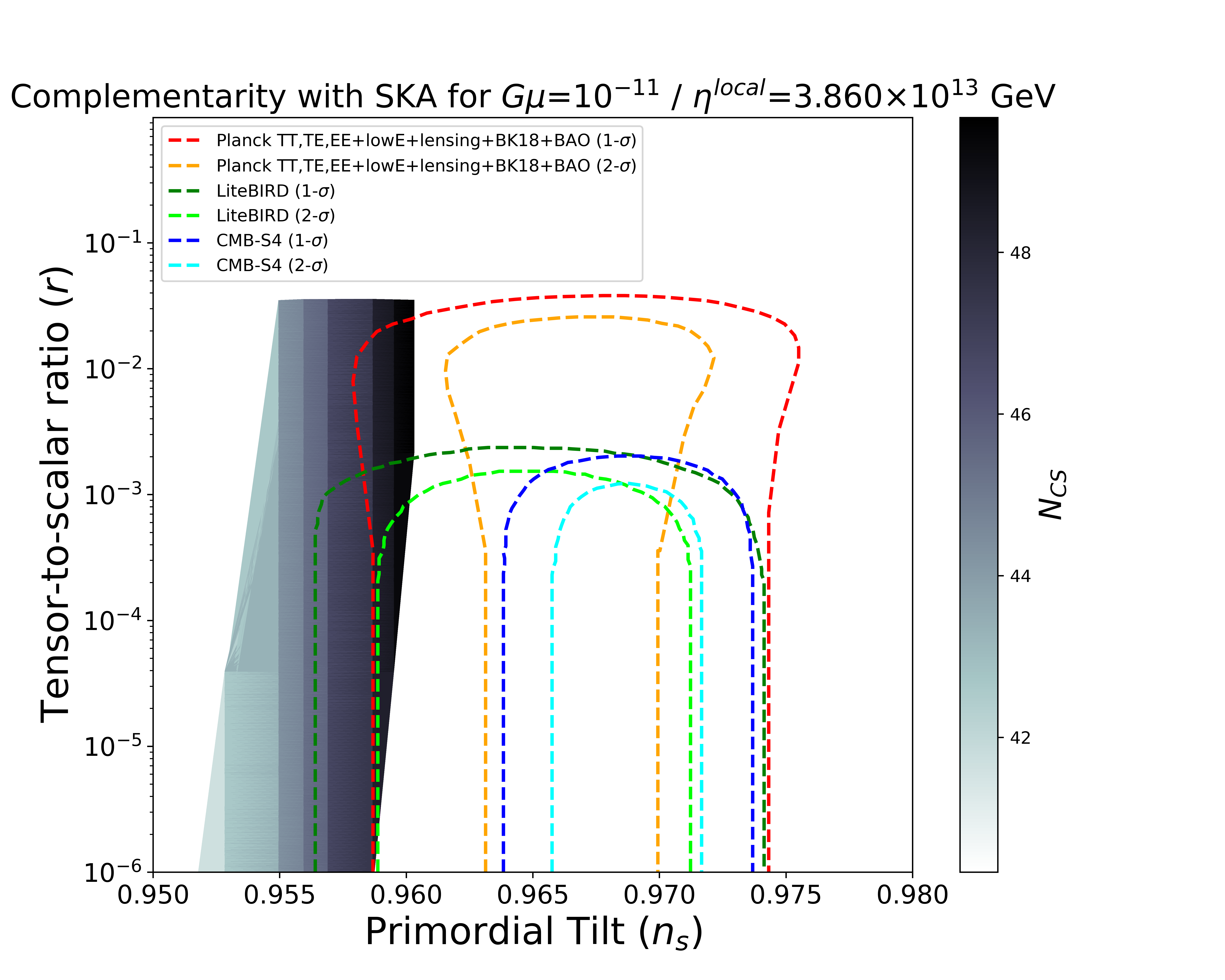}
        \label{fig:SKA Complement 1}
    \end{subfigure}
    \hfill
    \hfill
    \begin{subfigure}{0.4962\textwidth} 
        \centering
        \includegraphics[width=\linewidth]{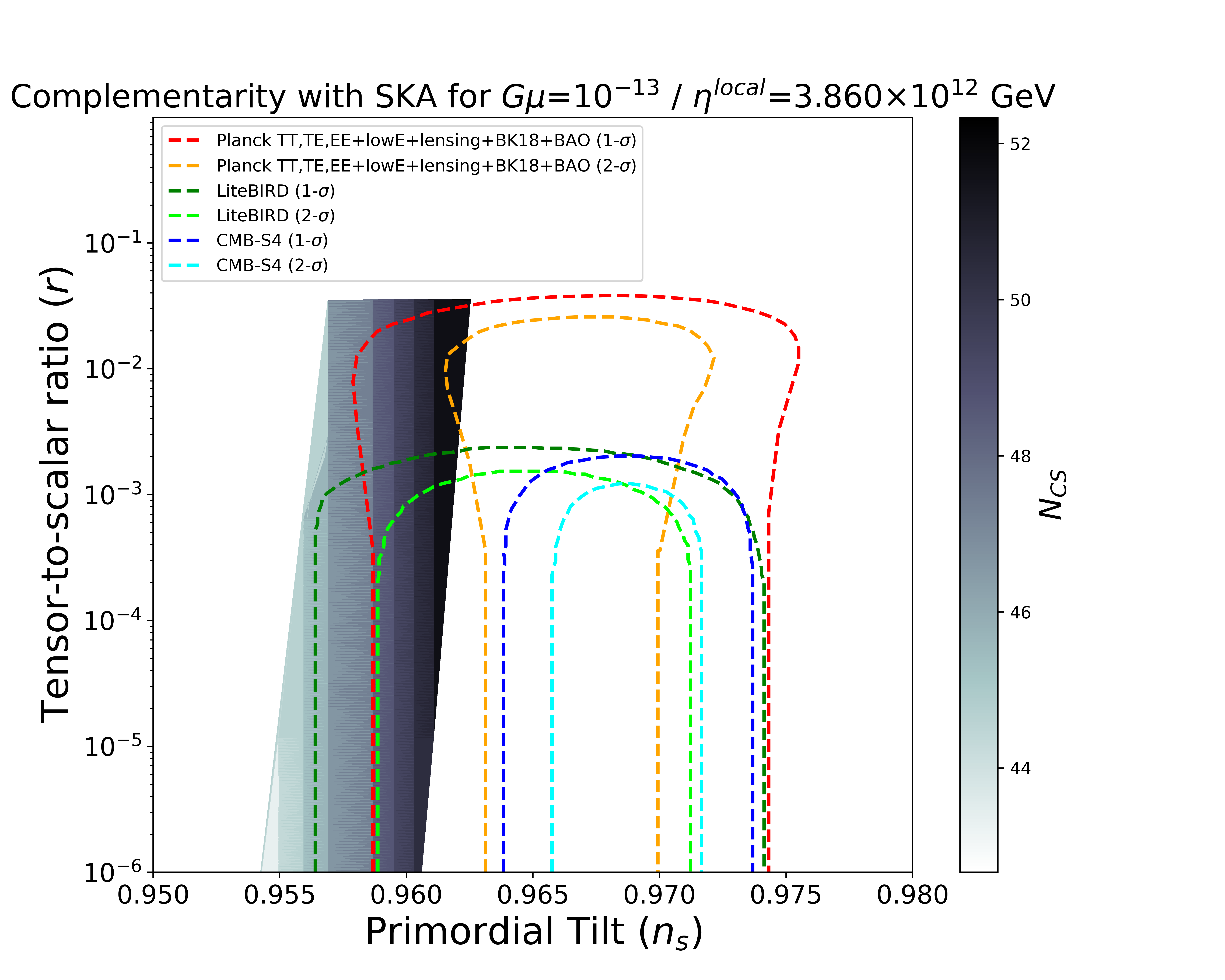}
        \label{fig:SKA Complement 2}
    \end{subfigure}

     \begin{subfigure}{0.4962\textwidth} 
        \centering
        \includegraphics[width=\linewidth]{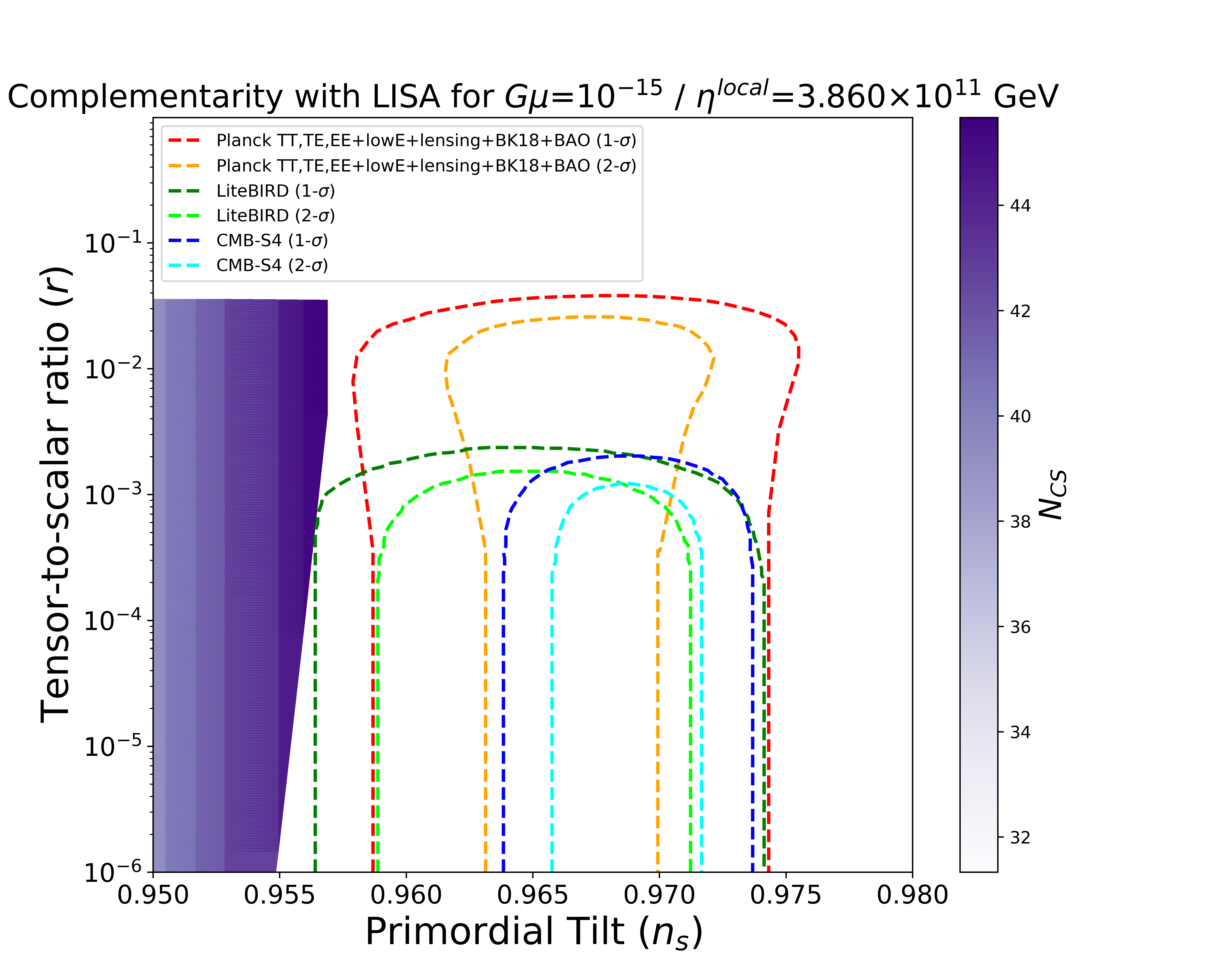}
        \label{fig:LISA Complement 1}
    \end{subfigure}
    \hfill
    \hfill
     \begin{subfigure}{0.4962\textwidth} 
        \centering
        \includegraphics[width=\linewidth]{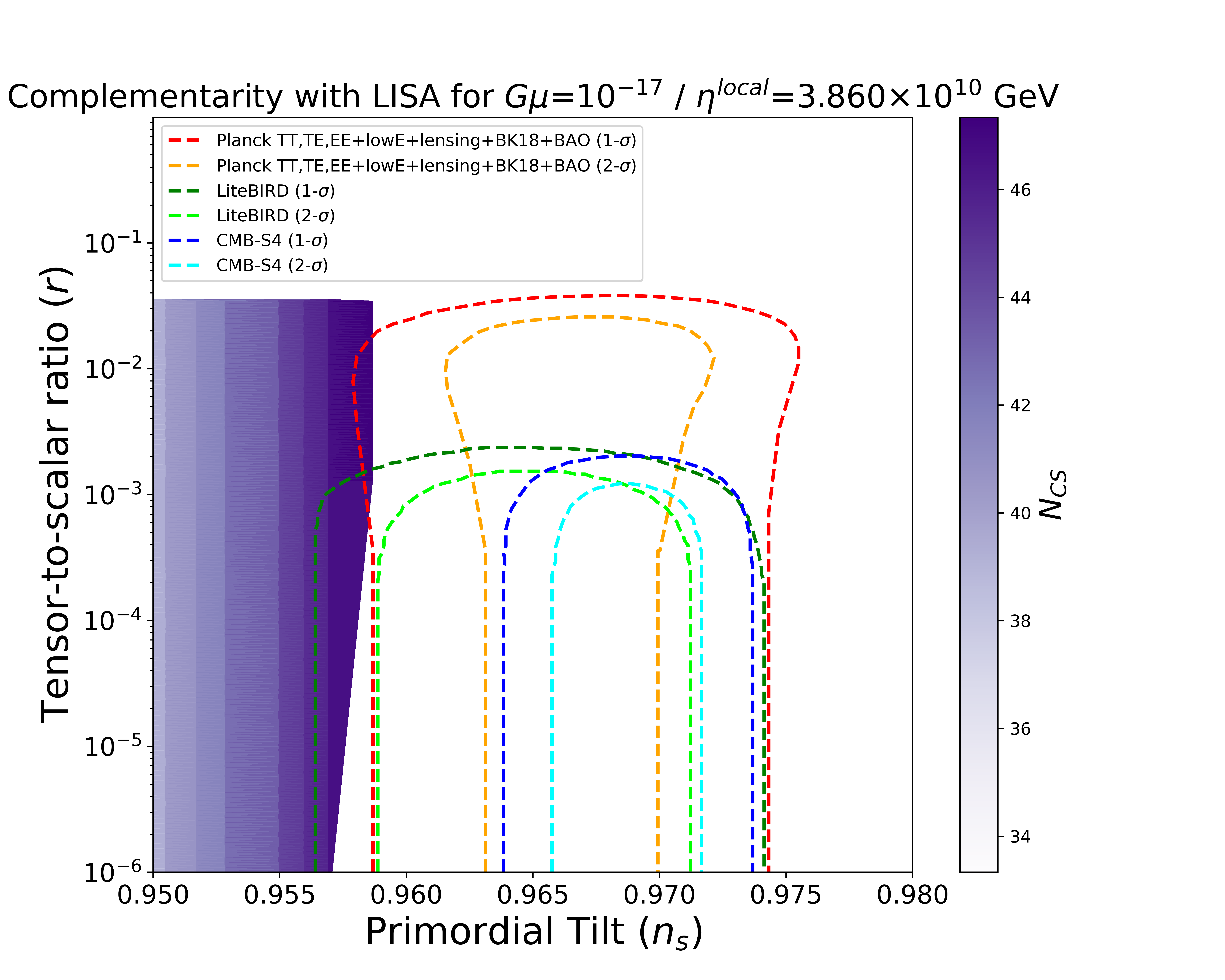}
        \label{fig:LISA Complement 2}
    \end{subfigure}

    \caption{\it Complementarity between CMB and GW detectors (SKA and LISA) in probing local cosmic strings of different string tensions impacted by inflation. In each of the scenarios above, we have considered inflation to be driven by the T-Model of alpha-attractor with $n=1$ and $\alpha=1$ and $N_{total}=65$. The colour bars on the $n_s-r$ plane indicate the potential $(n_s,r)-$values for which we can detect their corresponding GW spectra at a given detector.}
    \label{fig:Complementarity_CMB_GW 1}
\end{figure}

\begin{figure}[H] 
    \centering
    \begin{subfigure}{0.4962\textwidth} 
        \centering
        \includegraphics[width=\linewidth]{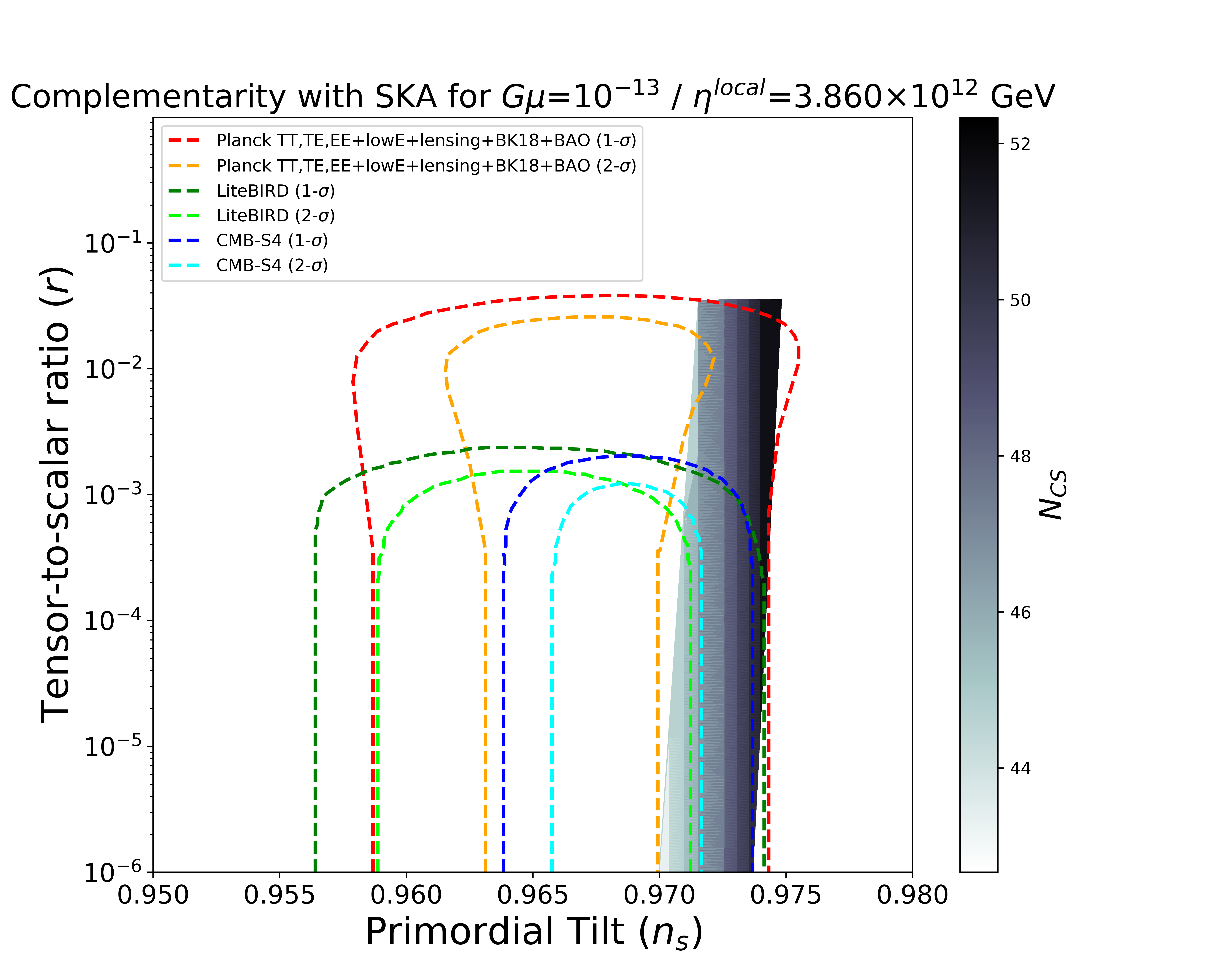}
        \label{fig:SKA Complement 0.1}
    \end{subfigure}
    \hfill
    \hfill
    \begin{subfigure}{0.4962\textwidth} 
        \centering
        \includegraphics[width=\linewidth]{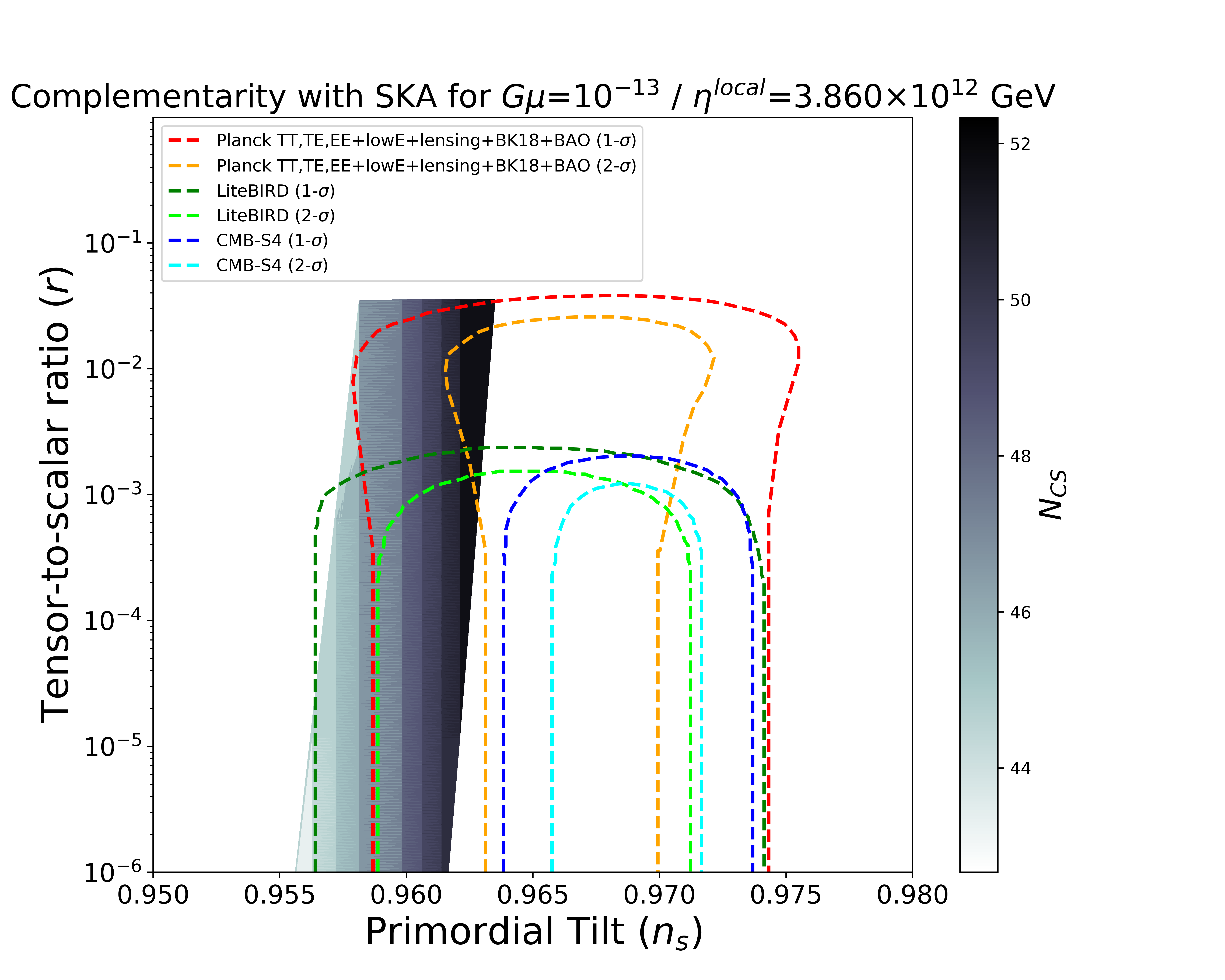}
        \label{fig:SKA Complement 0.8}
    \end{subfigure}

     \begin{subfigure}{0.4962\textwidth} 
        \centering
        \includegraphics[width=\linewidth]{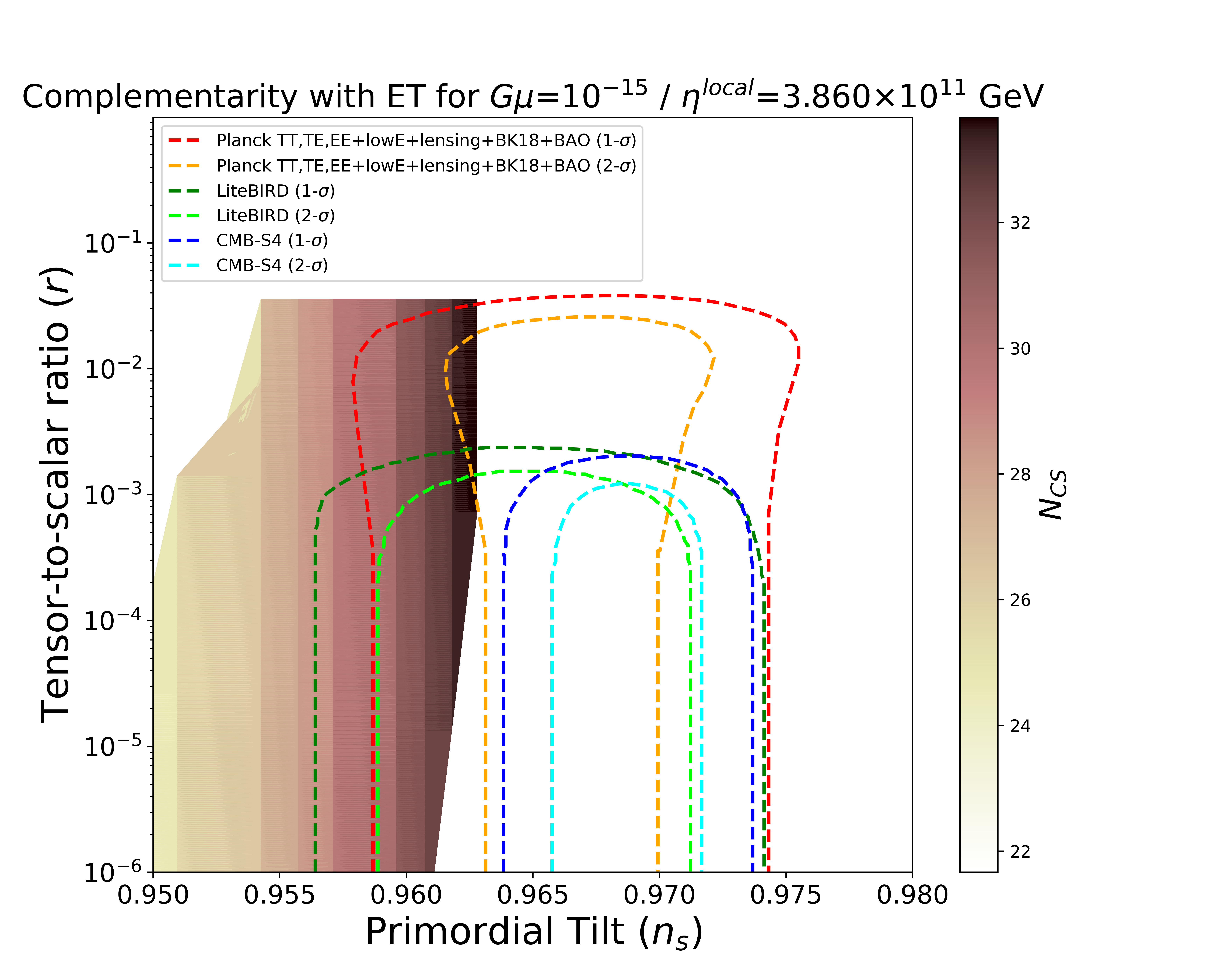}
        \label{fig: ET Complement 0.1}
    \end{subfigure}
    \hfill
    \hfill
     \begin{subfigure}{0.4962\textwidth} 
        \centering
        \includegraphics[width=\linewidth]{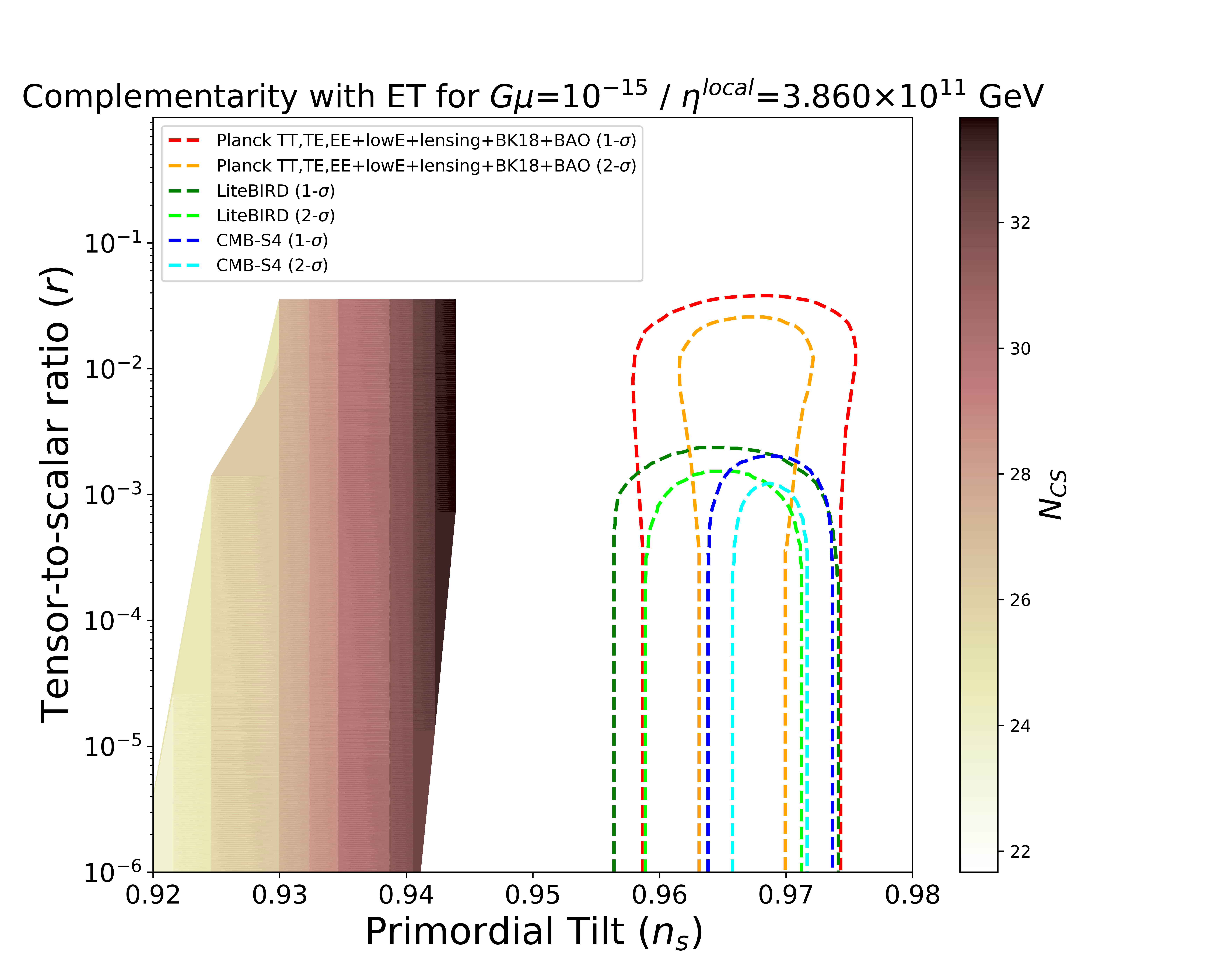}
        \label{fig: ET Complement 0.8}
    \end{subfigure}

      \begin{subfigure}{0.4962\textwidth} 
        \centering
        \includegraphics[width=\linewidth]{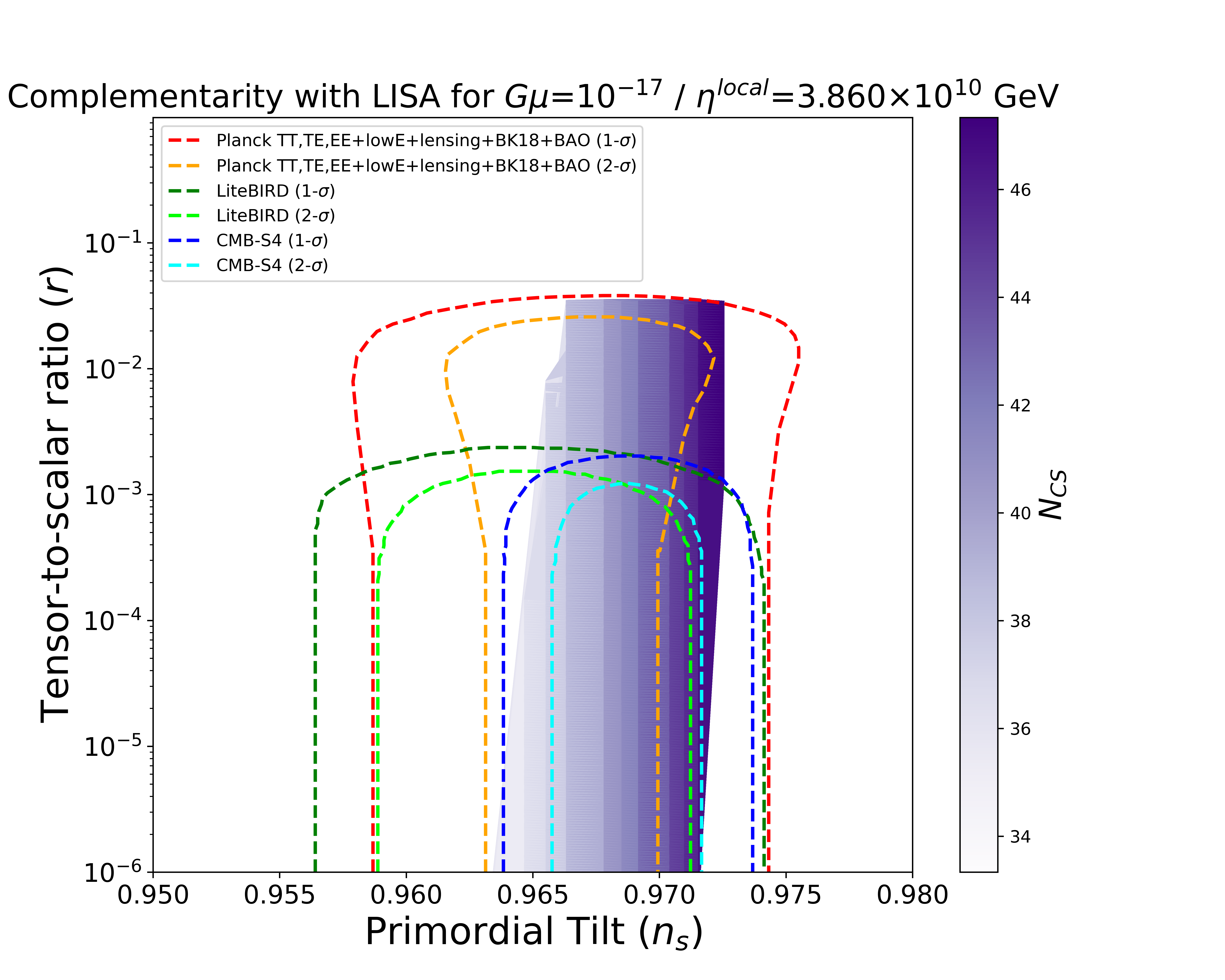}
        \label{fig:LISA Complement 0.1}
    \end{subfigure}
    \hfill
    \hfill
     \begin{subfigure}{0.4962\textwidth} 
        \centering
        \includegraphics[width=\linewidth]{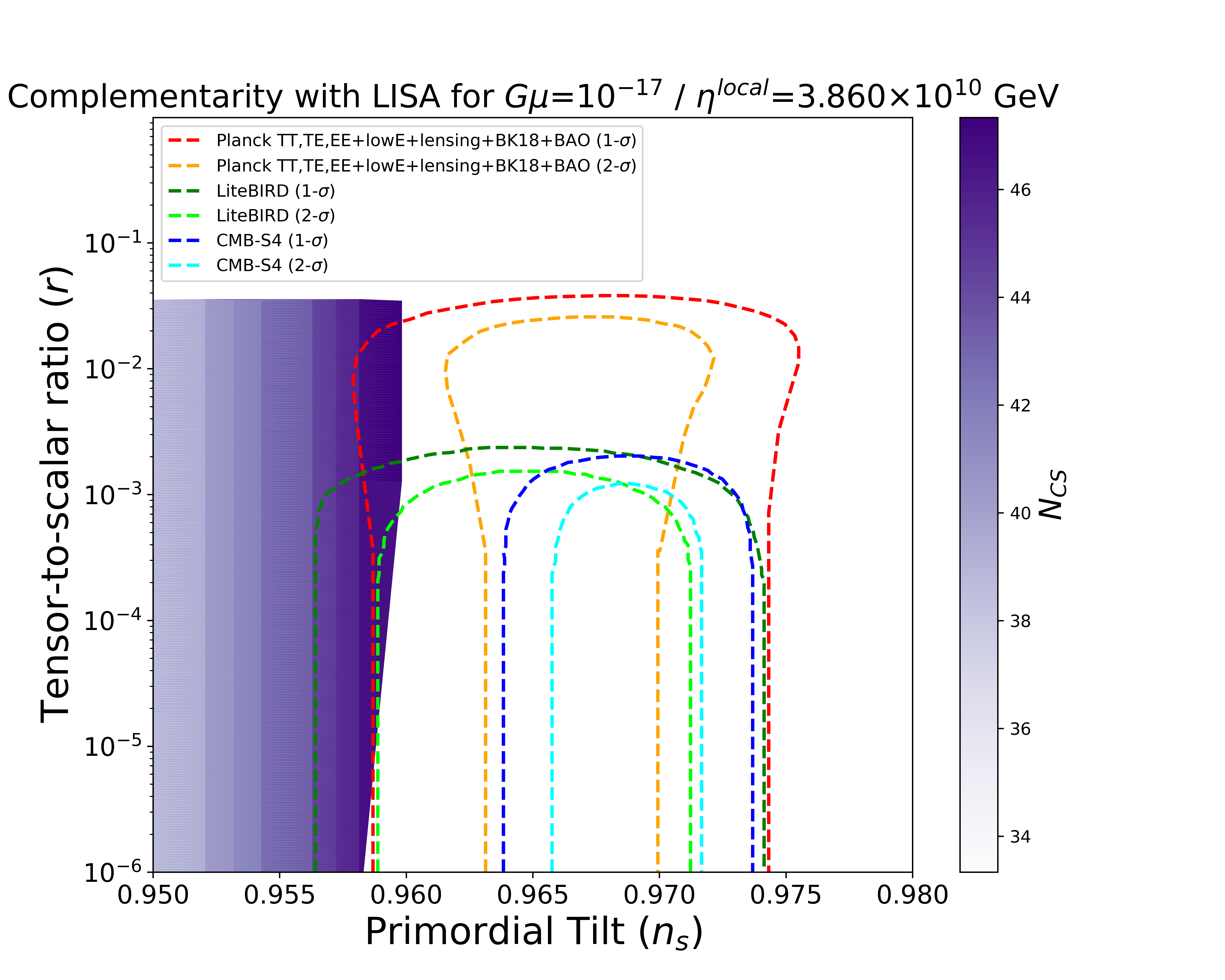}
        \label{fig:LISA Complement 0.8}
    \end{subfigure}

    \caption{\it Complementarity between CMB and GW detectors (SKA, ET and LISA) in probing local cosmic strings of different string tensions impacted by inflation. Here, we have considered inflation to be driven by the T-Model of alpha-attractor with $n=0.1$ (left panel) and $n=0.8$ (right panel). We have set $\alpha=10$ and $N_{total}=65$ for all the plots.}
    \label{fig:Complementarity_CMB_GW 2}
\end{figure}

\begin{figure}[H] 
    \centering
    \begin{subfigure}{0.4962\textwidth} 
        \centering
        \includegraphics[width=\linewidth]{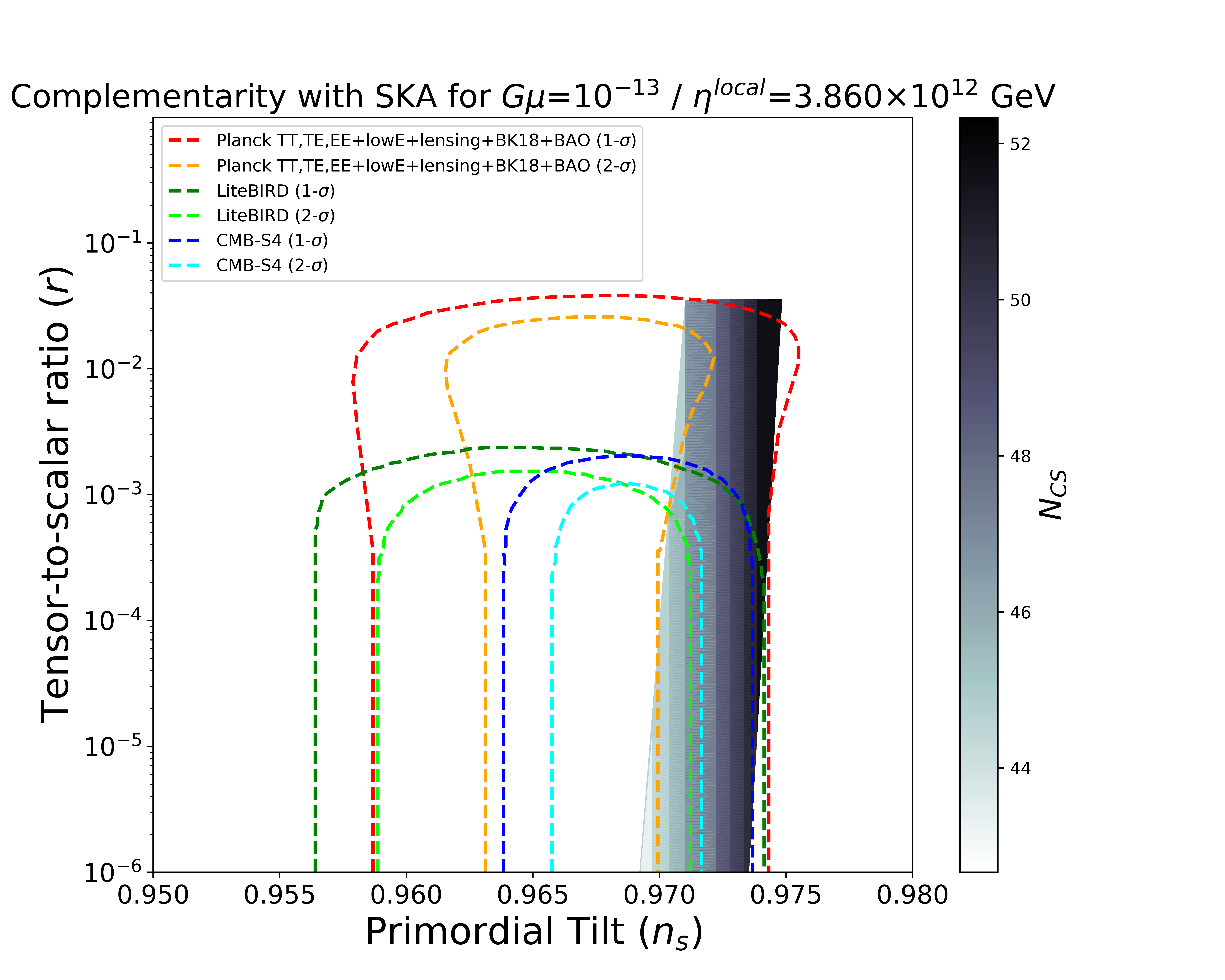}
        \label{fig: SKA Complement P1}
    \end{subfigure}
    \hfill
    \hfill
    \begin{subfigure}{0.4962\textwidth} 
        \centering
        \includegraphics[width=\linewidth]{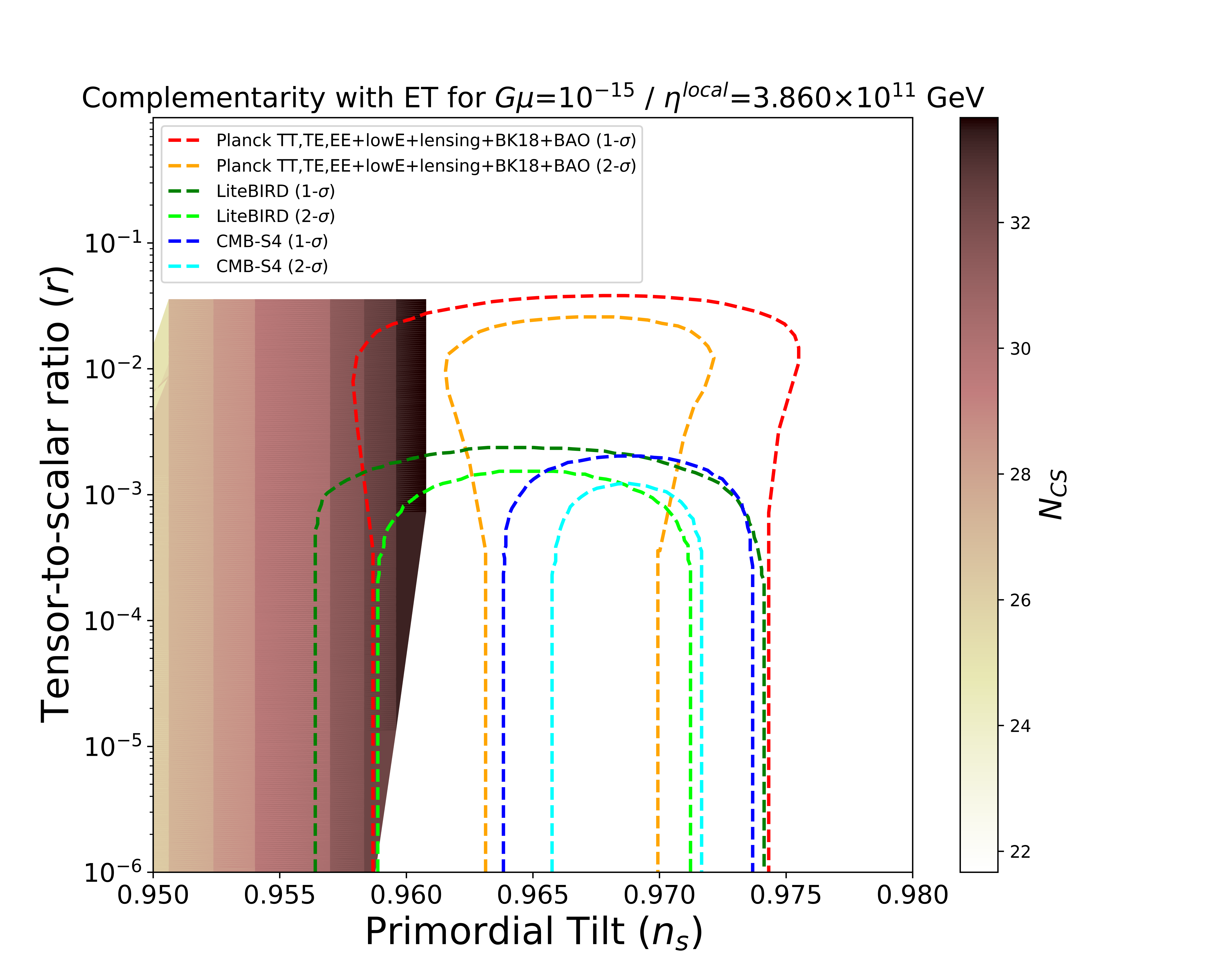}
        \label{fig: ET Complement P1}
    \end{subfigure}

     \begin{subfigure}{0.4962\textwidth} 
        \centering
        \includegraphics[width=\linewidth]{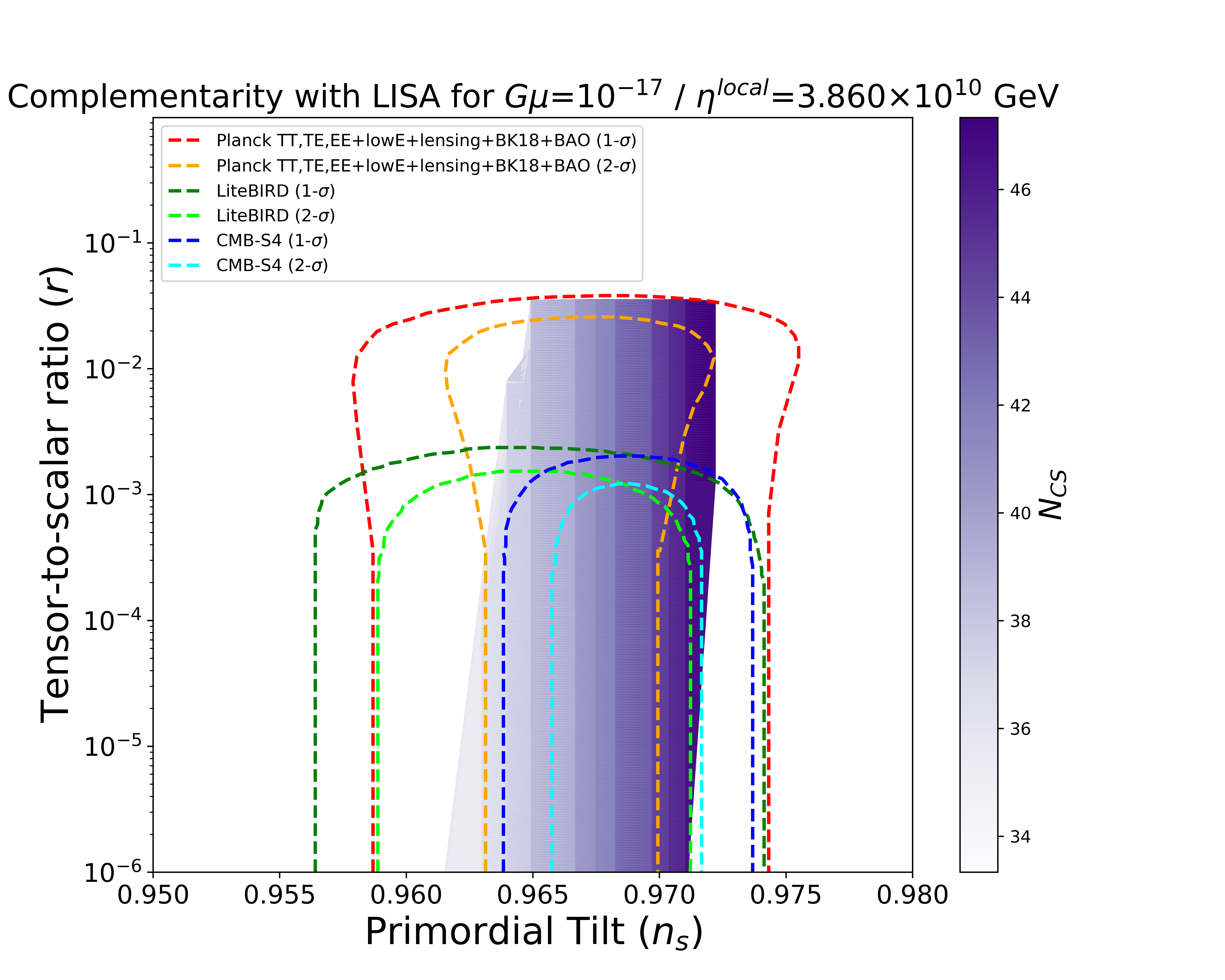}
        \label{fig: LISA Complement P1}
    \end{subfigure}
    \hfill
    \hfill
     \begin{subfigure}{0.4962\textwidth} 
        \centering
        \includegraphics[width=\linewidth]{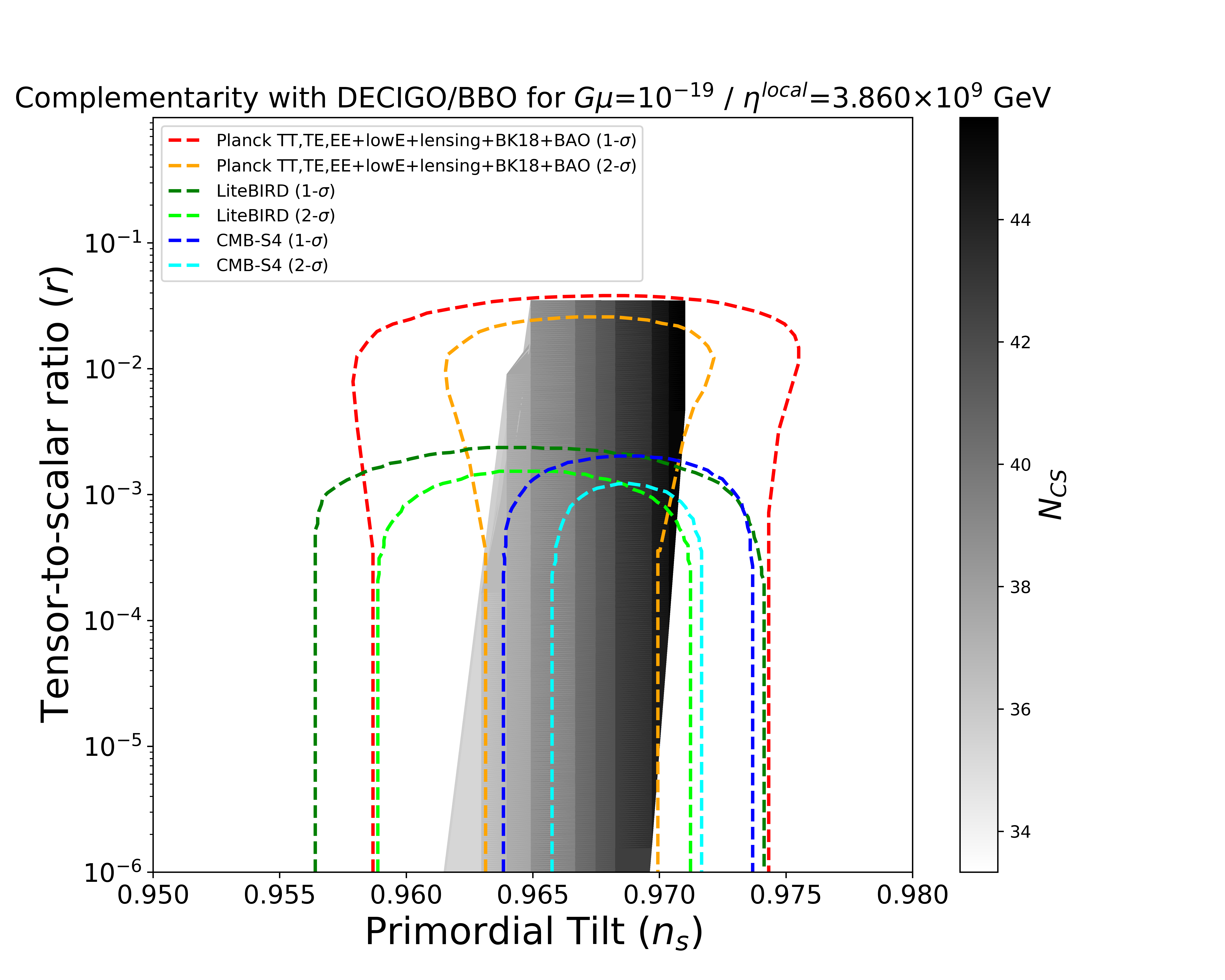}
        \label{fig: Decigo Complement P1}
    \end{subfigure}

    \caption{\it Complementarity between CMB and GW detectors (SKA, ET, LISA and DECIGO/BBO) in probing local cosmic strings of different string tensions impacted by inflation. We have considered inflation to be driven by the Polynomial attractor (Eq. \ref{Poly Eq1}) with $n=1$ and $N_{total}=65$ for all the plots.}
    \label{fig:Complementarity_CMB_GW 3}
\end{figure}

\begin{figure}[H] 
    \centering
    \begin{subfigure}{0.4962\textwidth} 
        \centering
        \includegraphics[width=\linewidth]{Plots/Fig29.png}
        \label{fig: LISA Complement P01}
    \end{subfigure}
    \hfill
    \hfill
    \begin{subfigure}{0.4962\textwidth} 
        \centering
        \includegraphics[width=\linewidth]{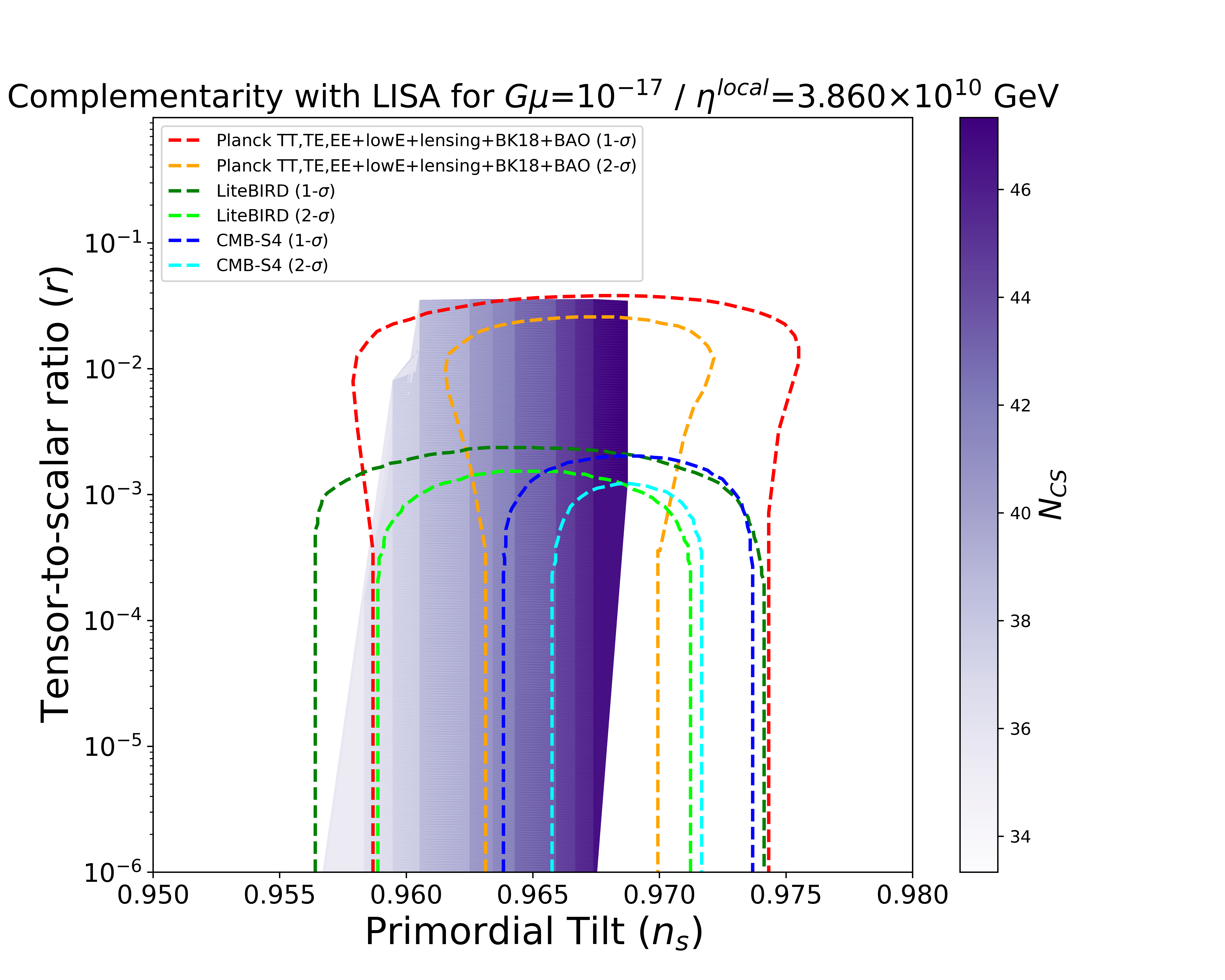}
        \label{fig: LISA Complement P2}
    \end{subfigure}

     \begin{subfigure}{0.4962\textwidth} 
        \centering
        \includegraphics[width=\linewidth]{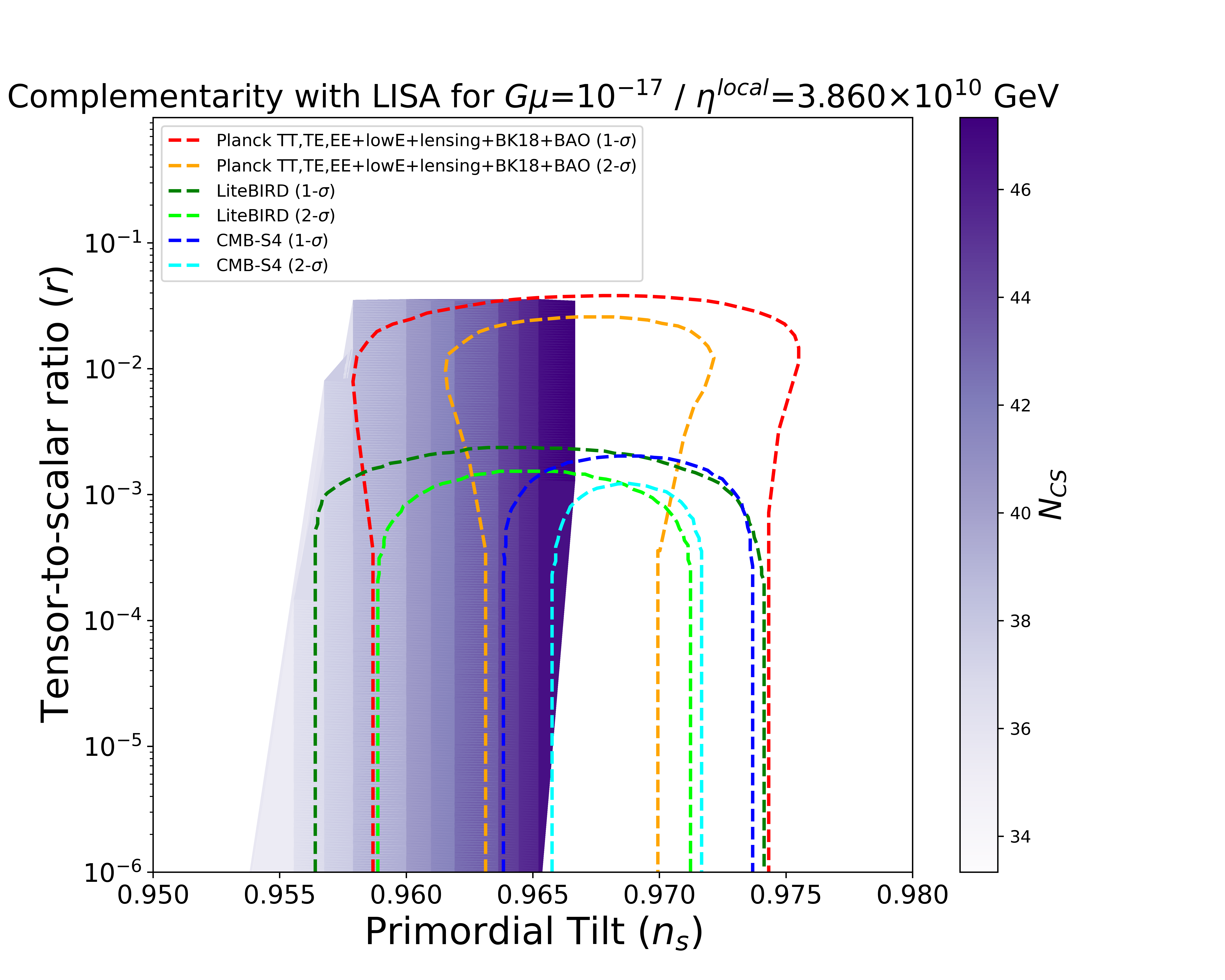}
        \label{fig: LISA Complement P3}
    \end{subfigure}
    \hfill
    \hfill
     \begin{subfigure}{0.4962\textwidth} 
        \centering
        \includegraphics[width=\linewidth]{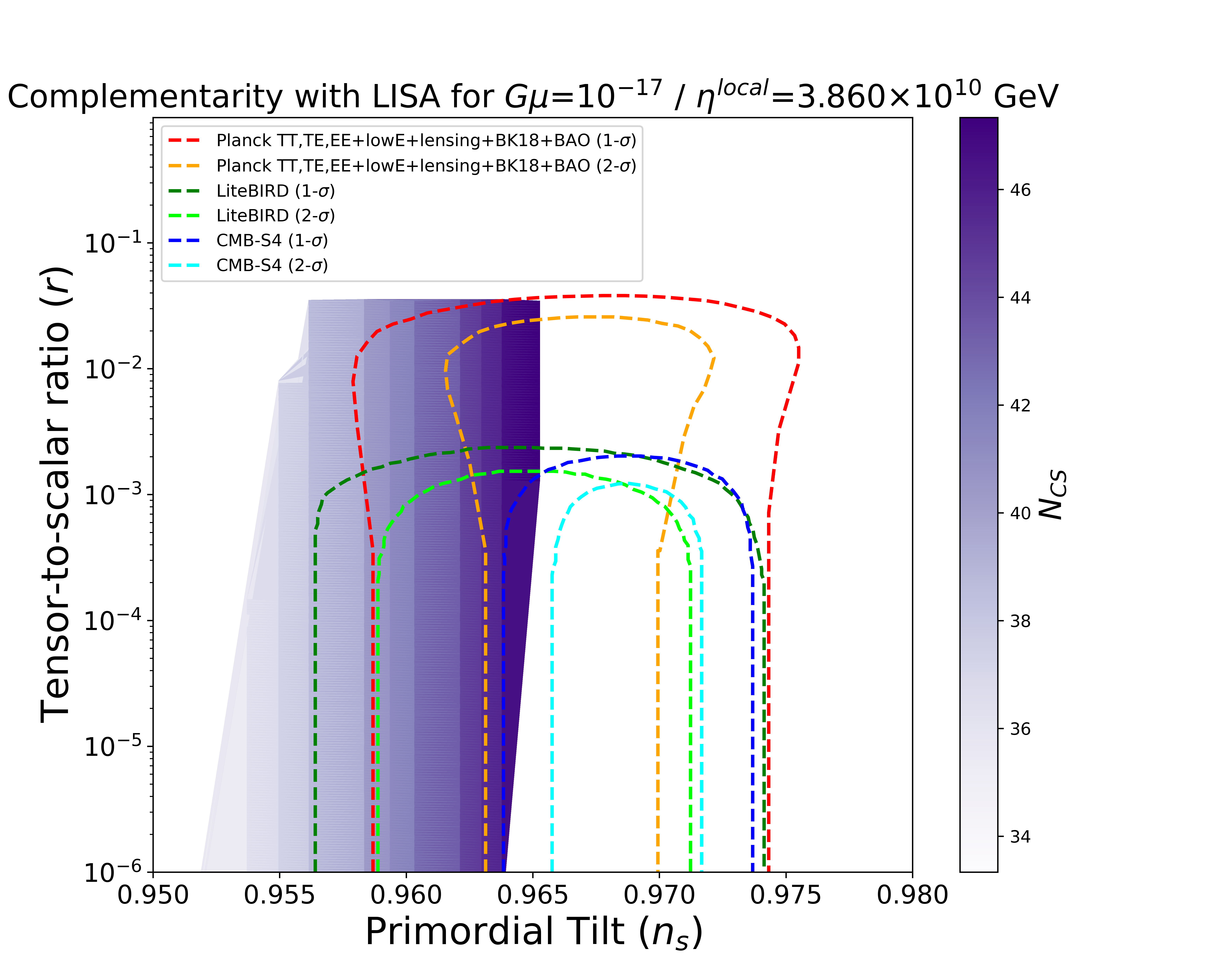}
        \label{fig: LISA Complement P4}
    \end{subfigure}

    \caption{\it Impact on the complementarity of CMB with LISA for local cosmic strings upon varying the exponent $n$ of the Polynomial Attractor Model (Eq. \ref{Poly Eq1}). We have $n=1$ (top-left), $n=2$ (top-right), $n=3$ (bottom-left), and $n=4$ (bottom-right). $N_{total}$ is fixed at 65 e-folds.}
    \label{fig:Complementarity_CMB_GW 4} 
\end{figure}

Figures \ref{fig:Complementarity_CMB_GW 1}, \ref{fig:Complementarity_CMB_GW 2}, \ref{fig:Complementarity_CMB_GW 3} and \ref{fig:Complementarity_CMB_GW 4} provide crucial information connecting inflationary physics to observable GW signatures. It captures the range of inflationary parameters ($n_s$, $r$) that enables cosmic strings (formed during inflation) to produce observable GWs which can be detected at the present and future observatories. Consider the top-left plot in Fig. \ref{fig:Complementarity_CMB_GW 4}, which considers the Polynomial attractor inflation model with $n=1$ (using Eq. (\ref{Poly Eq1}):
\begin{equation}\label{Poly n=1 model}
    V(\phi)=V_0\frac{\phi^{2}}{\phi^{2}+\mu^{2}}.
\end{equation}
The shaded region (in purple) maps the inflationary parameter space/($n_s$, $r$)-plane to detectable GW signals at LISA from cosmic strings formed during inflation. For a given value of ($n_s$, $r$), the associated $N_{CS}$ determines whether the CS network can produce detectable GWs within LISA's sensitivity. This provides a way to constrain inflationary models based on their capability to deliver the right balance of $N_{CS}$ and $G\mu$ / $\eta^{local}$ that can produce GW signals observable by the chosen detector. In other words, inflationary models predicting ($n_s$, $r$) values outside the shaded region can be ruled out as viable candidates owing to their inability to produce cosmic string GWs detectable by LISA. Moreover, the shaded region indicates the range of $N_{CS}$ values that LISA can probe, which is in the range $\sim 34$ to $47$ as indicated by the colour gradient of the shaded region. This range directly depends on the inflationary dynamics and formation epoch of the cosmic string network. While we have focused solely on LISA, other GW detectors can also provide complementary probes, such as SKA. Consider the top-right plot from Fig. \ref{fig:Complementarity_CMB_GW 1} which considers the inflationary scenario for T-Model with $n=1$ and $\alpha=1$. It highlights the corresponding parameter space for SKA, assuming $G\mu=10^{-13}$ (or alternatively $\eta^{local}=3.86\times10^{12}$ GeV). A larger string tension shifts the sensitivity region, enabling SKA to probe higher $N_{CS}$ values in the range $\sim 43$ to $52$, expanding the observational reach to models undetectable by LISA. For instance, consider the complementarity plot with SKA for string tension $G\mu=10^{-13}$ (top left) in Fig. \ref{fig:Complementarity_CMB_GW 3}. The shaded region in the $n_s-r$ plane corresponds to the allowed $n_s-r$ values that our Polynomial Attractor model (given by Eq. (\ref{Poly Eq1})) with $n=1$ can take. Any other $n_s-r$ value outside the shaded region will not be a viable model to drive inflation. One can also infer from this plot, the allowed $N_{CS}$ values which lies in the range $\sim 43$ to $52$. So if our cosmic string of tension $G\mu=10^{-13}$ has experienced any value of $N_{CS}$ in the above-mentioned range, its turning point frequency should lie within SKA's sensitivity. Additionally, we can predict the bounds on $N_{CS}$ at different GW detectors. For example, in Fig. \ref{fig:Complementarity_CMB_GW 3}, SKA can probe $N_{CS}$ in the range $\sim 43$ to $52$, LISA in the range $\sim 34$ to $47$, ET in the range $\sim 22$ to $33$ and DECIGO/BBO in the range $\sim 34$ to $45$.


\subsection{Global Cosmic Strings} \label{GW Strings Comp}
\begin{figure}[H] 
    \centering
    \begin{subfigure}{0.4962\textwidth} 
        \centering
        \includegraphics[width=\linewidth]{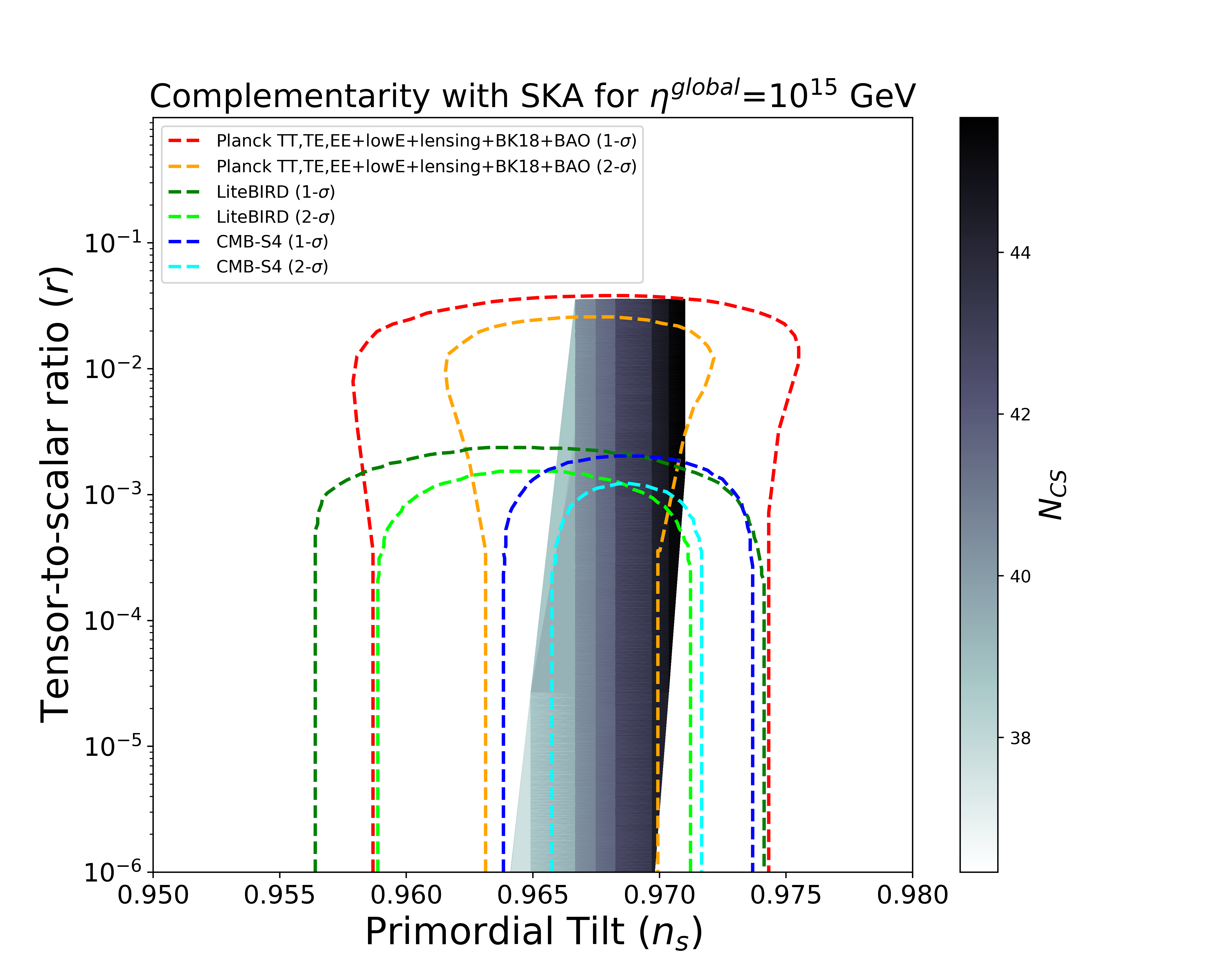}
        \label{fig: SKA Global Complement P01}
    \end{subfigure}
    \hfill
    \hfill
    \begin{subfigure}{0.4962\textwidth} 
        \centering
        \includegraphics[width=\linewidth]{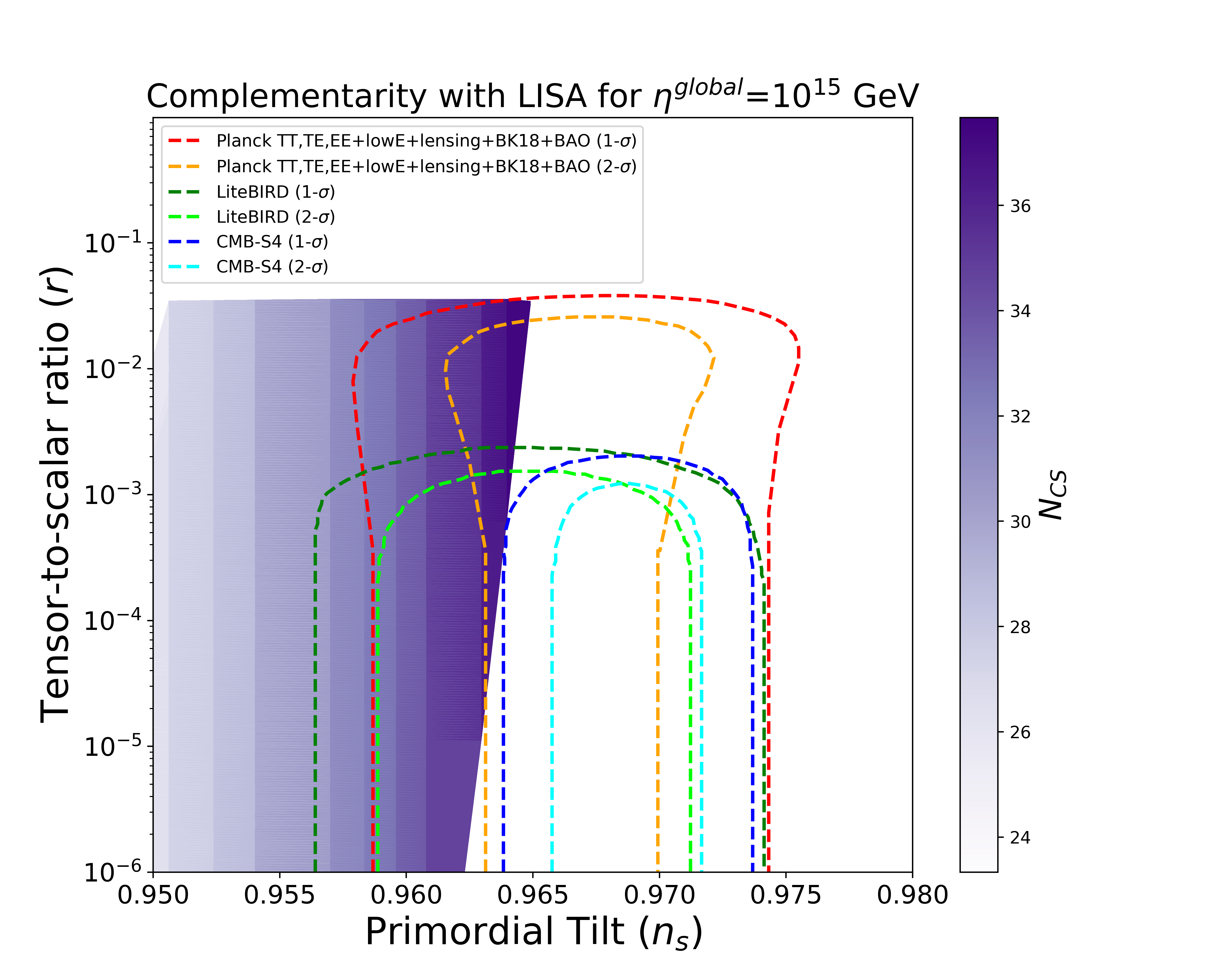}
        \label{fig: LISA Global Complement P01}
    \end{subfigure}

     \begin{subfigure}{0.4962\textwidth} 
        \centering
        \includegraphics[width=\linewidth]{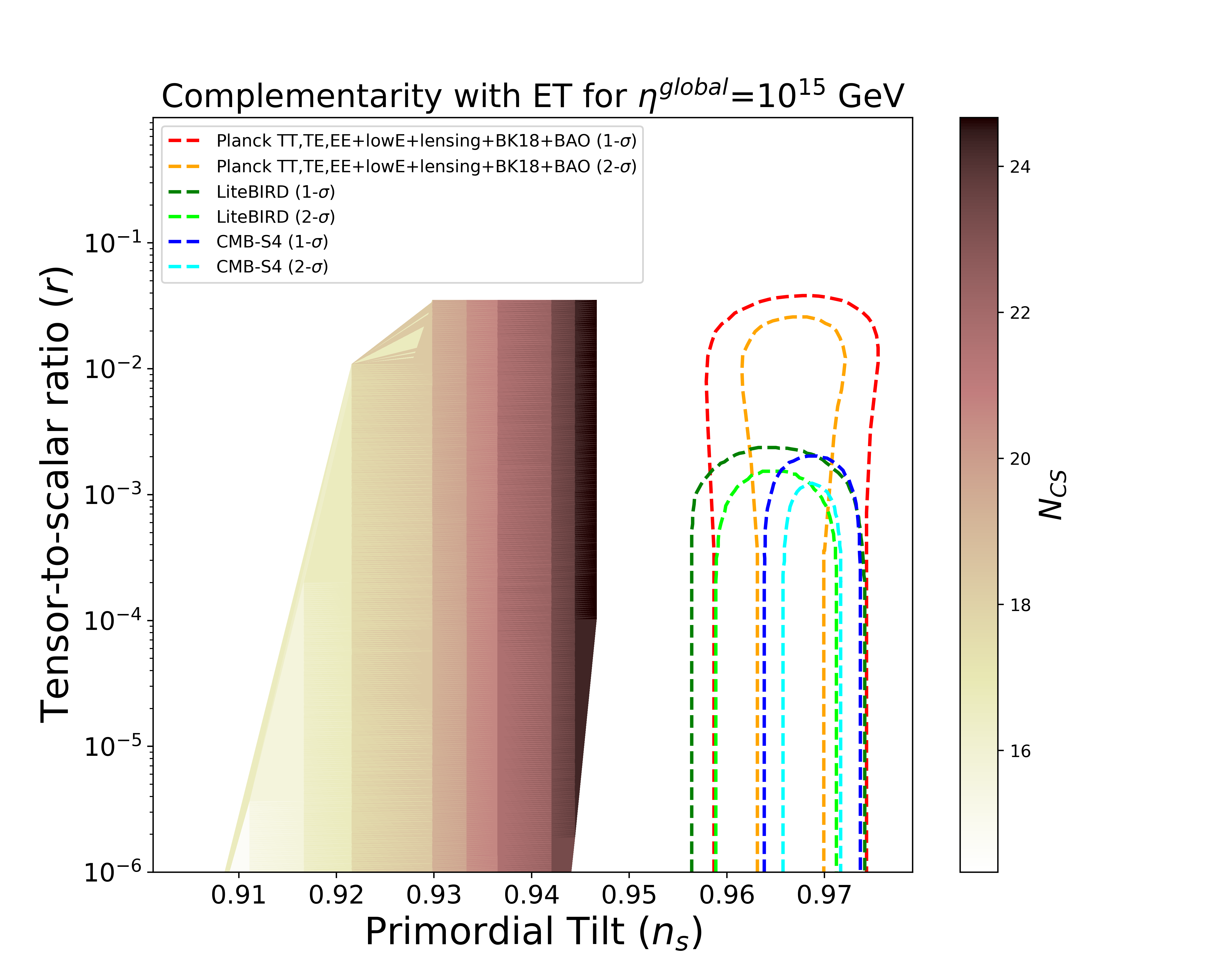}
        \label{fig: ET Global Complement P01}
    \end{subfigure}
    \hfill
    \hfill
     \begin{subfigure}{0.4962\textwidth} 
        \centering
        \includegraphics[width=\linewidth]{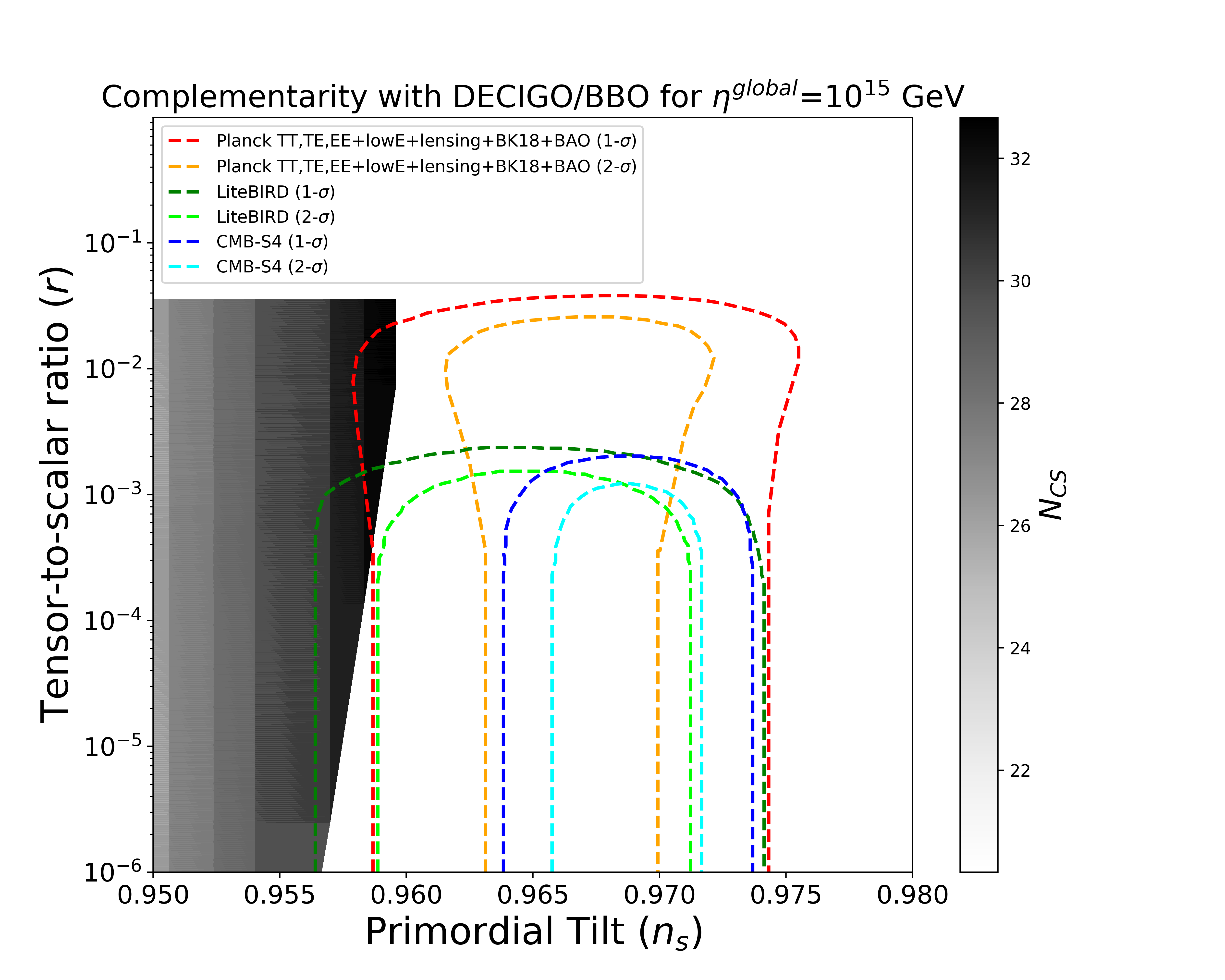}
        \label{fig: Decigo Global Complement P01}
    \end{subfigure}

    \caption{\it Complementarity between CMB and GW detectors (SKA, ET, LISA and DECIGO/BBO) in probing global cosmic strings having scalar field VEV at $\eta^{global}=10^{15}$ GeV. We have considered inflation to be driven by the Polynomial attractor (Eq. \ref{Poly Eq1}) with $n=1$ and $N_{total}=65$ for all the plots. } 
    \label{Global Complement GW 1}
\end{figure}

\begin{figure}[H] 
    \centering
    \begin{subfigure}{0.4962\textwidth} 
        \centering
        \includegraphics[width=\linewidth]{Plots/Fig34.png}
        \label{fig: SKA Global Complement P1}
    \end{subfigure}
    \hfill
    \hfill
    \begin{subfigure}{0.4962\textwidth} 
        \centering
        \includegraphics[width=\linewidth]{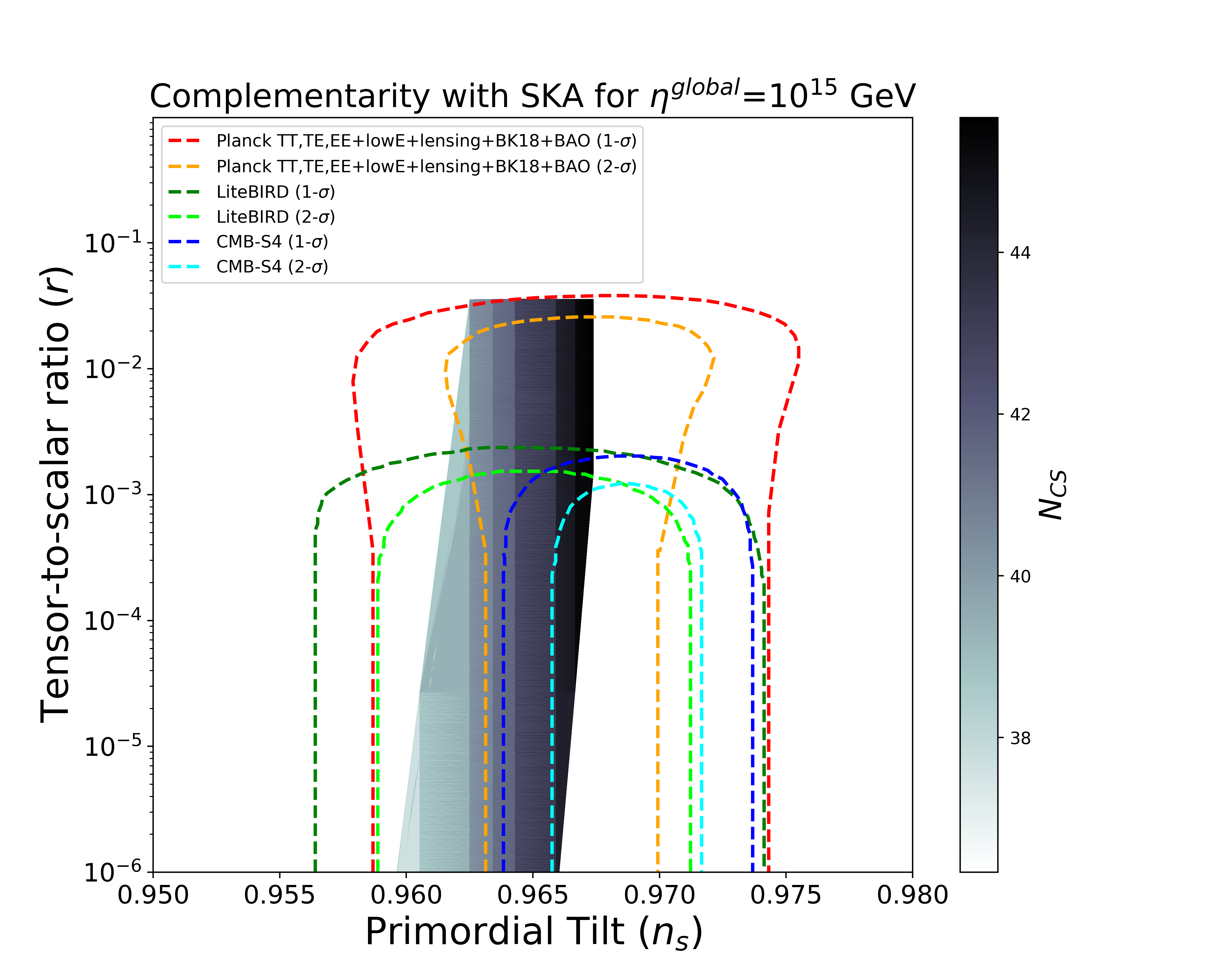}
        \label{fig: SKA Global Complement P2}
    \end{subfigure}

     \begin{subfigure}{0.4962\textwidth} 
        \centering
        \includegraphics[width=\linewidth]{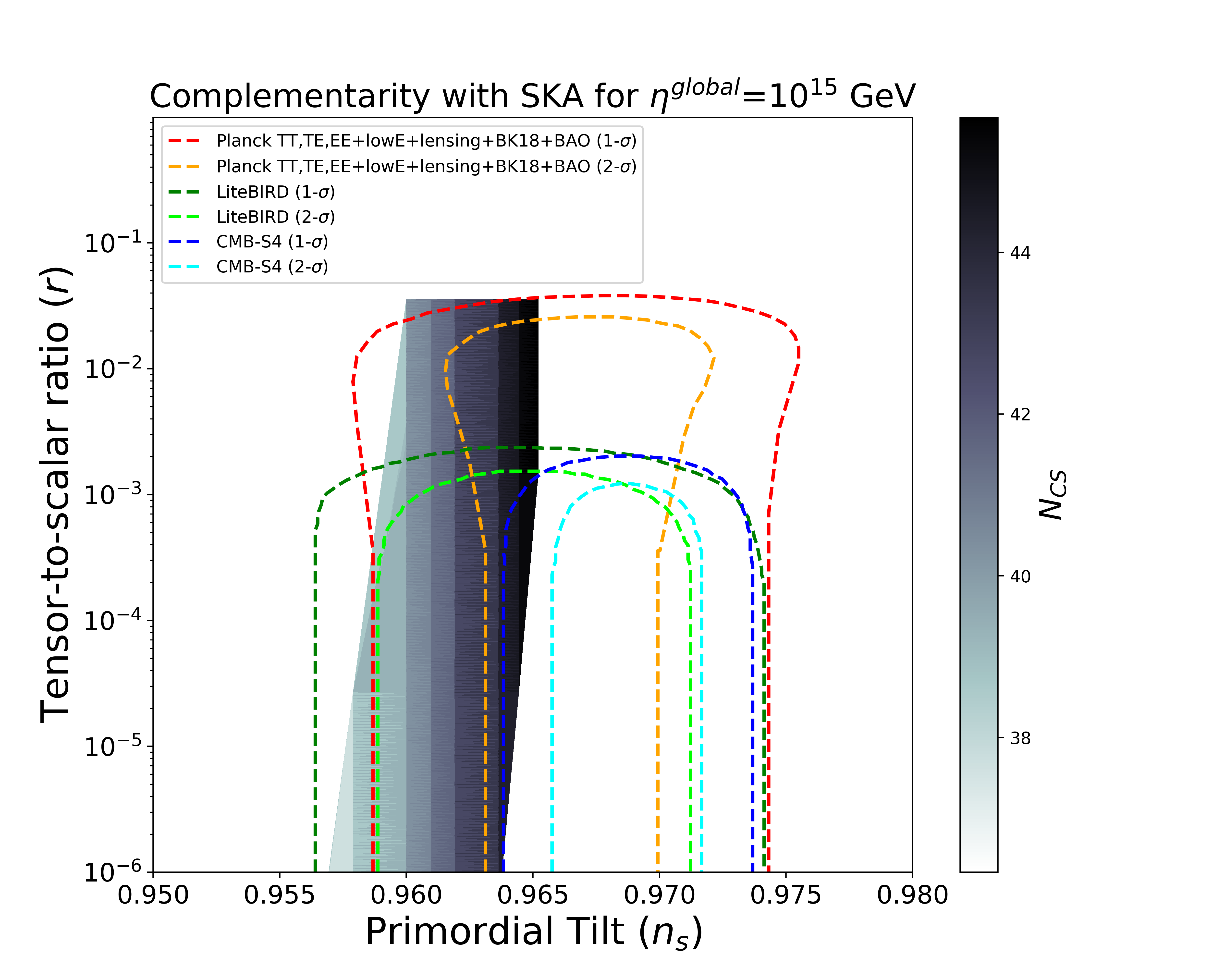}
        \label{fig: SKA Global Complement P3}
    \end{subfigure}
    \hfill
    \hfill
     \begin{subfigure}{0.4962\textwidth} 
        \centering
        \includegraphics[width=\linewidth]{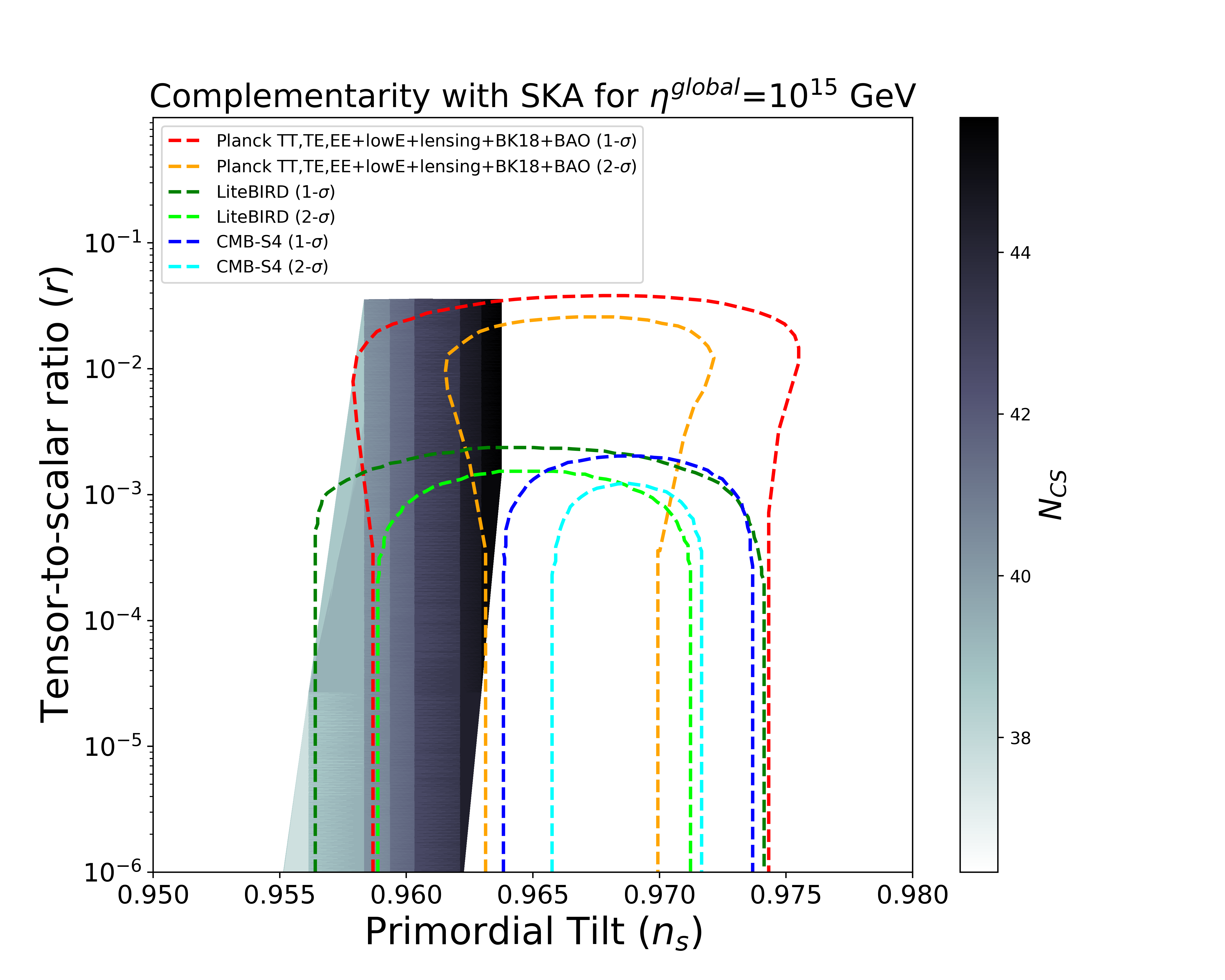}
        \label{fig: SKA Global Complement P4}
    \end{subfigure}

    \caption{\it Impact on the complementarity between CMB and SKA in probing global cosmic strings having scalar field vev  at $\eta^{global}=10^{15}$ GeV for different values of the exponent n of the Polynomial Attractor Model (Eq. \ref{Poly Eq1}). We have $n=1$ (top-left), $n=2$ (top-right), $n=3$ (bottom-left) and $n=4$ (bottom-right). $N_{total}$ is fixed at 65 e-folds.}
    \label{Global Complement GW 2}
\end{figure}

Analogous to our approach to finding the complementarity relation in local strings, we use Eqs. (\ref{global turn freq}), (\ref{re-entry temp}) and (\ref{einf to r}) to find the complementarity relation for global strings:
\begin{equation}\label{complementarity global}
    r = 3.20 \times 10^{-48} \times (f_{\Delta}\cdot \exp N_{CS})^4 
\end{equation}
$f_{\Delta}$ is in Hertz and $N_{CS}$ is the number of e-folds of inflation that the cosmic string has suffered. Interestingly enough, the complementarity relation for global strings is independent of the string tension. 
The shaded region in Figs. \ref{Global Complement GW 1} and \ref{Global Complement GW 2}, just like in the case for local strings, correspond to those combinations of $n_s$ and $r$ that can produce detectable GW signals from Global CS (formed during inflation) at the current and future detectors. This mapping from the ($n_s$, $r$)-plane to the observable GW signal solely depends on the dynamics of inflation and the e-folds experienced by the strings ($N_{CS}$), making it a more effective probe of inflationary physics, as it depends on fewer parameters. Consider the top left plot from Fig. \ref{Global Complement GW 1} which is based on the Polynomial attractor inflation model (see Eq. (\ref{Poly Eq1})) with $n=1$. It maps the $n_s-r$ parameter space to the observable GW signature at the SKA, assuming the VEV $\eta^{global}=10^{15}$ GeV. The gradient of the shaded region tells us that the SKA can probe $N_{CS}$ in the range $\sim 36$ to $46$. Similarly, from the other plots from the same figure, we can infer that LISA can probe $N_{CS}$ in the range $\sim 24$ to $37$, DECIGO/BBO in the range $\sim 20$ to $32$ and ET in the range $\sim 14$ to $24$. However, the parameter space of ET in our analysis is far beyond the reach of the current and future CMB experiments and hence can not be detected by them. But, if we can see observable GW signatures or in other words if we observe the turning point frequency of global strings with $\eta^{global}=10^{15}$ GeV at the ET, it would allow us to put new constraints on the $n_s-r$ parameter space. 

\section{Discussion and Conclusion} \label{Sec. 6}

In summary, we studied the impact of cosmic string formation during primordial inflation. We showed that due to the peculiar property of the cosmic strings not getting diluted away even with a period of inflation, they regrow to reach scaling and leave signatures on SGWB produced with specific primordial features which can be detectable in upcoming GW experiments. From such measurements, we were able to show that any given any model of inflation, the cosmological scenarios can be tested in a complementary manner with GW and CMB detectors in synergy, depending on the e-folds of inflation that CS experiences. Studying $\alpha-$attractor models of inflation, we identified the parameter space currently ruled out from Planck data and which will be within the reach of upcoming GW and CMB detectors. The main highlights of our analysis are as follows:

\begin{itemize}
    \item 
    The T-Model and Polynomial attractor models of inflation have similar $n_s-r$ predictions in the CMB detectors, especially in the large $N_e$ and small $\alpha$ limit. However, we showed that this degeneracy can be broken if cosmic strings are formed during inflation and suffer a few e-folds of inflation. We use this information to create our inflation model and determine the estimated GW spectrum from the cosmic strings (see Eqs. (\ref{complementarity local}) and (\ref{complementarity global}) and Figs. \ref{fig:1e-11 alpha10 n1 N70} to \ref{fig: Vary etas for GlobalStrings}.

    \item 
    Local or global Cosmic strings experiencing certain e-folds of inflation can potentially be detected in upcoming detectors such as LISA and ET, as shown in Figs. \ref{fig:1e-11 alpha10 n1 N70}–\ref{fig: Vary etas for GlobalStrings}. However, there is a limit to the maximum number of e-folds of inflation they can undergo, so that we can have observable GW signatures at the detectors. This number is controlled primarily by the string tension $G\mu$ determined by the vev of the scalar field $\eta^{local}$ and the energy scale of inflation or the tensor-to-scalar ratio $r$. Roughly, the maximum value of $N_{CS}$ decreases as the string tension is reduced (see the GW spectra in Fig. \ref{fig: Vary Gmus}). The maximum value of $N_{CS}$ for observable GW signals also decreases as we decrease the energy scale of inflation (see the GW spectra in Figs. \ref{fig:1e-11 alpha10 n1 N70}–\ref{fig:1e-11 alpha1e-4 n1 N70}). 
    

    \item The GW spectrum from local cosmic strings formed during inflation sees a departure from the flat plateau behaviour (the flat plateau can be seen in Fig. \ref{fig: Standard Local Spectrum} in the high-frequency regime). The extent of this deviation (see Figs. \ref{fig:1e-11 alpha10 n1 N70}–\ref{fig: Vary Gmus}) depends on the number of e-folds of inflation $N_{CS}$ that the strings experience: the greater the value of $N_{CS}$, the shorter such duration, the flat plateau region becomes, which is determined by the turning point frequency given by Eq. (\ref{turn freq}). 
    
    \item 
    The GW spectrum due to global cosmic string is suppressed as compared to local cosmic string due to a shorter global string loop lifetime. Fig. \ref{fig: Global Standard Spectrum} gives the full numerical spectrum due to global strings. Fig. \ref{fig: Vary etas for GlobalStrings} shows the GW spectrum for global strings that were formed during inflation. The turning point frequency given by Eq. \ref{global turn freq} for global strings, is independent of the string tension or the vev $\eta^{global}$, as compared to the turning point frequency from local strings given in Eq. \ref{turn freq}.

    \item 
    Eq. (\ref{complementarity local}) gives the complementary relation for local cosmic strings which depends on the turning point frequency $f_{\Delta}$, the number of e-folds of inflation the local cosmic strings suffer $N_{CS}$ and the string tension $G\mu$ determined by the vacuum expectation value of the scalar field $\eta^{local}$. Figs. \ref{fig:Complementarity_CMB_GW 1}-\ref{fig:Complementarity_CMB_GW 4} show the complementarity plots which form a bridge between inflationary observables and GW signatures. In addition to putting bounds on the $N_{CS}$ values, one can also use this method to impose constraints on the permissible $n_s-r$ values that the inflation model can take. For instance, LISA can probe $N_{CS} \sim 34 - 47$ irrespective of the value of the local string tension. For the Polynomial Attractor model, consider the complementary plots in Fig.\ref{fig:Complementarity_CMB_GW 4}: it allows us to constrain the spectral index $n_s$ within the range $0.962 \lesssim n_s \lesssim 0.972$ for $n=1$, $0.956 \lesssim n_s \lesssim 0.968$ for $n=2$, $0.954 \lesssim n_s \lesssim 0.965$ for $n=3$ and $0.963 \lesssim n_s \lesssim 0.964$ for $n=4$.


    \item Eq. (\ref{complementarity global}) gives the complementary relation for global cosmic strings and depends only on the turning point frequency $f_{\Delta}$ and the number of e-folds of inflation the local cosmic strings experience $N_{CS}$. Figs. \ref{Global Complement GW 1} and \ref{Global Complement GW 2} show the complementarity plots due to global cosmic strings. 
    Besides putting constraints on the values of $N_{CS}$ for the different GW detectors, using this technique (just like the case of local cosmic strings) we can also put constraints and probe regions of the $n_s-r$ parameter space that are within the reach of the current and future CMB experiments. As an example the top right plot in Fig. \ref{Global Complement GW 1} which shows the complementarity for global cosmic strings at the LISA, constrains the spectral index $n_s$ in the range $0.960 \lesssim n_s \lesssim 0.963$ for the Polynomial Attractor model with $n=1$. A part of the aforementioned range of spectral index (corresponding to $N_{CS} \sim 30-37$) is consistent with Planck and lies well within the reach of next generation  CMB experiments like CMB-S4 and LiteBIRD. 
\end{itemize}

If the features predicted in this paper are observed in the GW spectrum from cosmic strings, as well as with additional observations from CMB, we will be able to distinguish between various inflationary models that otherwise have degeneracies in the CMB predictions. It is worth noting that GW signals from cosmic strings together with CMB measurements may generate other signals that could be explored in the future, for instance dark radiation ($\Delta N_{\rm eff}$). Other indicators of the inflationary cosmology are predictions of non-gaussianity (see Ref.~\cite{Meerburg:2019qqi} for a recent review). These forms of various detection channels of the same new physics if executed will provide us with very valuable independent confirmations of our results. Moreover, this will also offer a unique and unprecedented opportunity for synergies between GW searches, CMB and large-scale structure and perhaps even understanding of cross-correlation among different length scales of the universe.

The detection of GW has opened up a new avenue for studying the early universe, which is complementary to other methods. Our analysis shows that cosmic strings can serve as excellent yardstick for probing the cosmic inflationary scenarios involving formation of such defects, should they exist. International GW detector networks planned for the future may allow us to explore such possibilities, which occur at the time of the Big Bang.

\section*{Acknowledgments}
Authors thank Anshuman Maharana and Francesco Muia for their invaluable guidance throughout this work and Ivonne Zavala, Gonzalo Villa, Koushik Dutta and Ahmad Moursy for comments on the manuscript. MB acknowledges the hospitality and support provided by the Visiting Students Program at the Harish-Chandra Research Institute (HRI), which made this research possible. MB also extends heartfelt thanks to Ivonne Zavala, Dibya Chakraborty and Swagat Mishra for insightful discussions and to Barnali Das, Nakshatra Gangopadhay, and Saikat Bera for their insightful inputs on coding and computational aspects during the initial stages of the project. DFM thanks the Research Council of Norway for their support and the resources provided by UNINETT Sigma2 -- the National Infrastructure for High-Performance Computing and Data Storage in Norway.

\section{Appendix: Short Review on Gravitation Waves from Cosmic Strings}
\label{Sec. 3}


In the standard Hot Big Bang Model of cosmology, the universe initially experiences a period of exponential expansion called inflation, followed by an era of radiation domination, then matter domination and more recently it has entered a stage of dark-energy domination. Support for this framework comes mainly from the CMB observations and the accurate predictions of the Big Bang Nucleosynthesis (BBN) but using these we cannot probe cosmic temperatures above $\mathcal{O}(\text{MeV})$.

This is where Primordial Gravitational Waves (PGWs) come to the rescue. Due to the weakness of gravity, GWs decouple from the rest of matter and radiation
components in the universe, upon production \cite{Caprini:2018mtu}. One may do a quantitative estimate of the decoupling by comparing the interaction rate of the gravitons $\Gamma_{int}(T)$ to the expansion rate $H(T)$ \cite{Caprini:2018mtu}:
\begin{equation}
    \left(\frac{\Gamma_{int}(T)}{H(T)}\right)_{\text{graviton}}\sim\frac{G^2T^5}{T^2/M_{Pl}}=\left(\frac T{M_{Pl}}\right)^3,
\end{equation}
where $M_{P l}$ is the Planck mass, $G=1 / M_{P l}^2$ is the Newton constant, $H(T) \sim T^2 / M_{P l}$ is the Hubble parameter during the radiation dominated era, and $\Gamma_{int}(T)=n \sigma v$, is the interaction rate of the processes that maintain equilibrium, where number density of particles $n \sim T^3$, cross-section $\sigma \sim G^2 T^2$ and $v \sim 1$. The interaction process in question remains in thermal equilibrium if $\Gamma_{\mathrm{int}}\gg H$. If, however, $\Gamma_{\mathrm{int}}<H$ then the graviton decouples from the thermal bath and the corresponding particle species is said to freeze out. This quantitative estimate shows that the GW interaction rate is lower than the Hubble parameter, essentially at any temperature in the universe $T<M_{P l}$. Thus, GWs propagate freely in the early universe, immediately after they are generated, which enables them to carry unique information about the processes that produced them, and therefore about the state of the universe at epochs and energy scales unreachable by any terrestrial experiments.

There exists a number of cosmological sources that contribute to the primordial GW background, but of particular interest to us are the cosmic strings (CS), which act as a long-lasting source of GWs from the time of their production, presumably very early on, until today. The resulting frequency spectrum therefore encodes information from the almost entire cosmic history of our universe, and could possibly reveal precious details about the physics of the primordial universe that are far beyond the reach of presently available observational probes of the universe, mostly based on electromagnetic and neutrino emissions.

A phase transition in the early universe corresponds to a process of spontaneous symmetry-breaking from a symmetric phase (false vacuum) to a broken phase (true vacuum). This is typically driven by scalar field(s) acquiring a non-zero vacuum expectation value within a certain vacuum manifold $\mathcal{M}$. If $\mathcal{M}$ satisfies certain conditions, cosmic defects may be produced as remnants of a phase transition \cite{Vilenkin:2000jqa, Hindmarsh:1994re, Kibble:1976sj}. In particular, if the vacuum manifold has a nontrivial homotopy group $\pi_n(\mathcal{M}) \neq \mathcal{I}$, defects may arise in such manifolds. Domain walls arise when the vacuum manifold has disconnected components or $\pi_0(\mathcal{M})\neq \mathcal{I}$. The existence of non-shrinkable or non-contractible loops or a non-trivial first homotopy group, $\pi_1(\mathcal{M})\neq \mathcal{I}$, indicates the presence of cosmic strings in the vacuum manifold. $\pi_2(\mathcal{M})\neq \mathcal{I}$ denotes the presence of monopoles, and $\pi_3(\mathcal{M})\neq \mathcal{I}$ implies the existence of textures. Table \ref{table 1} summarises all possible defects that may occur when the vacuum manifold is topologically nontrivial. 

\begin{table}[!h] 
\centering
\caption{The topology of the vacuum manifold $\mathcal{M}$ determines the nature of the topological defect.}
\begin{tabular}{lllll}
\hline
$\pi_0(\mathcal{M}) \neq \mathcal{I}$ & $\mathcal{M}$ disconnected & Domain Walls \\
$\pi_1(\mathcal{M}) \neq \mathcal{I}$ & non-contractible loops in $\mathcal{M}$ & Cosmic Strings \\
$\pi_2(\mathcal{M}) \neq \mathcal{I}$ & non-contractible 2-spheres in $\mathcal{M}$ & Monopoles \\
$\pi_3(\mathcal{M}) \neq \mathcal{I}$ & non-contractible 3-spheres in $\mathcal{M}$ & Textures \\
\hline
\end{tabular}
\label{table 1}
\end{table}

\subsection{The gravitational-wave spectrum from Local Cosmic Strings} 

Cosmic Strings (CS) may originate as fundamental objects in String Theory \cite{WITTEN1985243, Copeland:2009ga, Polchinski:2004ia} as well as topological defects in field theory during a symmetry-breaking phase transition. We will focus on cosmic strings that were formed in the early universe owing to the spontaneous breaking of a local or global $U(1)$ symmetry, which are also known as field-theory strings. The simplest model in which cosmic strings can arise is a $U(1)$ local gauge or global theory in the context of a complex scalar field $\varphi$. The Lagrangian density for such a theory is given by, 
\begin{equation}
    \mathcal{L}=D_\mu\varphi^*D^\mu\varphi-\frac{1}{4}F_{\mu\nu}F^{\mu\nu}-V(\varphi),
\end{equation}
where $D_\mu\varphi=\partial_\mu\varphi+\mathrm{i}eA_\mu\varphi $ is the covariant derivative and $F_{\mu\nu}=\partial_\mu A_\nu-\partial_\nu A_\mu $ is the field tensor. 
$V(\varphi)$ is the Mexican hat potential given by,
\begin{equation}
    V(\varphi)=\frac{1}{2}\lambda(\varphi^*\varphi-\frac{1}{2}\eta^2)^2,
\end{equation}
where $\lambda$ is the self-quartic coupling and $\eta$ is the vacuum expectation value (VEV) of the scalar field. 
Cosmic strings are essentially field configurations at the top of this Mexican hat potential. The CS has a negligible thickness (with widths of order $1/(\sqrt{\lambda}\eta$)), and hence, their dynamics in an expanding universe are well described by the Nambu-Goto action, a zero-width approximation. The mass per unit length $\mu$ of the local CS is given by \cite{Gouttenoire:2019kij}, 
\begin{equation}
    \mu=2\pi n \eta^2,
\end{equation}
where $\eta \equiv \eta^{local}$ is the vacuum expectation value (VEV) of the scalar field constituting the local CS, and n is the (integer-valued) winding number of the string. Only $n=1$ is considered to be a stable configuration \cite{PhysRevLett.62.1948} \cite{Ghoshal:2023sfa}. 
The local CS network starts forming at, 
\begin{equation}\label{local CS network form}
    T_{\mathrm{CS}}\simeq\eta^{local}\simeq10^{11} \mathrm{GeV}\left(\frac{G\mu}{10^{-15}}\right)^{1/2},
\end{equation}
which is the temperature of the $U(1)$-breaking phase transition. The dimensionless parameter $G\mu$ is called the string tension. 


Straight infinitely long strings are stable against decay owing to their topological nature and hence have very little contribution to the GW spectra compared to cosmic string loops. Local CS loops, formed by intercommutation events of CS, are free of any topological charge and contribute significantly to GW emission \cite{PhysRevD.31.3052, PhysRevD.45.1898, Vilenkin:2000jqa}. The GW radiation power for CS loops is given by,
\begin{equation}
    P_{\mathrm{GW}}=\Gamma G\mu^2,
\end{equation}
where $\Gamma$ is the total GW emission efficiency of loops and is determined from the Nambu-Goto simulation and found to be $\Gamma\simeq70$ \cite{BLANCOPILLADO2018392}.

The GW spectrum from Local CS Loops observed today reads as follows \cite{Gouttenoire:2019kij}:
\begin{equation}
    \Omega_{\mathrm{GW}}(f) \equiv \frac{f}{\rho_c}\left|\frac{d \rho_{\mathrm{GW}}}{d f}\right|=\sum_k \Omega_{\mathrm{GW}}^{(k)}(f),
\end{equation}

where

\begin{equation}\label{Local GW1}
    \begin{aligned}\Omega_{\mathrm{GW}}(f)=\sum_{k}\frac{1}{\rho_{c}}\int_{t_{\mathrm{osc}}}^{t_{0}}d\tilde{t}\int_{0}^{1}d\alpha\cdot \Theta\left[t_{i}-\frac{l_{*}}{\alpha}\right]\cdot\Theta[t_{i}-t_{\mathrm{osc}}]\cdot\left[\frac{a(\tilde{t})}{a(t_{0})}\right]^{4}\cdot P_{\mathrm{GW}}^{(k)}\times\\\times\left[\frac{a(t_{i})}{a(\tilde{t})}\right]^{3}\cdot\mathcal{P}_{\mathrm{loop}}(\alpha)\cdot\frac{dt_{i}}{df}\cdot\frac{dn_{\mathrm{loop}}}{dt_{i}}\end{aligned}
\end{equation}

The first Heaviside function $\Theta\left[t_{i}-\frac{l_{*}}{\alpha}\right]$ implies that loops smaller than the critical length $l_{*}$ will not contribute to the GW spectrum because their main decay channel is through massive particle production. $l_* \equiv l_c$ for cusps, and $l_k$ for kinks:

\begin{equation}
    l_c \equiv \beta_c \: \frac{\mu^{-1/2}}{(\Gamma G \mu)^2}; \quad
    l_k \equiv \beta_k \: \frac{\mu^{-1/2}}{\Gamma G \mu},
    \label{eq:cusp_kink_lengths}
\end{equation}

where $\beta_c$ and $\beta_k$ are $\mathcal{O}(1)$ numbers. The second Heaviside function $\Theta[t_{i}-t_{\mathrm{osc}}]$ eliminates all those loops that were formed before the network formation (given by Eq. (\ref{local CS network form})) or which form during the friction dominated epoch, 
\begin{equation}
    T \lesssim T_{\mathrm{fric}}=\frac{4\times10^9 \mathrm{GeV}}{\beta}\left(\frac{g_*}{100}\right)^{1/2}\left(\frac{G\mu}{10^{-11}}\right),
\end{equation}
and $t_{\mathrm{osc}}=\operatorname{Max}\left[t_{fric}, t_F\right]$, where $t_F$ is the time of CS network formation, defined as $\sqrt{\rho_{tot}\left(t_F\right)} \equiv \mu$ where $\rho_{tot}$ is the total energy density of the universe. In the presence of friction, the string motion is damped at a high temperature until time $t_{fric}$.
$dn_{\mathrm{loop}}/dt_i$ denotes the rate of loop formation with distribution size $\mathcal{P}_{\mathrm{loop}}(\alpha)$. They initially redshift as $a^{-3}$ before radiating GW with power $P_{\mathrm{GW}}^{(k)}$ which dilute as $a^{-4}$. Using these simplifications, we can write Eq. (\ref{Local GW1}) as \cite{Gouttenoire:2019kij}:

\begin{equation}\label{Local GW2}
    \Omega_{\mathrm{GW}}^{(k)}(f)=\frac{1}{\rho_{c}}\cdot\frac{2k}{f}\cdot\frac{\mathcal{F}_{\alpha}}{\alpha(\alpha+\Gamma G\mu)}\int_{t_{\mathrm{osc}}}^{t_{0}}d\tilde{t} \frac{C_{\mathrm{eff}}(t_{i})}{t_{i}^{4}}\left[\frac{a(\tilde{t})}{a(t_{0})}\right]^{5}\left[\frac{a(t_{i})}{a(\tilde{t})}\right]^{3}\Theta(t_{i}-t_{\mathrm{osc}})\Theta(t_{i}-\frac{l_{*}}{\alpha}),
\end{equation}

$\alpha$ is the loop length at the time of formation and $\mathcal{F}_{\alpha}$ denotes the fraction of loops that forms with size $\alpha$. $t_i$ is the time of loop production which is dependent on the emission time $\tilde{t}$ and the observed frequency today, 
\begin{equation}
    t_i(f,\tilde{t})=\frac{1}{\alpha+\Gamma G\mu}\left[\frac{2k}{f}\frac{a(\tilde{t})}{a(t_0)}+\Gamma G\mu \tilde{t}\right],
\end{equation}
where $t_0$ is the present time. $C_{eff}(t_i)$ is the loop-production efficiency given by,
\begin{equation}
    \tilde C_{eff}\equiv\sqrt{2} C_{eff}(t)=\frac{\tilde{c} \bar{v}(t)}{\xi^3(t)},
\end{equation}
where $\tilde{c}=0.23\pm0.04$ is a phenomenological parameter quantifying the loop chopping efficiency. $\bar{v}$ is the root-mean-square speed of the string loops and $\xi$ is defined as $\xi\equiv L/t$, where $L$ is the correlation length of the string.

\begin{figure}[!h]
    \centering
    \includegraphics[width=\linewidth]{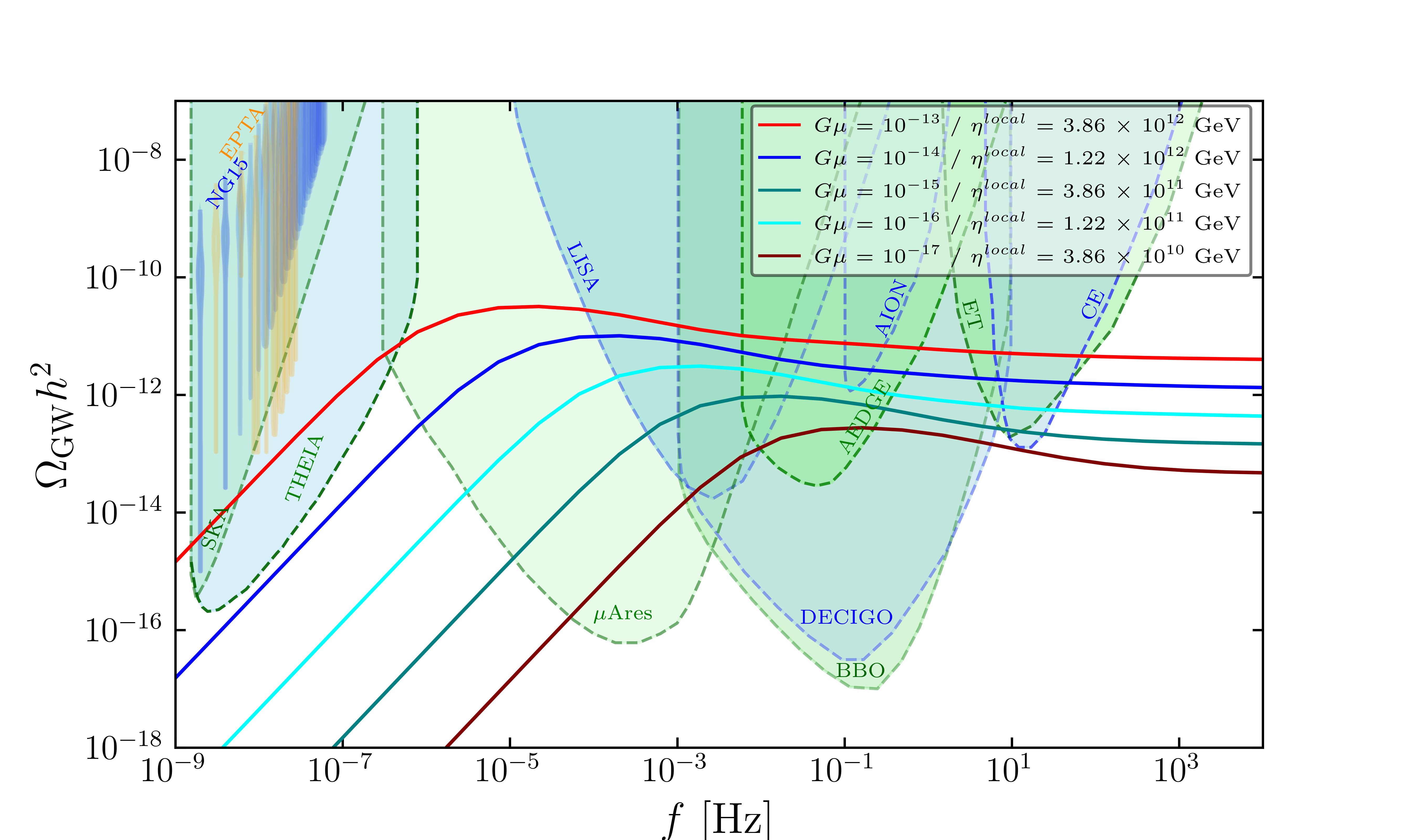}
    \caption{\it Variation of the SGWB spectra from local cosmic strings with respect to the string tension G$\mu$ assuming they were formed after inflation and estimated using Eq. (\ref{Local GW2}). $\eta^{local}$ denotes the VEV of the scalar field constituting the local cosmic string.}
    \label{fig: Standard Local Spectrum}
\end{figure}

Figure \ref{fig: Standard Local Spectrum} shows the full GW spectrum \footnote{We have used the publicly available code \url{https://github.com/moinulrahat/CosmicStringGW} \cite{Fu:2024rsm} for plotting GW spectrum.} due to local CS computed using Eq. (\ref{Local GW2}) assuming a $\Lambda$CDM universe and the strings were formed after inflation ends. The spectrum can be broadly classified into a flat high-frequency regime and the radiation-to-matter transition at the low-frequency regime. Higher and higher frequencies correspond to earlier and earlier in time scales, i.e., it corresponds to a radiation-dominated era. The local CS loops that were produced during the radiation-dominated epoch and begun GW emission have a nearly flat GW spectrum, with the GW amplitude being expressed as,
\begin{equation}
    \Omega_{\mathrm{std}}^{\mathrm{CS}}h^2\simeq\Omega_rh^2\mathcal{G}(\tilde{T}_{\mathrm{M}})\left(\frac{\eta}{M_{\mathrm{pl}}}\right),
\end{equation}
where $\Omega_r h^2 \simeq 4.2 \times 10^{-5}$ \cite{ParticleDataGroup:2020ssz}. $\tilde{T_M}$ is the temperature of the universe at the time of maximum GW emission $\tilde{t_M}$,
\begin{equation}
    \tilde{t_M}=\frac{\alpha}{2\Gamma G\mu} t_i.
\end{equation}
The deviation from the nearly flat spectrum happens due to a change in the number of relativistic degrees of freedom captured in,

\begin{equation}
\mathcal{G}(T) \equiv\left(\frac{g_*(T)}{g_*\left(T_0\right)}\right)\left(\frac{g_* s\left(T_0\right)}{g_{* s}(T)}\right)^{4 / 3}=0.39\left(\frac{106.75}{g_*(T)}\right)^{1 / 3}.
\end{equation}

\subsection{The gravitational-wave spectrum from Global Cosmic Strings}


The main feature that distinguishes global CS from local CS is their coupling to a Goldstone boson field that gives rise to long-range interactions \cite{Vilenkin:2000jqa}. The dynamical differences between global and local CS are not great on cosmological scales, except for the fact that global CS can decay into Goldstone bosons. Just like the local CS, the global CS will also obey the Nambu-Goto equations of motion. The Global CS network is formed below the temperature of the U(1)-breaking phase transition,
\begin{equation}
    T_{form}^g \simeq \eta.
\end{equation}

The GW spectrum due to global string loops is given by \cite{Gouttenoire:2019kij},

\begin{equation}
    \Omega_{\mathrm{GW}}^{\mathrm{g}}(f)\equiv\frac{f}{\rho_{c}}\left|\frac{d\rho_{\mathrm{GW}}^{\mathrm{g}}}{df}\right|=\sum_{k}\Omega_{\mathrm{GW}}^{(k),\mathrm{g}}(f),
\end{equation}
where,
\begin{equation}\label{global cs standard}
    \Omega_{\mathrm{GW}}^{(k), \mathrm{g}}(f)=\frac{1}{\rho_{c}}\cdot\frac{2k}{f}\cdot\frac{\mathcal{F}_{\alpha} \Gamma^{(k)}G\mu_{\mathrm{g}}^{2}}{\alpha(\alpha+\Gamma G\mu_{\mathrm{g}}+\kappa)}\int_{t_{F}}^{t_{0}}d\tilde{t} \frac{C_{\mathrm{eff}}^{\mathrm{g}}(t_{i}^{\mathrm{g}})}{(t_{i}^{\mathrm{g}})^{4}}\left[\frac{a(\tilde{t})}{a(t_{0})}\right]^{5}\left[\frac{a(t_{i}^{\mathrm{g}})}{a(\tilde{t})}\right]^{3}\Theta(t_{i}^{\mathrm{g}}-t_{F}).
\end{equation}
Here, $t_{i}^{\mathrm{g}}$ is the time of formation of global cosmic string loops and is related to the GW emission time $\tilde{t}$ as,
\begin{equation}
    t_i^\mathrm{g}=\frac{\Gamma G\mu_\mathrm{g}+\kappa}{\alpha+\Gamma G\mu_\mathrm{g}+\kappa}\tilde{t},
\end{equation}
where,
\begin{equation}
    \kappa\equiv\frac{\Gamma_{\mathrm{Gold}}}{2\pi \ln\left(\eta t\right)}.
\end{equation}
The loop-production efficiency for global strings, similar to the local strings, is given by,
\begin{equation}
    C_{\mathrm{eff}}^\mathrm{g}=\tilde{c} \bar{v}/\xi^3,
\end{equation}
where $\xi$ is the correlation length of the string. The string tension $\mu_g$ is defined as,
\begin{equation}
    \mu_{\mathrm{g}}\equiv\mu_{1} \ln\left(\frac{H^{-1}}{\delta}\right)\simeq\mu_{1} \ln\left(\eta t\right),\quad\mathrm{with}\quad\mu_{1}\equiv2\pi\eta^{2},
\end{equation}
where $\eta \equiv \eta^{global}$ is the vacuum expectation value of the scalar field and $\delta \sim \eta^{-1}$ is the string thickness. The CMB imposes strong constraints on string tensions,
\begin{equation}
    G\mu_\mathrm{g}(CMB)=2\pi\left(\frac{\eta^{global}}{M_\mathrm{pl}}\right)^2\log(\eta^{global}\cdot t_{CMB})\lesssim10^{-7}
\end{equation}
\begin{equation}
    \implies\quad\eta^{global}\lesssim1.4\times10^{15} GeV,
\end{equation}
where $t_{CMB} = 374$ kyr.

\begin{figure}[!h]
    \centering
    \includegraphics[width=\linewidth]{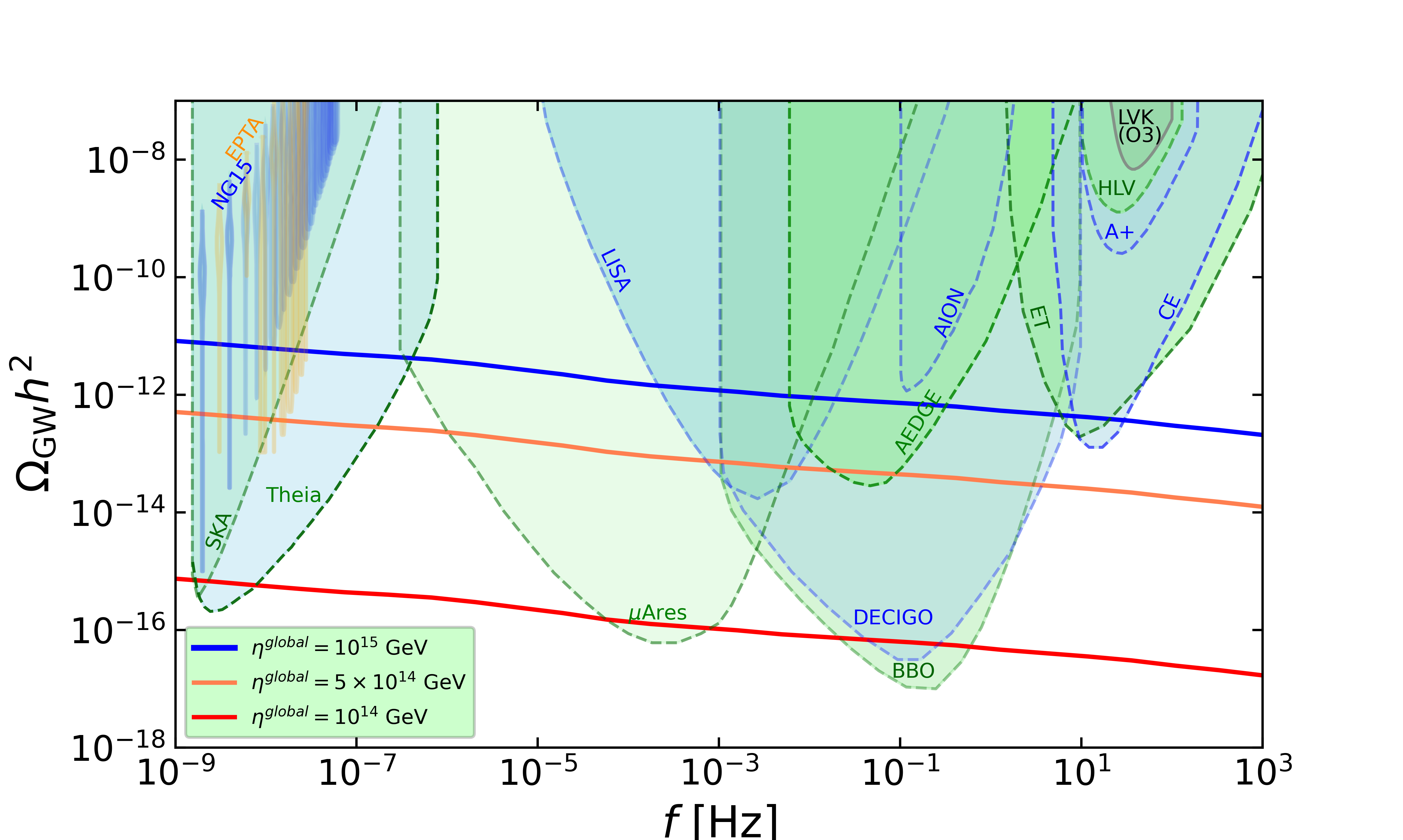}
    \caption{\it GW Spectra from global strings computed using Eq. (\ref{global cs standard}). $\eta^{global}$ denotes the VEV of the scalar field constituting the global cosmic string.}
    \label{fig: Global Standard Spectrum}
\end{figure}

Using Eq. (\ref{global cs standard}) we numerically evaluate the full GW spectra across decades of frequency ranges, arising due  a Global CS as shown in figure \ref{fig: Global Standard Spectrum}. In this case, we considered up to 1000 $k$-modes for the numerical evaluation of this final spectrum. One of the key and significant difference between the cosmology involving the local and global string loop dynamics is that global string loops decay away within $\sim1$ Hubble time after its formation. This arises due to the additional emission of massless (pseudo-)Goldstone bosons besides GW emission. Because of this, the amplitude of the final GW spectrum from a network of global strings is suppressed by a factor of $\frac{\eta}{M_{pl}}$ accounting for the shorter global string loop lifetime. GW emission from the global string loop occurs almost as soon as it is formed, $\tilde{t} \sim t_i$. Moreover, the GW spectrum is also logarithmically enhanced by a factor of $\log^3$ due to greater cosmic string tension in the network that formed,
\begin{equation}
    \Omega_{\mathrm{std}}^{\mathrm{CS}}h^{2}\sim\Omega_{r}h^{2}\mathcal{G}(\tilde{T}_{\mathrm{M}})\left(\frac{\eta^{global}}{M_{\mathrm{pl}}}\right)^{3}\log^{3}\left(\eta^{global}\tilde{t}_{\mathrm{M}}\right).
\end{equation}

A more detailed description of local and global strings can be found in Refs. \cite{Gouttenoire:2019kij, Chang:2021afa, Vilenkin:2000jqa}.


\medskip

\subsection{Gravitational Wave Detectors}
In figures \ref{fig: Standard Local Spectrum} and \ref{fig: Global Standard Spectrum} we have depicted the present day and future planned GW experimental sensitivity reaches of the following categories:
\begin{itemize}

    \item \textbf{Pulsar Timing Arrays (PTAs)}: Square Kilometre Array (SKA) \cite{Carilli:2004nx, Weltman:2018zrl, Janssen:2014dka}, EPTA \cite{EPTA:2015gke, EPTA:2015qep}, NANOGrav \cite{McLaughlin:2013ira, Aggarwal:2018mgp, NANOGRAV:2018hou, NANOGrav:2020bcs}.

    \item \textbf{Recasts of star surveys}: THEIA/GAIA \cite{Garcia-Bellido:2021zgu}

    \item \textbf{Ground based interferometers}: LIGO/VIRGO \cite{LIGOScientific:2016aoc, LIGOScientific:2017bnn, LIGOScientific:2016sjg, LIGOScientific:2017vox, LIGOScientific:2017vwq, LIGOScientific:2017ycc}, aLIGO/aVIRGO \cite{LIGOScientific:2014pky, VIRGO:2014yos, LIGOScientific:2019lzm}, Einstein Telescope (ET) \cite{Punturo:2010zz, Hild:2010id}, Cosmic Explorer (CE) \cite{LIGOScientific:2016wof, Reitze:2019iox}, LIGO-VIRGO-KAGRA (LVK) \cite{LIGOScientific:2022sts, KAGRA:2021kbb}, AION \cite{Badurina:2019hst, Badurina:2021rgt, Graham:2016plp, Graham:2017pmn}.

    \item \textbf{Space based interferometer}: LISA \cite{LISA:2017pwj, Baker:2019nia}, BBO \cite{Crowder:2005nr, Corbin:2005ny}, DECIGO/U-DECIGO \cite{Yagi:2011wg, Kudoh:2005as, Seto:2001qf, Nakayama:2009ce, Kawamura:2020pcg}, $\mu$-Ares \cite{Sesana:2019vho}, AEDGE \cite{Badurina:2019hst, AEDGE:2019nxb}
\end{itemize}
We also show present and future CMB experiments which measures the amount of dark radiation ($\Delta N_{\rm eff}$) which are used, later on in our discussion:
\begin{itemize}
    \item \textbf{CMB polarization}: Planck 2018 \cite{Planck:2018vyg}, BICEP/Keck \cite{BICEP:2021xfz}, LiteBIRD \cite{Hazumi:2019lys}, CMB-S4 \cite{CMB-S4:2016ple}. 
\end{itemize}

\bibliography{Bibliography}
\bibliographystyle{JHEP}

\end{document}